\documentclass[format=acmsmall, review=false, screen=true]{acmart}
\usepackage[T1]{fontenc} 
\usepackage[utf8]{inputenc}
\usepackage{amsmath}
\usepackage{bm} 
\usepackage{graphicx}
\usepackage{tikz}
\usepackage{tcolorbox}
\usepackage{pdflscape}
\usepackage{rotating}
\usepackage{fancyhdr} 
\usepackage{color, colortbl}
\usepackage{longtable}
\usepackage{multirow}
\usepackage{enumitem}
\usepackage{fontawesome}
\usepackage{balance}
\usepackage{tablefootnote}
\usepackage{colortbl}
\tcbuselibrary{skins}
\tcbuselibrary{listingsutf8}
\usepackage[export]{adjustbox}
\usepackage{framed}
\usepackage{arydshln}
\usepackage{ragged2e}
\newcolumntype{M}[1]{>{\centering\arraybackslash}m{#1}}
\newcolumntype{N}{@{}m{0pt}@{}}
\usepackage{hyperref}
\usepackage{booktabs}
\usepackage{tabularx}
\usepackage{array}
\usepackage{ragged2e}

\hypersetup
{
 colorlinks = true, 
 urlcolor   = blue, 
 linkcolor  = blue, 
 citecolor  = blue 
}

\usepackage{graphicx}
\usepackage{svg}
\usepackage{subfig}
\usepackage{array}
\definecolor{mybeige}{HTML}{FFF7F3}
\definecolor{myoffwhite}{HTML}{F1F1F1}
\definecolor{mydarkpurple}{HTML}{49006A}
\definecolor{mypurple}{HTML}{99017B}
\definecolor{mydarkpink}{HTML}{E23E99}
\definecolor{mypink}{HTML}{F767A1}
\definecolor{mypink2}{HTML}{F769A1}
\definecolor{mylightpink}{HTML}{F994B1}
\definecolor{mysalmon}{HTML}{FCC8C3}
\definecolor{mylightsalmon}{HTML}{FBBABD}
\definecolor{lightgray}{rgb}{0.9, 0.9, 0.9}  
\definecolor{skyblue}{rgb}{0.529, 0.808, 0.922}   
\newcommand{\fakesection}[2][1em]{\par\addvspace{#1}\noindent\textit{\textbf{#2}}}
\AtBeginDocument{%
  }

\setcopyright{acmlicensed}
\copyrightyear{2026}
\acmYear{2026}
\acmDOI{XXXXXXX.XXXXXXX}

\acmJournal{TOSEM}
\acmVolume{0}
\acmNumber{0}
\acmArticle{0}
\acmMonth{0}

\newif\ifcleanversion
\cleanversiontrue
\ifcleanversion
    \newcommand{\rev}[1]{#1}
    
    \newcommand{\todo}[1]{}
    
\else
    \newcommand{\rev}[1]{\textcolor{blue}{#1}}
    
    \newcommand{\todo}[1]{{\color{red}\textbf{[TODO: #1]}}}
    
\fi

\begin{document}
\title[Designing LLM-based Multi-Agent Systems for SE Tasks: Quality Attributes, Design Patterns and Rationale]{Designing LLM-based Multi-Agent Systems for Software Engineering Tasks: Quality Attributes, Design Patterns and Rationale}

\author{Yangxiao Cai}
\orcid{0009-0007-7892-6611}
\affiliation{
  \institution{School of Computer Science, Wuhan University}
  \city{Wuhan}
  \country{China}
}
\email{yangxiaocai@whu.edu.cn}

\author{Ruiyin Li}
\orcid{0000-0001-8536-4935}
\affiliation{
  \institution{School of Computer Science, Wuhan University}
  \city{Wuhan}
  \country{China}
}
\email{ryli_cs@whu.edu.cn}

\author{Peng Liang}
\orcid{0000-0002-2056-5346}
\affiliation{
  \institution{School of Computer Science, Wuhan University}
  \city{Wuhan}
  \country{China}
}
\email{liangp@whu.edu.cn}

\author{Mojtaba Shahin}
\orcid{0000-0002-9081-1354}
\affiliation{%
  \institution{School of Computing Technologies, RMIT University}
  \city{Melbourne}
  \country{Australia}
}
\email{mojtaba.shahin@rmit.edu.au}

\author{Zengyang Li}
\orcid{0000-0002-7258-993X}
\affiliation{%
  \institution{School of Computer Science, Central China Normal University}
  \city{Wuhan}
  \country{China}
}
\email{zengyangli@ccnu.edu.cn}

\author{Aldeida Aleti}
\orcid{0000-0002-1716-690X}
\affiliation{%
  \institution{Department of Software Systems \& Cybersecurity, Monash University}
  \city{Melbourne}
  \country{Australia}
}
\email{aldeida.aleti@monash.edu}

\author{Chenxing Zhong}
\orcid{0000-0001-7298-1926}
\affiliation{%
  \institution{School of Computer Science and Engineering, Nanjing University of Science and Technology}
  \city{Nanjing}
  \country{China}
}
\email{chenxingzhong@njust.edu.cn}

\renewcommand{\shortauthors}{Cai et al.}

\begin{abstract}
As the complexity of Software Engineering (SE) tasks continues to escalate, Multi-Agent Systems (MASs) have emerged as a focal point of research and practice due to their autonomy and scalability. Furthermore, through leveraging the reasoning and planning capabilities of Large Language Models (LLMs), the application of LLM-based MASs in the field of SE is garnering increasing attention. However, there is no dedicated study that systematically explores the design of LLM-based MASs, including the Quality Attributes (QAs) on which designers mainly focus, the design patterns used by designers, and the rationale guiding the design of LLM-based MASs for SE tasks. 
To this end, we conducted a study to identify the QAs that LLM-based MASs for SE tasks focus on, the design patterns used in the MASs, and the design rationale for the MASs. We collected 106 papers on LLM-based MASs for SE tasks as the source. Our study shows that: (1) \textit{Code Generation} is the most common SE task solved by LLM-based MASs among ten identified SE tasks, (2) \textit{Functional Suitability} is the QA on which designers of LLM-based MASs pay the most attention, (3) \textit{Role-Based Cooperation} is the design pattern most frequently employed among 16 patterns used to construct LLM-based MASs, and (4) \textit{Improving the Quality of Generated Code} is the most common rationale behind the design of LLM-based MASs. Building on the study results, we presented the implications for the design of LLM-based MASs to support SE tasks.
\end{abstract}

\ccsdesc[500]{Software and its engineering~Designing software} 

\keywords{Multi-Agent, Software Design, Large Language Model, Software Engineering Task}

\maketitle

\section{Introduction}\label{sec:Introduction}
While traditional Multi-Agent Systems (MASs) face limitations in handling complex decision-making tasks \citep{dorri2018agent}, Large Language Models (LLMs) can enhance the reasoning \citep{huang2023acl} and planning capabilities of MASs~\citep{shi2024sigir}. LLM-based MASs consist of multiple autonomous agents that collaborate through communication and responsibility specialization to tackle complex tasks given by users and simulate problem-solving environments \citep{ijcai2024agent}. LLM-based MASs have demonstrated significant potential in addressing Software Engineering (SE) tasks and have become a focal point of research and practice in SE \citep{He2025survey}. Many researchers have designed LLM-based MASs to address specific SE tasks \citep{He2025survey}. For example, Hong et al. \citep{Hong2023metagpt} proposed an LLM-based MAS named MetaGPT, which is capable of automatically developing an entire software system. Jin et al. \citep{Jin2024mare} proposed an LLM-based MAS named MARE (Multi-Agent Collaboration Framework for Requirements Engineering), which could complete the process of requirements engineering. 

Effective design is crucial to the success of LLM-based MASs, as it directly shapes key quality attributes, including reliability, maintainability, scalability, and safety, and profoundly influences how agents collaborate (e.g., volunteering, conformity, and destructive actions), which in turn affect the overall task-solving efficiency among agents in MASs \citep{chen2024agentverse}. Therefore, a principled understanding of design rationale is essential not only for constructing high-quality MASs but also for navigating trade-offs among competing quality attributes (e.g., correctness vs. efficiency) \citep{atam1998}. Despite the increasing interest and adoption of LLM-based MASs for various SE tasks, no existing study has systematically investigated the specific Quality Attributes (QAs), design patterns, and underlying design rationale that inform their construction. To address this gap, our \textbf{goal} is to identify QAs and design patterns that \rev{designers} prioritize when building LLM-based MASs, thereby highlighting best practices and informing future system designs. Moreover, explicitly articulating the rationale behind design choices is essential for ensuring transparency in design decision-making and facilitating the transfer of design knowledge.

\rev{In this study, we conducted an empirical analysis of 106 papers to investigate the design of LLM-based MASs for SE tasks. We examined the SE tasks addressed, the QAs considered, the design patterns employed, and the reported design rationale. Our \textbf{study results} show that, among the included studies, (1) \textit{Code Generation} is the most commonly addressed SE task among the ten identified task categories; (2) \textit{Functional Suitability} is the most frequently considered QA; (3) \textit{Role-Based Cooperation} is the most frequently adopted design pattern among the 16 identified patterns; and (4) \textit{Improving the Quality of Generated Code} is the most commonly reported design rationale.}

\rev{We further examined how quality concerns, pattern usage, and reported design rationale vary across SE tasks, as well as the associations between QAs and design patterns. These analyses informed six implications for designing LLM-based MASs according to the target SE tasks and their quality requirements. The \textbf{main contributions} of this work are as follows:}

\begin{itemize}
    \item \rev{We characterized the SE tasks addressed by LLM-based MASs and the QAs considered in their design.}
    \item \rev{We identified and categorized the design patterns employed to construct LLM-based MASs for SE tasks and the design rationale reported in the included studies.}
    \item \rev{We established mappings among SE tasks, QAs, design patterns, and design rationale, and derived six design implications based on the observed associations.}
\end{itemize}

\textbf{Paper Organization}: Section~\ref{sec:RelatedWork} reviews the related work. Section~\ref{sec:Study Design} presents the Research Questions (RQs) and the research process conducted in this study. Section~\ref{sec:Results} presents the results of this study. Section~\ref{sec:Discussions} interprets the study results and discusses their implications. Section~\ref{sec:Threats} outlines the potential threats to validity, and Section~\ref{sec:Conclusion} concludes this work with future research directions.

\section{Related Work}\label{sec:RelatedWork}
In this section, we first review LLM-based MASs, specifically constructed to address SE tasks (Section~\ref{mas2.1}), and introduce studies that explore the characteristics of LLM-based MASs (Section~\ref{mas2.2}). Then, we present the recent literature surveys that focus on LLM-based MASs (Section~\ref{mas2.3}). We compare recent studies with our work to highlight the research gap (Section~\ref{mas2.4}).

\subsection{LLM-based MASs for SE Tasks}\label{mas2.1}
As LLMs continue to evolve at a rapid pace, SE researchers and \rev{designers} are increasingly eager to employ the reasoning and decision-making abilities of LLMs to support SE tasks. To better exploit the potential of LLMs in complex, multi-step SE workflows (particularly those requiring coordination), LLM-based MASs have been designed to address SE tasks, such as requirements engineering, code generation, and fault localization. 
Arora et al. \citep{arora2023requirement} developed an LLM-based MAS aimed at enhancing the efficiency and accuracy of requirements engineering tasks. This MAS reveals the significant potential of applying LLMs to requirements elicitation, analysis, specification, and validation. 
Li et al. \citep{Li2025MAAD} proposed \textit{MAAD}, a knowledge-driven MAS specifically built for automated architecture design. \textit{MAAD} assigns the architecture design task to four role-dedicated agents with external knowledge bases, and the four agents collaboratively generate architecture design by decomposing a given Software Requirements Specifications (SRS) into corresponding artifacts. 
Dong et al. \citep{dong2024generation} designed an LLM-based MAS based on a self-collaboration framework to achieve high-quality code generation. Their framework enables the MAS to solve complex repository-level tasks that are not readily solved by a single LLM agent. 
Pan et al. \citep{pan2025codecor} proposed a novel optimization pipeline for LLM-based MASs based on self-generated textual feedback, named \textit{CodeCoR}, which has a multi-phase workflow and iterative code repair progress. Its self-reflective mechanism can evaluate and refine the outputs of each agent. 
Some LLM-based MASs are also designed to complete specific SE tasks, such as code translation and release management. 
Yuan et al. \citep{yuan2024transagent} designed an LLM-based MAS named \textit{TRANSAGENT} to improve code translation accuracy through the collaboration of four specialized agents, which could be used in software migration, system refactoring, and cross-platform development. By aligning the execution behavior of the source and target programs, \textit{TRANSAGENT} effectively localizes faulty code blocks, thereby narrowing the error-fixing scope and reducing debugging complexity.
Besides MASs tailored for specific SE tasks, an increasing number of studies have constructed LLM-based MASs to support holistic SE tasks such as end-to-end development and end-to-end maintenance.
Qian et al. \citep{qian2024chatdev} developed an LLM-based MAS named \textit{ChatDev}, which assigns multiple agents different roles in the software development lifecycle, such as design, coding, and testing to develop a complete software system. 


\subsection{Characteristics on LLM-based MASs}\label{mas2.2}
With the gradual adoption of LLM-based MASs in various SE tasks, many researchers have conducted empirical studies to examine their characteristics, including system architectures, inter-agent coordination mechanisms, and evaluation protocols.
Shen et al. \citep{shen2025optimizing} designed a two-stage textual feedback optimization pipeline for LLM-based MASs focusing on software development. In the first stage, the MAS leverages its own failure explanations to identify underperforming agents. In the second stage, targeted prompt adjustments are applied to individual agents based on these explanations. The two-stage textual feedback pipeline provides a practical architecture pattern for runtime diagnosis and agent-level adaptation to LLM-based MASs for software development.
Cemri et al. \citep{cemri2025multiagent} selected seven open-source LLM-based MASs, and collected over 200 dialogues and execution traces, which contain the complete interaction records among all agents within the same MAS. Their analysis results show that MAS failures arise not only from limitations of the integrated LLMs but also from the design of MASs. They finally proposed a two-tier taxonomy of failure modes in LLM-based MASs.
Sarkar et al. \citep{sarkar2025survey} explored the enhancement of communication reliability and scalability in LLM-based MASs through the application of classical design patterns, with a specific focus on the Model Context Protocol (MCP). The study delved into the transition from single-agent to MASs, identifying key communication challenges such as architectural ambiguity, coordination misalignment, and task validation.
Bouzenia et al. \citep{bouzenia2025study} conducted a focused empirical analysis of thought-action-result workflows produced by three contemporary LLM-based MASs. They focused on a manually annotated corpus of 120 sampled workflows unified into canonical sequences. The authors parsed heterogeneous logs, computed trajectory-level metrics, mined frequent action subsequences, and performed open coding of semantic relations among thoughts, actions, and results. They concluded that detecting trajectory ``smells'' and enforcing alignment checks or self-critique loops are promising directions to improve agent robustness. 

\subsection{Surveys on LLM-based MASs}\label{mas2.3}
Researchers have also conducted surveys focusing on the architecture, capabilities, and limitations of LLM-based MASs.
Liu et al. \citep{liu2024agent} conducted a systematic literature review to investigate the architectural challenges of designing foundation model-based agents. Based on the review of 57 selected academic and industrial sources between 2023 and 2024, the authors identified 18 reusable architectural patterns that support key agent capabilities such as goal-seeking, planning, explainability, and accountability. These patterns were synthesized into a structured design pattern catalog, accompanied by a decision model to guide \rev{designers} in selecting appropriate patterns. Their study provides holistic design knowledge for building reliable and scalable LLM-based MASs, contributing to both research and practice in MAS architectures.
Liu et al. \citep{Liu2024survey} collected 106 papers before November 11th, 2024, using keyword-based searches and snowballing from DBLP and other sources to explore what SE tasks can be solved by LLM-based agent systems and the architectural components of these systems. They categorized the works from both SE and agent perspectives, covering tasks like requirements engineering, code generation, testing, and debugging, as well as agent architectures such as planning, memory, perception, and multi-agent coordination. Their survey identified the capabilities, design patterns, and open research challenges on LLM-based agent frameworks, offering valuable insights for the design and application of LLM-based agents in SE.
Chen et al. \citep{Chen2024survey} collected 125 papers published in top artificial intelligence conferences and some unpublished yet valuable papers from arXiv in 2023 and 2024 to provide a comprehensive and up-to-date review of LLM-based MASs. Their research categorized LLM-MAS applications into three domains: solving complex tasks, simulating specific scenarios, and evaluating generative agents. For each category, the survey discussed representative systems, open-source resources, and evaluation benchmarks. They also identified key challenges for the development of LLM-based MASs, including hallucinations, limited long-context handling, communication inefficiency, and a lack of objective evaluation metrics.
Yan et al. \citep{yan2025self} provided a comprehensive review of LLM-based MASs by focusing on the pivotal role of communication between agents. They collected 68 papers before May 2025, including domain-specific research papers and technical papers on communication protocols, agent behaviors, and applications of LLM-based MASs in different areas. They introduced a structured framework that integrates system-level communication with system-internal communication to explore how agents interact, negotiate, and achieve collective intelligence.
Yu et al. \citep{yu2025survey} conducted a systematic survey to investigate trustworthiness in LLM-based agents, selecting 175 representative papers from top AI, security, and NLP conferences and journals published since 2023. The authors introduced a unified conceptual framework that categorizes threats and countermeasures along two dimensions: internal and external trustworthiness.
He et al. \citep{He2025survey} investigated 71 papers before November 14th, 2024, indexed in DBLP, and conducted a systematic literature review complemented by empirical case studies. They assessed and characterized the current state and potential of LLM-based MASs for SE tasks, and proposed a comprehensive research agenda aimed at strengthening individual agent competence and optimizing inter-agent collaboration.
\rev{Wang et al. \citep{Wang2024agent} collected 115 papers after filtering and quality assessment to systematically investigate how LLM-based agents have been applied in SE tasks. Based on the reviewed studies, they proposed a general conceptual framework for LLM-based agents in SE consisting of three key modules: perception, memory, and action, and further discussed multi-agent collaboration as a mechanism through which multiple agents cooperate to accomplish complex SE tasks. Their survey also identified major challenges and future opportunities, including multimodal perception, external knowledge support, hallucination mitigation, and the efficiency of multi-agent collaboration, providing a comprehensive overview of how LLM-based agents are being integrated into SE.}

\subsection{Conclusive Summary}\label{mas2.4}
With the advances in the reasoning and decision-making capabilities of LLMs, plenty of LLM-based MASs have been designed to solve specific SE tasks. These MASs have been applied to various SE activities, such as requirements analysis, architecture design, code generation, and testing, demonstrating strong potential in simulating collaborative workflows, enabling role specialization, and enhancing automation and decision-making in software development. In addition, existing studies and surveys have investigated the design principles, agent architectures, communication mechanisms, and evaluation strategies of LLM-based MASs. These efforts provide foundational insights for constructing more robust and generalizable MAS frameworks for SE applications. 

\rev{Table~\ref{tab:Comparison} compares our study with three closely related reviews in terms of agent architecture, research scope, SE tasks, QAs, design patterns, design rationale, and relationships among these dimensions. Existing reviews characterize agent components, roles, workflows, and collaboration mechanisms and address quality concerns~\citep{He2025survey,Liu2024survey,Wang2024agent}. Our study complements these contributions by classifying QAs according to ISO/IEC 25010:2023 and organizing design patterns and design rationale into separate taxonomies for LLM-based MASs in SE. We further examine mappings between SE tasks and QAs and between QAs and design patterns, alongside separate task-level distributions of design patterns and design rationale.}

\rev{Together, these analyses provide a QA-oriented characterization of LLM-based MAS designs across SE tasks. The findings can help designers identify relevant quality concerns, select appropriate design patterns, and articulate the rationale for their choices in light of task-specific requirements.} 

\begin{table}[htbp]
\centering
\caption{\rev{Comparison of our study with existing reviews}}
\label{tab:Comparison}

\footnotesize
\setlength{\tabcolsep}{4pt}
\renewcommand{\arraystretch}{1.15}

\begin{tabularx}{\linewidth}{
@{}
>{\centering\arraybackslash}m{2.05cm}
>{\centering\arraybackslash}X
>{\centering\arraybackslash}X
>{\centering\arraybackslash}X
>{\centering\arraybackslash}X
@{}
}
\toprule

\textbf{\#}
& \textbf{Our Study}
& \textbf{Review}~\citep{He2025survey}
& \textbf{Review}~\citep{Liu2024survey}
& \textbf{Review}~\citep{Wang2024agent}
\\

\midrule

\textbf{Study Type}
& Empirical study
& Systematic review and case studies
& Comprehensive survey
& Literature survey
\\

\midrule

\textbf{agent architecture}
& LLM-based MASs for SE
& LLM-based MASs for SE
& LLM-based agents for SE
& LLM-based agents for SE
\\

\midrule

\textbf{Main Scope}
& Quality-driven design patterns and rationale of LLM-based MASs
& LLM-based MAS applications across SDLC stages and research agenda
& Design and use of LLM-based agents across software engineering tasks
& Perception, memory, and action of LLM-based agents
\\

\midrule

\textbf{SE Tasks}
& Taxonomy of SE tasks addressed by LLM-based MASs
& SDLC-based taxonomy
& SDLC-based taxonomy
& SE tasks are used as examples to illustrate agent capabilities
\\

\midrule

\textbf{Quality Attributes}
& Taxonomy of quality attributes based on ISO/IEC 25010:2023
& Coverage of robustness, reliability, scalability, efficiency, maintainability, and privacy
& Coverage of robustness, security, efficiency, and reusability
& Coverage of efficiency, compatibility, and system integrity
\\

\midrule

\textbf{Design Patterns}
& Taxonomy of design patterns used in LLM-based MASs
& Agent roles, process models, and collaboration 
& Agent component, role, and collaboration 
& A component-level agent framework based on perception, memory, and action
\\

\midrule

\textbf{Design Rationale}
& Separate taxonomy of design rationale
& No 
& No 
& No
\\

\midrule

\textbf{Relationships among Analysis Dimensions}
& Mapping relationship across SE tasks, quality attributes, design patterns, and design rationale
& No
& No
& No
\\

\bottomrule
\end{tabularx}
\end{table}

\section{Study Design}\label{sec:Study Design}
In this section, we present the goal and Research Questions (RQs) (Section~\ref{sec:RQs}) formulated to investigate the design of LLM-based MASs for SE tasks. Besides, we describe the process and criteria for data collection (Section~\ref{sec:datacollcetion}) and detail the procedures for data extraction (Section~\ref{sec:dataextraction}) and data analysis (Section~\ref{sec:dataanalysis}).

\subsection{Research Questions}\label{sec:RQs}
\begin{figure*}[htbp]
	\centering
	\includegraphics[width=\textwidth]{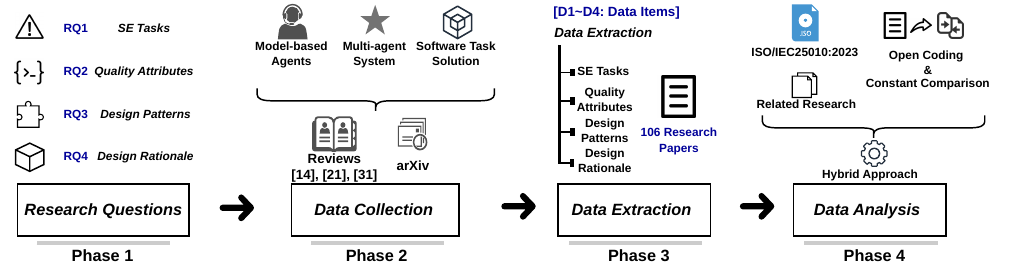}
	\caption{\rev{Overview of the research process}}
	\label{fig:Overview_of_research_process}
\end{figure*}

The \textbf{goal} of this study is to understand how LLM-based MASs for SE tasks are designed. To this end, we formulated four Research Questions (RQs): the SE tasks addressed by LLM-based MASs (RQ1), the quality attributes prioritized by their designers (RQ2), and the design patterns (RQ3) and rationale (RQ4) guiding their construction. The four RQs are addressed by following the research process illustrated in Figure~\ref{fig:Overview_of_research_process}.

\rev{To systematically investigate the design of LLM-based MASs for SE, we decompose the research goal into four complementary dimensions, each corresponding to one extracted data item. First, RQ1 examines the application dimension by identifying the SE tasks addressed by existing LLM-based MASs. Based on the identified SE tasks, RQ2 investigates the quality dimension by examining the quality attributes considered in the design of MASs for various SE tasks. RQ3 then focuses on the architectural dimension by identifying the design patterns adopted to achieve these quality attributes. Finally, RQ4 examines the rationale dimension by revealing the motivations and considerations underlying the corresponding design choices. Together, the four data items form a analytical chain that connects SE tasks, quality attributes, design patterns, and design rationale. This supports a comprehensive understanding of what LLM-based MASs are designed to accomplish, which quality concerns guide their design, how these concerns are addressed through reusable architectural solutions, and why the corresponding design choices are adopted.}

\begin{tcolorbox}[colback=gray!20, colframe=gray]
\textbf{RQ1: What SE tasks are addressed by LLM-based MASs?}
\end{tcolorbox}
\textbf{\textit{Rationale}}: As mentioned in Section \ref{sec:Introduction}, LLM-based MASs have demonstrated significant potential in addressing various SE tasks. The answer to this RQ can provide an overall view of the SE tasks addressed by LLM-based MASs, thereby informing and guiding the direction of future research in this emerging domain.

\begin{tcolorbox}[colback=gray!20, colframe=gray]
\textbf{RQ2: What quality attributes are considered when designing LLM-based MASs for SE tasks?}
\end{tcolorbox}
\textbf{\textit{Rationale}}: Through this RQ, we aim to explore the quality attributes that designers of LLM-based MASs for SE tasks prioritize, thereby identifying the specific considerations that should be handled when designing such systems for different SE tasks.

\begin{tcolorbox}[colback=gray!20, colframe=gray]
\textbf{RQ3: What design patterns are employed for building LLM-based MASs for SE tasks?}
\end{tcolorbox}
\textbf{\textit{Rationale}}: Design patterns are reusable solutions that help balance the quality attributes of MASs. In this study, we identified the design patterns employed in building LLM-based MASs. By analyzing these patterns, we aim to uncover effective approaches that can be adopted when designing LLM-based MASs for SE tasks.

\begin{tcolorbox}[colback=gray!20, colframe=gray]
\textbf{RQ4: What is the design rationale for constructing LLM-based MASs for SE tasks?}
\end{tcolorbox}
\textbf{\textit{Rationale}}: This RQ aims to understand the rationale behind designing LLM-based MASs for SE tasks. By examining the design rationale, we seek to reveal the guiding principles that designers can follow when constructing LLM-based MASs to support various SE tasks. 

\subsection{Data Collection}\label{sec:datacollcetion}
To obtain reliable insights into the design of LLM-based MASs for SE tasks, we relied on academic papers as our main data source. 
\rev{We began data collection with the papers covered in three recent surveys by He et al. \citep{He2025survey}, Liu et al. \citep{Liu2024survey}, and Wang et al. \citep{Wang2024agent}. We selected these surveys because they are highly relevant to the scope of our study: Liu et al.~\citep{Liu2024survey} and Wang et al.~\citep{Wang2024agent} systematically reviewed LLM-based agents for SE, while He et al. specifically focused on LLM-based MASs for SE. Collectively, these surveys provide broad and complementary coverage of existing studies on the application and design of LLM-based agents and MASs in SE. Building on their curated literature sets also enables us to leverage studies that have already undergone systematic search and relevance screening, thereby reducing the risk of omitting representative works in this rapidly evolving research area.}
\rev{To further improve the timeliness and comprehensiveness of our dataset, we used arXiv \citep{arXiv} as a supplementary data source rather than as a replacement for the survey-based literature collection. LLM- and agent-related research evolves rapidly, and recent studies are frequently disseminated through arXiv before formal publication. Therefore, arXiv enables us to capture emerging studies that may not yet have been covered by the selected surveys. We conducted a keyword search in the Computer Science category of arXiv using the query (``\texttt{large language model}'' AND ``\texttt{agent}'') and restricted the retrieved papers to those tagged with ``\texttt{cs.SE}'' and available up to November 14, 2024. This combination of survey-based literature collection and targeted arXiv search allows us to balance the coverage of established studies with the timely inclusion of recent developments.} 
\rev{Note that we used three literature surveys~\citep{He2025survey,Liu2024survey,Wang2024agent} to identify candidate studies. We also consulted the accompanying GitHub literature repositories for the surveys by Liu et al.~\citep{Liu2024survey} and Wang et al.~\citep{Wang2024agent}. These repositories continued to be updated after the surveys' respective search cutoff dates, and we considered the additional studies listed in them at the time of our data collection, subject to our inclusion criteria. Furthermore, some candidate studies were available only as preprints at the surveys' respective search cutoff dates but had been formally published by the time of our data collection. For these studies, we retrieved the published versions and used them for data extraction. Consequently, some included studies first became publicly available after November 14, 2024, while others were available earlier as preprints but have formal publication dates after that date.}
\rev{Table~\ref{tab:DataSources} provides an overview of the four sources used for data collection, together with their respective start and cutoff dates.}
We set the following criteria to select relevant papers for our research:

\begin{enumerate}[label=(\arabic*)]
\item The paper must introduce at least one multi-agent system.
\item The agent(s) introduced in the paper must leverage LLMs to implement their functions.
\item The agent(s) introduced in the paper must have addressed at least one SE task.
\end{enumerate}

We finally obtained \rev{106} papers (listed in Table~\ref{tab:included-studies} in the Appendix: Included Studies) that served as the source for data extraction and analysis. The information from these selected papers was recorded in an MS file~\citep{dataset}.

\begin{table}[htbp]
\centering
\caption{\rev{Data collection sources and their time period of the included studies}}
\label{tab:DataSources}

\footnotesize
\setlength{\tabcolsep}{4pt}
\renewcommand{\arraystretch}{1.15}

\begin{tabularx}{\linewidth}{
@{}
>{\raggedright\arraybackslash}p{1.8cm}
>{\centering\arraybackslash}X
>{\centering\arraybackslash}X
>{\centering\arraybackslash}X
>{\centering\arraybackslash}X
@{}
}
\toprule

& \rev{\textbf{Review}~\citep{He2025survey}}
& \rev{\textbf{Review}~\citep{Liu2024survey}}
& \rev{\textbf{Review}~\citep{Wang2024agent}}
& \rev{\textbf{arXiv}}
\\

\midrule

\rev{\textbf{Start date}}
& \rev{November, 2022}
& \rev{No start date limit}
& \rev{2020}
& \rev{No start date limit}
\\

\midrule

\rev{\textbf{Cutoff date}}
& \rev{November 14, 2024}
& \rev{November 11, 2024}
& \rev{June 17, 2024}
& \rev{November 14, 2024}
\\

\bottomrule
\end{tabularx}
\end{table}

\subsection{Data Extraction}\label{sec:dataextraction}
To answer the four RQs presented in Section \ref{sec:RQs}, we defined a set of data items for data extraction, as shown in Table~\ref{Data Items and RQs}. The data items D1$\sim$D4 are respectively used to extract information regarding SE tasks, QAs, design patterns, and design rationale, corresponding to RQ1$\sim$RQ4. These data items can be extracted from any section of the collected papers in our dataset, including the title, abstract, and main text. To ensure the completeness of the extracted information, we manually performed data extraction from the included papers. 

\begin{table}[htbp]
\footnotesize
\caption{Data items extracted and their corresponding RQs}
\label{Data Items and RQs}

\setlength{\tabcolsep}{4pt}

\begin{tabular}{
m{0.4cm}<{\centering}
>{\raggedright\arraybackslash}m{1.1cm}
>{\raggedright\arraybackslash}m{9.3cm}
m{1.35cm}<{\centering}
m{0.45cm}<{\centering}}
\hline

\textbf{\#}
& \textbf{Data Item}
& \textbf{Description}
& \rev{\textbf{Cardinality}}
& \textbf{RQ}
\\ \hline

D1
& SE Task
& \textit{The SE task addressed by the LLM-based MAS proposed in a paper.}
& \rev{Single}
& RQ1
\\ \hline

D2
& Quality Attribute
& \textit{The quality attribute(s) considered in the design of the LLM-based MAS proposed in a paper.}
& \rev{Multiple}
& RQ2
\\ \hline

D3
& Design Pattern
& \textit{The design pattern(s) employed in the design of the LLM-based MAS proposed in a paper.}
& \rev{Multiple}
& RQ3
\\ \hline

D4
& Design Rationale
& \textit{The design rationale that supports the design of the LLM-based MAS proposed in a paper.}
& \rev{Multiple}
& RQ4
\\ \hline

\end{tabular}
\end{table}

\rev{To improve the transparency and consistency of the data extraction process, we further establish explicit coding rules for D1--D4. As summarized in Table~\ref{tab:coding_protocol}, the coding protocol specifies the criteria under which evidence should be included or excluded and clarifies how potentially ambiguous cases should be handled. These rules operationalize the data item definitions in Table~\ref{Data Items and RQs} and are applied when interpreting evidence extracted from the primary studies.}

\begin{table}[htbp]
\footnotesize
\caption{\rev{Coding protocol for the extracted data items}}
\label{tab:coding_protocol}

\setlength{\tabcolsep}{3.5pt}

\begin{tabular}{
m{0.45cm}<{\centering}
>{\raggedright\arraybackslash}m{3.9cm}
>{\raggedright\arraybackslash}m{3.9cm}
>{\raggedright\arraybackslash}m{4.1cm}}
\hline

\textbf{\#}
& \textbf{Inclusion Rule}
& \textbf{Exclusion Rule}
& \textbf{Boundary Case}
\\ \hline

D1
& \rev{Include the primary SE task that the LLM-based MAS is explicitly designed to address.}
& \rev{Do not treat auxiliary activities that merely support the primary SE task as separate SE tasks.}
& \rev{When multiple closely related tasks jointly constitute a higher-level SE task, code the corresponding composite task. For example, Fault Localization and Program Repair can jointly be coded as End-to-End Software Maintenance.}
\\ \hline

D2
& \rev{Include a QA when the paper provides explicit evidence that the corresponding quality concern is considered in the design of the LLM-based MAS.}
& \rev{Exclude general claims of improvement when the reported evidence is insufficient to identify a specific QA.}
& \rev{When the same design decision provides evidence for multiple QAs, all supported QAs can be recorded.}
\\ \hline

D3
& \rev{Include a design pattern when the study describes a recurring structural, interaction, coordination, or behavioral solution instantiated in the MAS design.}
& \rev{Exclude isolated implementation details that do not constitute a reusable design solution for organizing or coordinating the MAS.}
& \rev{Similar implementations are coded separately when they exhibit substantively different agent structures or interaction mechanisms.}
\\ \hline

D4
& \rev{Include design rationale when the study provides evidence explaining the motivation or intended benefit underlying a design decision of the LLM-based MAS.}
& \rev{Exclude descriptions that only state what the system does or how it is structured without indicating why the corresponding design is adopted.}
& \rev{When the paper provides evidence of multiple distinct motivations for the same design decision, record each motivation as design rationale for that paper.}
\\ \hline

\end{tabular}
\end{table}

\subsubsection{Pilot Data Extraction}\label{subsec:pilotdataextraction}

To examine whether all data items could be correctly extracted from the papers included in our dataset, we conducted a pilot data extraction. We randomly selected five papers from our dataset to perform the pilot data extraction. The first author carried out the extraction independently and discussed the extraction results with the second and third authors. \rev{To ensure the rigor and reliability of the pilot data extraction process, we adopted the negotiated agreement approach \cite{negotiated2013} to review and validate the extraction results. Any disagreements or ambiguities among the first three authors were resolved through collaborative discussion until consensus was reached on the final extraction results. We chose this approach because the extraction of QAs and design rationale involves substantial contextual interpretation that may not be adequately captured by predefined coding rules alone. Rather than treating initial differences among coders solely as measurement error, negotiated agreement enabled the researchers to examine ambiguous cases, refine their interpretation of the extraction criteria, and establish a shared understanding before finalizing the extracted data.}

\rev{The pilot extraction further revealed that SE task categories do not naturally form a mutually exclusive and collectively exhaustive taxonomy at a single level of abstraction. Some tasks exhibit hierarchical or functional overlap. For example, \textit{Fault Localization} and \textit{Program Repair} can be studied as individual SE tasks, while they may also constitute successive activities within an \textit{End-to-End Software Maintenance} process. Therefore, the SE task taxonomy was used as an analytical abstraction for a consistent classification in this study rather than as a strictly disjoint decomposition of SE activities.}

The results indicate that the four data items listed in Table~\ref{Data Items and RQs} can be extracted from our dataset.
Based on the results obtained from the pilot data extraction, we reached a consensus and established the following rules for the formal data extraction:

\rev{(1) Each paper was assigned one \textit{SE Task} label for analysis of every included study. When a system addresses a single identifiable task, it was assigned to the corresponding task category. When multiple closely related tasks jointly form an end-to-end SE process, we assigned the system to the corresponding higher-level composite category. For example, a system integrating \textit{Fault Localization} and \textit{Program Repair }as part of a complete maintenance workflow was categorized as \textit{End-to-End Software Maintenance}. This progress improves the consistency of task-level aggregation and avoids counting the same paper multiple times. However, it abstracts away some fine-grained information about the individual sub-tasks performed within an end-to-end system.}
(2) If multiple \textit{Quality Attributes} are considered in the design of the LLM-based MAS in a paper, we record all the \textit{Quality Attributes}.
(3) If multiple \textit{Design Patterns} are employed in the design of the LLM-based MAS in a paper, we record all the \textit{Design Patterns}.
(4) If more than one \textit{Design Rationale} supports the design of the LLM-based MAS in a paper, we record all the \textit{Design Rationale}.

\subsubsection{Formal Data Extraction}\label{subsec:formaldataextraction}
After the pilot extraction, the first author independently conducted the formal data extraction based on the data items listed in Table~\ref{Data Items and RQs}. If any issues arose during the formal data extraction process, the first author discussed these issues with the second and third authors to resolve the problems. \rev{After five rounds of discussion, the first author completed the formal data extraction.} Once the first author completed the formal data extraction, the results of the extraction were reviewed by the second and third authors. Any inconsistencies were discussed among the first three authors to reach a consensus. The three authors conducted \rev{three} rounds of reviews and revisions on the formal data extraction results to obtain the final results. The formal data extraction results were recorded in an MS Excel file and a MAXQDA file, which are publicly available in our dataset~\citep{dataset}.

\subsection{Data Analysis}\label{sec:dataanalysis}
After completing the data extraction, we conducted data analysis to answer the four RQs in Section \ref{sec:RQs}. We categorized the \textit{SE tasks} to answer RQ1 \rev{by examining the SE tasks that the LLM-based MASs in the included papers were explicitly designed to address. We assigned the task labels using the coding protocol in Table~\ref{tab:coding_protocol}, assigning each paper a single task category and using composite categories for end-to-end workflows that integrate multiple closely related tasks.} To answer RQ2, we categorized the considered \textit{quality attributes} based on ISO/IEC 25010:2023 \citep{ISOIEC25010}. To address RQ3, we used a hybrid approach to classify the employed \textit{design patterns}. We first categorized the \textit{design patterns} according to the category of architecture patterns for LLM-based agents provided by Liu et al. \citep{liu2024agent}. When no matching architecture pattern existed in that category, we subsequently employed the Open Coding and Constant Comparison methods for categorization, which are commonly used for qualitative data analysis in SE studies \citep{stol2016grounded}. Additionally, we employed the Open Coding and Constant Comparison methods to determine the category of \textit{design rationale} for answering RQ4. \rev{For RQ1, we categorized the SE tasks addressed by the LLM-based MASs in the included papers following the coding rules in Table~\ref{tab:coding_protocol}. For RQ2 and RQ3, we primarily adopted existing classification schemes and standards from prior studies and industry standards to improve the consistency of the data analysis and facilitate comparison with existing research. However, during the analysis of RQ3, we identified several emerging design patterns that could not be adequately represented by the existing classifications. We therefore additionally applied Open Coding and Constant Comparison to derive new categories through a bottom-up process. In contrast, no widely accepted taxonomy is currently available for the design rationale of LLM-based MASs for SE. Therefore, RQ4 was analyzed entirely using a bottom-up qualitative coding approach.}

The steps we conducted for data analysis are as follows:
(1) The first author read the full text of each paper in the dataset and identified the \textit{SE task} that the LLM-based MAS aims to address, the \textit{quality attributes} considered in the design of the LLM-based MAS, the \textit{design patterns} employed in the design of the LLM-based MAS, and the \textit{design rationale} supporting the design of the LLM-based MAS for SE tasks. 
(2) For the data items that used the Open Coding and Constant Comparison methods, the first author compared the descriptions to identify their similarities and differences, then grouped similar descriptions following a bottom-up approach, which led to the formation of higher-level categories. 
(3) Whenever uncertainties about the code descriptions arose, the first author discussed these uncertainties with the second and third authors to reach a consensus. Due to the iterative nature of Constant Comparison, the coding schemes were progressively refined through repeated comparison, consolidation, and revision of categories. 
(4) The second and third authors reviewed and verified the preliminary analysis results. If any questions or discrepancies arose, the first three authors discussed them to reach a final consensus and resolve the conflicts. 

\rev{In total, the first author conducted five rounds of discussion and iterative refinement with the second and third authors to review, revise, and finalize the data analysis results. To provide an audit trail of this process, Table~\ref{tab:rq4_audit_trail} summarizes the evolution of the coding schemes from their initial versions to the finalized categories. During iterative refinement, the number of SE Task categories increased from 8 to 10, and the number of Design Pattern categories increased from 13 to 16 as additional categories were identified. The number of Design Rationale categories decreased from 10 to 8 through the merging of conceptually similar categories. For Quality Attributes, the initial coding covered 9 categories and 21 subcategories, while the final coding covered 8 categories and 19 subcategories following the review and revision of the initial coding. These QA counts refer to the categories represented in our coding results under ISO/IEC 25010:2023~\citep{ISOIEC25010}.}

\begin{table}[htbp]
\footnotesize
\centering
\caption{\rev{Evolution of the coding schemes during the qualitative analysis}}
\label{tab:rq4_audit_trail}
\begin{tabular}{p{2.1cm}p{2.6cm}p{2.6cm}p{5.0cm}}
\hline
\textbf{Data Item} &
\textbf{Initial Scheme} &
\textbf{Final Scheme} &
\textbf{Evolution} \\
\hline

SE Task &
8 categories &
10 categories &
New task categories were identified through more precise characterization of the SE tasks. \\
\hline

Quality Attribute &
9 categories, 21 subcategories &
8 categories, 19 subcategories &
Categories and subcategories considered unsuitable were excluded during refinement. \\
\hline

Design Pattern &
13 categories &
16 categories &
Additional recurring design patterns were identified during iterative analysis. \\
\hline

Design Rationale &
10 categories &
8 categories &
Conceptually similar design rationale were merged during refinement. \\
\hline

\end{tabular}
\end{table}

\section{Results}\label{sec:Results}
In this section, we report the results of the four RQs formulated in Section \ref{sec:RQs}.
For the taxonomies of \textit{SE Task}, \textit{Design Pattern}, and \textit{Design Rationale}, we provide a one-tier categorization. For the taxonomy of \textit{Quality Attribute}, we provide a two-tier categorization according to ISO/IEC 25010:2023 \citep{ISOIEC25010}.

\subsection{Category of SE Tasks Addressed (RQ1)}\label{sec:RQ1_results}
\rev{Table~\ref{SE Tasks solved by LLM-based MASs}} presents the taxonomy of SE tasks extracted from our dataset~\citep{dataset}. It can be observed that \textit{Code Generation} \rev{(46.2\%)} is the most common SE task that LLM-based MASs address. In addition, LLM-based MASs are also used to support \rev{\textit{End-to-End Software Development} (10.4\%) and \textit{Fault Localization} (9.4\%)}. The remaining LLM-based MASs are used to facilitate \rev{\textit{End-to-End Software Maintenance} (8.5\%), \textit{Program Repair} (7.5\%), \textit{Software Testing} (6.6\%), \textit{Code Review} (5.7\%), \textit{Requirements Engineering} (3.8\%), \textit{Code Translation} (0.9\%), and \textit{Release Management} (0.9\%).}

\rev{The reported percentages represent the proportions of the 106 included papers classified under each SE task category. As described in Section~\ref{subsec:pilotdataextraction}, each included study was assigned a single SE task label. When multiple closely related tasks jointly constitute a higher-level SE process, the study was categorized under the corresponding composite SE task. For example, an MAS integrating \textit{Fault Localization} and \textit{Program Repair} within a complete maintenance workflow was categorized as \textit{End-to-End Software Maintenance}, then the paper was not included in the counts for the separate SE task categories of \textit{Fault Localization} and \textit{Program Repair}.} 


{\tiny
\begin{table}[htbp]
\centering
\renewcommand{\arraystretch}{1.3}
\caption{\rev{SE Tasks addressed by LLM-based MASs}} \label{SE Tasks solved by LLM-based MASs}
\begin{tabular}
{>{\centering\arraybackslash}m{2.4cm}
                       >{\raggedright}m{4.3cm}
                       m{1.5cm}<{\centering}
                       >{\raggedright\arraybackslash}m{4.3cm}}
\hline
\textbf{SE Tasks}                                           & \textbf{Example} 
& \textbf{Count(\%)}   & \textbf{Studies}           \\ \hline
Code Generation                                      & \textit{To enable LLMs to handle these real-world repo-level code generation, we present CODEAGENT, a novel LLM-based agent framework that employs external tools for effective repo-level code generation.}~\hyperlink{S12}{[S12]}    
& \rev{49 (46.2\%)}           & \hyperlink{S1}{[S1]}, \hyperlink{S7}{[S7]}, \hyperlink{S9}{[S9]}, \hyperlink{S11}{[S11]}, \hyperlink{S12}{[S12]}, \hyperlink{S15}{[S15]}, \hyperlink{S20}{[S20]}, \hyperlink{S23}{[S23]}, \hyperlink{S24}{[S24]}, \hyperlink{S27}{[S27]}, \hyperlink{S29}{[S29]}, \hyperlink{S33}{[S33]}, \hyperlink{S34}{[S34]}, \hyperlink{S36}{[S36]}, \hyperlink{S37}{[S37]}, \hyperlink{S38}{[S38]}, \hyperlink{S40}{[S40]}, \hyperlink{S41}{[S41]}, \hyperlink{S42}{[S42]}, \hyperlink{S45}{[S45]}, \hyperlink{S46}{[S46]}, \hyperlink{S51}{[S51]}, \hyperlink{S52}{[S52]}, \hyperlink{S53}{[S53]}, \hyperlink{S54}{[S54]}, \hyperlink{S58}{[S58]}, \hyperlink{S61}{[S61]}, \hyperlink{S62}{[S62]}, \hyperlink{S63}{[S63]}, \hyperlink{S64}{[S64]}, \hyperlink{S67}{[S67]}, \hyperlink{S68}{[S68]}, \hyperlink{S71}{[S71]}, \hyperlink{S73}{[S73]}, \hyperlink{S74}{[S74]}, \hyperlink{S76}{[S76]}, \hyperlink{S78}{[S78]}, \hyperlink{S79}{[S79]}, \hyperlink{S82}{[S82]}, \hyperlink{S83}{[S83]}, \hyperlink{S84}{[S84]}, \hyperlink{S86}{[S86]}, \hyperlink{S87}{[S87]}, \hyperlink{S92}{[S92]}, \hyperlink{S94}{[S94]}, \hyperlink{S96}{[S96]}, \hyperlink{S98}{[S98]}, \hyperlink{S100}{[S100]}, \hyperlink{S105}{[S105]}
 \\ \hline
End-to-End Software Development                      & \textit{MetaGPT encodes Standardized Operating Procedures (SOPs) into prompt sequences for more streamlined workflows, thus allowing agents with human-like domain expertise to verify intermediate results and reduce errors.}~\hyperlink{S81}{[S81]}   
& \rev{11 (10.4\%)}            & \hyperlink{S5}{[S5]}, \hyperlink{S13}{[S13]}, \hyperlink{S21}{[S21]}, \hyperlink{S57}{[S57]}, \hyperlink{S65}{[S65]}, \hyperlink{S81}{[S81]}, \hyperlink{S93}{[S93]}, \hyperlink{S97}{[S97]}, \hyperlink{S99}{[S99]}, \hyperlink{S101}{[S101]}, \hyperlink{S106}{[S106]}
\\ \hline
Fault Localization                                   & \textit{To address  the limitation, this paper presents AGENTFL, a multi-agent system based on ChatGPT for automated fault localization.}~\hyperlink{S4}{[S4]}        
& \rev{10 (9.4\%)}            & \hyperlink{S4}{[S4]}, \hyperlink{S16}{[S16]}, \hyperlink{S19}{[S19]}, \hyperlink{S25}{[S25]}, \hyperlink{S26}{[S26]}, \hyperlink{S35}{[S35]}, \hyperlink{S49}{[S49]}, \hyperlink{S75}{[S75]}, \hyperlink{S77}{[S77]}, \hyperlink{S103}{[S103]}
 \\ \hline
End-to-End Software Maintenance                      & \textit{In this paper, we introduce MarsCode Agent , a novel framework that leverages LLMs to automatically identify and repair bugs in software code. MarsCode Agent combines the power of LLMs with advanced code analysis techniques to accurately localize faults and generate patches.}~\hyperlink{S31}{[S31]}                          
& \rev{9 (8.5\%)}            & \hyperlink{S2}{[S2]}, \hyperlink{S3}{[S3]}, \hyperlink{S28}{[S28]}, \hyperlink{S31}{[S31]}, \hyperlink{S32}{[S32]}, \hyperlink{S39}{[S39]}, \hyperlink{S59}{[S59]}, \hyperlink{S60}{[S60]}, \hyperlink{S104}{[S104]}
  \\ \hline
Program Repair                                       & \textit{In this paper, we propose an automated approach for solving GitHub issues to autonomously achieve  program improvement.}~\hyperlink{S8}{[S8]}     
& \rev{8 (7.5\%)}            & \hyperlink{S8}{[S8]}, \hyperlink{S14}{[S14]}, \hyperlink{S17}{[S17]}, \hyperlink{S50}{[S50]}, \hyperlink{S69}{[S69]}, \hyperlink{S70}{[S70]}, \hyperlink{S89}{[S89]}, \hyperlink{S90}{[S90]}
\\ \hline
Software Testing                                     & \textit{The proposed system mainly consists of three LLM-based agents responsible for action  planning, state checking and parameter selecting, respectively, and two additional modules for state sensing and case rewriting.}~\hyperlink{S43}{[S43]}          
& \rev{7 (6.6\%)}            & \hyperlink{S43}{[S43]}, \hyperlink{S47}{[S47]}, \hyperlink{S66}{[S66]}, \hyperlink{S80}{[S80]}, \hyperlink{S85}{[S85]}, \hyperlink{S91}{[S91]}, \hyperlink{S102}{[S102]}
 \\ \hline
Code Review                                          & \textit{In this paper, we present a novel approach to improving software quality and efficiency through a Large Language Model (LLM)-based model designed to review code and identify potential issues.}~\hyperlink{S6}{[S6]}   
& \rev{6 (5.7\%)}            & \hyperlink{S6}{[S6]}, \hyperlink{S10}{[S10]}, \hyperlink{S44}{[S44]}, \hyperlink{S55}{[S55]}, \hyperlink{S72}{[S72]}, \hyperlink{S88}{[S88]}
 \\ \hline
Requirements Engineering                             & \textit{These agents engage in product experience scenarios, through explaining their actions, observations, and challenges. Subsequent agent interviews and analysis uncover valuable user needs, including latent ones.}~\hyperlink{S18}{[S18]}    
& \rev{4 (3.8\%)}            & \hyperlink{S18}{[S18]}, \hyperlink{S30}{[S30]}, \hyperlink{S48}{[S48]}, \hyperlink{S95}{[S95]}
 \\ \hline
Code Translation                                     & \textit{In this work, we propose a novel LLM-based multi-agent system TRANSAGENT, which enhances LLM-based code translation by fixing the syntax errors and semantic errors with the synergy between four LLM-based agents.}~\hyperlink{S56}{[S56]}
& \rev{1 (0.9\%)}            & \hyperlink{S56}{[S56]}
 \\ \hline
Release Management                                 & \textit{GoNoGo represents an efficient and user-friendly LLM-based solution currently employed in our industrial partner’s company to assist with software release decision-making}~\hyperlink{S22}{[S22]}
& \rev{1 (0.9\%)}            & \hyperlink{S22}{[S22]}
 \\ \hline
\end{tabular}
\end{table}
\begin{minipage}{13.5cm} 

\end{minipage}
}

\fakesection[1em]{Code Generation} \rev{(49, 46.2\%)} denotes the SE task of automatically producing executable source code through coordinated LLM-based MASs. Because coding follows well-defined structures and rules, and large public datasets that support the fine-tuning and evaluation of models for code generation (e.g., MBPP \citep{austin2021mbpp}) are readily available, this facilitates building, refining, and evaluating MASs that perform code generation tasks. Consequently, code generation is the most common SE task addressed by MASs for SE tasks. For example, Zhao et al. proposed an LLM-based MAS named \textit{VisionCoder} that ``\textit{offers a cost-effective and efficient solution for code generation}'' by ``\textit{establishing a tree-structured thought distribution and development mechanism}''~\hyperlink{S51}{[S51]}. 

\fakesection[1em]{End-to-End Software Development} \rev{(11, 10.4\%)}
refers to build a software system by orchestrating LLM-based MASs. By organizing specialized agents around development activities such as requirements elicitation, software design, implementation, and testing, designers of MASs can establish a comprehensive and continuous software development pipeline. For example, Hong et al. proposed an LLM-based MAS named \textit{MetaGPT} that ``\textit{encodes Standardized Operating Procedures (SOPs) into prompt sequences for more streamlined workflows}'' to equip the system with ``\textit{human-like domain expertise}'' in software development~\hyperlink{S81}{[S81]}.

\fakesection[1em]{Fault Localization} \rev{(10, 9.4\%)} designates the task of using LLM-based MASs to pinpoint locations in a codebase that are responsible for observed failures and erroneous behavior. In an LLM-based MAS, various agents specialize in complementary tasks such as static analysis and fault candidate filtering. The iterative interactions between these agents enable cross-validation of evidence used to infer which code statement is responsible for the failure and consistent tracing of failure-inducing code statements. The combination of the specialization of agent roles, the coordination among agents, and the semantic reasoning capabilities of LLMs allows LLM-based MASs to produce accurate and explainable fault localization results. For instance, Qin et al. proposed an LLM-based MAS called \textit{AGENTFL} that simulates ``\textit{the behavior of a human developer}'' to ``\textit{localize bugs for an entire project}''~\hyperlink{S4}{[S4]}. 

\fakesection[1em]{End-to-End Software Maintenance} \rev{(9, 8.5\%)} is the task of automating the lifecycle of software maintenance activities, from detecting issues to producing and applying code changes, with LLMs providing analysis of software artifacts and the automated modification of existing software systems. By delegating responsibilities to specialized agents (e.g., fault localization, program repair, test generation, and verification), designers of LLM-based MASs can construct automated, modular, and scalable end-to-end maintenance workflows. For example, Liu et al. proposed an LLM-based MAS named \textit{MarsCode Agent} that ``\textit{leverages LLMs to automatically identify and repair bugs in software code}''~\hyperlink{S31}{[S31]}.

\fakesection[1em]{Program Repair} \rev{(8, 7.5\%)} refers to the automated process of correcting errors and defects in software systems to restore or enhance their functionality using LLM-based MASs. By leveraging multiple specialized agents to jointly analyze program logic, data flow, and semantic constraints, program repair can be carried out automatically, leading to improvements in repair accuracy and reliability. For example, Moon et al. proposed \textit{COFFEEPOTS}, an LLM-based MAS capable of ``\textit{automatically correcting critical errors generated from code LLMs}'' through ``\textit{feedback-driven Preference-Optimized Tuning and Selection}''~\hyperlink{S69}{[S69]}.

\fakesection[1em]{Software Testing} \rev{(7, 6.6\%)} covers automated evaluation of a system and its components using LLM-based MASs to verify conformance with requirements and to uncover deviations from expected behavior. When LLM-based agents cooperate in test case generation, prioritization, oracle construction, and automated test execution, MASs can deliver adaptive and efficient test suites that comprehensively enhance software quality, mitigate potential risks, and bolster confidence in the system under test. For instance, Yoon et al. proposed an LLM-based MAS named \textit{DROIDAGENT}, which is ``\textit{an autonomous GUI testing agent for Android, for semantic, intent-driven automation of GUI testing}''~\hyperlink{S47}{[S47]}.

\fakesection[1em]{Code Review} \rev{(6, 5.7\%)} is the automated examination of source code by LLM-based MASs to explore defects, check adherence to coding standards, and suggest improvements to code quality. Through the use of agents that perform style checking, correctness validation, semantic analysis, and reviewer suggestion generation, LLM-based MASs can provide consistent and context-aware feedback that accelerates the code review process. For instance, Tang et al. proposed an LLM-based MAS named \textit{CodeAgent} aimed at ``\textit{automating code review}''~\hyperlink{S44}{[S44]}.

\fakesection[1em]{Requirements Engineering} \rev{(4, 3.8\%)} involves a set of activities for eliciting, specifying, validating, and managing software requirements by leveraging LLM-based MASs to analyze requirements and maintain traceability between requirements and related artifacts. By allocating agents to tasks including elicitation, conflict identification, prioritization, and validation, designers can establish an automated and collaborative workflow for requirements engineering that ensures traceability and consistency between requirements and related artifacts. For example, Jin et al. proposed an LLM-based MAS named \textit{MARE} that ``\textit{leverages collaboration among LLMs throughout the entire RE process}'' to obtain ``\textit{high-quality requirements specifications}''~\hyperlink{S30}{[S30]}.

\fakesection[1em]{Code Translation} \rev{(1, 0.9\%)} denotes the task of transforming source code from one programming language into functionally equivalent code in another language with the assistance of LLM-based MASs. 
Agents coordinate over translation outputs, including candidate target-language code, syntactic structure (e.g., AST) of candidate code, API mappings, and execution results from generated tests to ensure the quality of translated code. For example, Yuan et al. proposed an LLM-based MAS named \textit{TRANSAGENT} that ``\textit{leverages parallel data to train models for automated code translation}''~\hyperlink{S56}{[S56]}.

\fakesection[1em]{Release Management} \rev{(1, 0.9\%)} denotes the analysis of release-related data to determine whether the software system has reached release readiness and to provide the corresponding release decision. By employing an MAS to automatically plan analytical steps and to perform analysis over software artifacts, the generation of release decisions can be automated. For example, Khoee et al. proposed an LLM-based MAS named \textit{GoNoGo} which ``\textit{has demonstrated significant improvements in the software release decision-making process besides saving time, improving accessibility, reducing reliance on specialized analysts, and accelerating overall workflow}''~\hyperlink{S22}{[S22]}.

\subsection{Category of Quality Attributes Considered (RQ2)}\label{sec:RQ2_results}
\begin{figure*}[htbp]
	\centering
	\includegraphics[width=\linewidth]{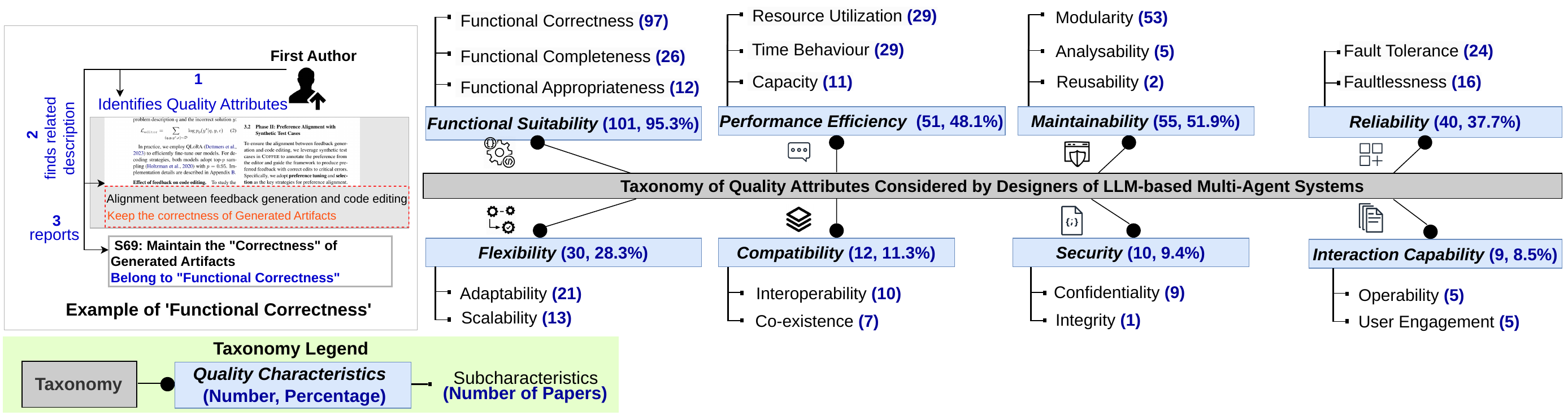}
	\caption{\rev{Taxonomy of Quality Attributes considered in the design of LLM-based MASs for SE Tasks}}
	\label{fig:Results of RQ2}
\end{figure*}

As mentioned in Section \ref{sec:RQs}, LLM-based MASs for SE tasks involve various Quality Attributes (QAs) that designers consider during the construction and implementation of these MASs. In this section, we report the results of RQ2 by following the categorization of QAs in ISO/IEC 25010:2023 \citep{ISOIEC25010}. Figure~\ref{fig:Results of RQ2} and Table~\ref{quality-attributes-papers} present the results of RQ2, in which \textit{Functional Suitability} \rev{(95.3\%)} is the QA that is most commonly considered in the design of LLM-based MASs to address SE tasks. Meanwhile, \textit{Maintainability} \rev{(51.9\%)} and \textit{Performance Efficiency} \rev{(48.1\%)} are also considered by a number of LLM-based MASs. The remaining QAs considered are \textit{Reliability} \rev{(37.7\%)}, \textit{Flexibility} \rev{(28.3\%)}, \textit{Compatibility} \rev{(11.3\%)}, \textit{Security} \rev{(9.4\%)}, and \textit{Interaction Capability} \rev{(8.5\%)}.

{\tiny
\begin{table}[htbp]
\centering
\renewcommand{\arraystretch}{1.3}
\caption{\rev{Quality Attributes considered in the design of LLM-based MASs for SE Tasks}}
\label{quality-attributes-papers}
\begin{tabular}{>{\centering\arraybackslash}m{3.0cm}
                  >{\centering\arraybackslash}m{2.6cm}
                  >{\raggedright\arraybackslash}m{6.9cm}} 
\hline
\textbf{Quality Attribute}              & \textbf{Sub\_Quality Attribute} & \textbf{Studies} \\ \hline
\multirow{3}{*}{\begin{minipage}[c][5\baselineskip][c]{3.0cm} \centering Functional Suitability \end{minipage}} & Functional Correctness          & \hyperlink{S1}{[S1]}, \hyperlink{S2}{[S2]}, \hyperlink{S3}{[S3]}, \hyperlink{S4}{[S4]}, \hyperlink{S5}{[S5]}, \hyperlink{S7}{[S7]}, \hyperlink{S8}{[S8]}, \hyperlink{S9}{[S9]}, \hyperlink{S11}{[S11]}, \hyperlink{S12}{[S12]}, \hyperlink{S13}{[S13]}, \hyperlink{S14}{[S14]}, \hyperlink{S15}{[S15]}, \hyperlink{S16}{[S16]}, \hyperlink{S17}{[S17]}, \hyperlink{S19}{[S19]}, \hyperlink{S20}{[S20]}, \hyperlink{S22}{[S22]}, \hyperlink{S23}{[S23]}, \hyperlink{S24}{[S24]}, \hyperlink{S25}{[S25]}, \hyperlink{S26}{[S26]}, \hyperlink{S27}{[S27]}, \hyperlink{S28}{[S28]}, \hyperlink{S29}{[S29]}, \hyperlink{S30}{[S30]}, \hyperlink{S31}{[S31]}, \hyperlink{S32}{[S32]}, \hyperlink{S33}{[S33]}, \hyperlink{S34}{[S34]}, \hyperlink{S36}{[S36]}, \hyperlink{S38}{[S38]}, \hyperlink{S39}{[S39]}, \hyperlink{S40}{[S40]}, \hyperlink{S41}{[S41]}, \hyperlink{S42}{[S42]}, \hyperlink{S43}{[S43]}, \hyperlink{S44}{[S44]}, \hyperlink{S46}{[S46]}, \hyperlink{S47}{[S47]}, \hyperlink{S48}{[S48]}, \hyperlink{S49}{[S49]}, \hyperlink{S50}{[S50]}, \hyperlink{S51}{[S51]}, \hyperlink{S52}{[S52]}, \hyperlink{S53}{[S53]}, \hyperlink{S54}{[S54]}, \hyperlink{S55}{[S55]}, \hyperlink{S56}{[S56]}, \hyperlink{S57}{[S57]}, \hyperlink{S58}{[S58]}, \hyperlink{S59}{[S59]}, \hyperlink{S61}{[S61]}, \hyperlink{S62}{[S62]}, \hyperlink{S63}{[S63]}, \hyperlink{S64}{[S64]}, \hyperlink{S65}{[S65]}, \hyperlink{S66}{[S66]}, \hyperlink{S67}{[S67]}, \hyperlink{S68}{[S68]}, \hyperlink{S69}{[S69]}, \hyperlink{S70}{[S70]}, \hyperlink{S71}{[S71]}, \hyperlink{S72}{[S72]}, \hyperlink{S73}{[S73]}, \hyperlink{S74}{[S74]}, \hyperlink{S75}{[S75]}, \hyperlink{S76}{[S76]}, \hyperlink{S77}{[S77]}, \hyperlink{S78}{[S78]}, \hyperlink{S79}{[S79]}, \hyperlink{S80}{[S80]}, \hyperlink{S81}{[S81]}, \hyperlink{S82}{[S82]}, \hyperlink{S83}{[S83]}, \hyperlink{S84}{[S84]}, \hyperlink{S85}{[S85]}, \hyperlink{S86}{[S86]}, \hyperlink{S87}{[S87]}, \hyperlink{S88}{[S88]}, \hyperlink{S89}{[S89]}, \hyperlink{S90}{[S90]}, \hyperlink{S91}{[S91]}, \hyperlink{S92}{[S92]}, \hyperlink{S93}{[S93]}, \hyperlink{S94}{[S94]}, \hyperlink{S95}{[S95]}, \hyperlink{S96}{[S96]}, \hyperlink{S97}{[S97]}, \hyperlink{S98}{[S98]}, \hyperlink{S99}{[S99]}, \hyperlink{S100}{[S100]}, \hyperlink{S101}{[S101]}, \hyperlink{S102}{[S102]}, \hyperlink{S103}{[S103]}, \hyperlink{S104}{[S104]}, \hyperlink{S105}{[S105]}
 \\ \cline{2-3} 
& Functional Completeness          & \hyperlink{S6}{[S6]}, \hyperlink{S12}{[S12]}, \hyperlink{S18}{[S18]}, \hyperlink{S21}{[S21]}, \hyperlink{S24}{[S24]}, \hyperlink{S29}{[S29]}, \hyperlink{S30}{[S30]}, \hyperlink{S34}{[S34]}, \hyperlink{S44}{[S44]}, \hyperlink{S47}{[S47]}, \hyperlink{S54}{[S54]}, \hyperlink{S55}{[S55]}, \hyperlink{S62}{[S62]}, \hyperlink{S64}{[S64]}, \hyperlink{S73}{[S73]}, \hyperlink{S76}{[S76]}, \hyperlink{S79}{[S79]}, \hyperlink{S80}{[S80]}, \hyperlink{S83}{[S83]}, \hyperlink{S84}{[S84]}, \hyperlink{S85}{[S85]}, \hyperlink{S91}{[S91]}, \hyperlink{S93}{[S93]}, \hyperlink{S94}{[S94]}, \hyperlink{S97}{[S97]}, \hyperlink{S99}{[S99]}
 \\ \cline{2-3}
& Functional Appropriateness       & \hyperlink{S22}{[S22]}, \hyperlink{S34}{[S34]}, \hyperlink{S36}{[S36]}, \hyperlink{S41}{[S41]}, \hyperlink{S49}{[S49]}, \hyperlink{S55}{[S55]}, \hyperlink{S67}{[S67]}, \hyperlink{S69}{[S69]}, \hyperlink{S73}{[S73]}, \hyperlink{S93}{[S93]}, \hyperlink{S94}{[S94]}, \hyperlink{S106}{[S106]}
 \\ \hline
\multirow{3}{*}{\begin{minipage}[c][6\baselineskip][c]{3.0cm} \centering Performance Efficiency \end{minipage}} & Resource Utilization            & \hyperlink{S2}{[S2]}, \hyperlink{S4}{[S4]}, \hyperlink{S6}{[S6]}, \hyperlink{S8}{[S8]}, \hyperlink{S18}{[S18]}, \hyperlink{S19}{[S19]}, \hyperlink{S20}{[S20]}, \hyperlink{S25}{[S25]}, \hyperlink{S28}{[S28]}, \hyperlink{S35}{[S35]}, \hyperlink{S36}{[S36]}, \hyperlink{S37}{[S37]}, \hyperlink{S39}{[S39]}, \hyperlink{S40}{[S40]}, \hyperlink{S45}{[S45]}, \hyperlink{S46}{[S46]}, \hyperlink{S47}{[S47]}, \hyperlink{S48}{[S48]}, \hyperlink{S49}{[S49]}, \hyperlink{S58}{[S58]}, \hyperlink{S63}{[S63]}, \hyperlink{S64}{[S64]}, \hyperlink{S68}{[S68]}, \hyperlink{S81}{[S81]}, \hyperlink{S87}{[S87]}, \hyperlink{S88}{[S88]}, \hyperlink{S94}{[S94]}, \hyperlink{S104}{[S104]}, \hyperlink{S105}{[S105]}
 \\ \cline{2-3} 
& Time Behavior                   & \hyperlink{S2}{[S2]}, \hyperlink{S3}{[S3]}, \hyperlink{S4}{[S4]}, \hyperlink{S8}{[S8]}, \hyperlink{S10}{[S10]}, \hyperlink{S13}{[S13]}, \hyperlink{S21}{[S21]}, \hyperlink{S22}{[S22]}, \hyperlink{S27}{[S27]}, \hyperlink{S39}{[S39]}, \hyperlink{S45}{[S45]}, \hyperlink{S46}{[S46]}, \hyperlink{S47}{[S47]}, \hyperlink{S49}{[S49]}, \hyperlink{S55}{[S55]}, \hyperlink{S56}{[S56]}, \hyperlink{S58}{[S58]}, \hyperlink{S60}{[S60]}, \hyperlink{S65}{[S65]}, \hyperlink{S73}{[S73]}, \hyperlink{S80}{[S80]}, \hyperlink{S81}{[S81]}, \hyperlink{S82}{[S82]}, \hyperlink{S87}{[S87]}, \hyperlink{S90}{[S90]}, \hyperlink{S91}{[S91]}, \hyperlink{S94}{[S94]}, \hyperlink{S104}{[S104]}, \hyperlink{S106}{[S106]}
 \\ \cline{2-3}
& Capacity                       & \hyperlink{S18}{[S18]}, \hyperlink{S26}{[S26]}, \hyperlink{S38}{[S38]}, \hyperlink{S45}{[S45]}, \hyperlink{S48}{[S48]}, \hyperlink{S57}{[S57]}, \hyperlink{S59}{[S59]}, \hyperlink{S85}{[S85]}, \hyperlink{S88}{[S88]}, \hyperlink{S89}{[S89]}, \hyperlink{S94}{[S94]}
 \\ \hline
\multirow{3}{*}{\begin{minipage}[c][3\baselineskip][c]{3.0cm} \centering Maintainability \end{minipage}}        & Modularity                      & \hyperlink{S2}{[S2]}, \hyperlink{S4}{[S4]}, \hyperlink{S5}{[S5]}, \hyperlink{S6}{[S6]}, \hyperlink{S7}{[S7]}, \hyperlink{S11}{[S11]}, \hyperlink{S14}{[S14]}, \hyperlink{S15}{[S15]}, \hyperlink{S16}{[S16]}, \hyperlink{S18}{[S18]}, \hyperlink{S21}{[S21]}, \hyperlink{S26}{[S26]}, \hyperlink{S29}{[S29]}, \hyperlink{S30}{[S30]}, \hyperlink{S31}{[S31]}, \hyperlink{S32}{[S32]}, \hyperlink{S36}{[S36]}, \hyperlink{S38}{[S38]}, \hyperlink{S43}{[S43]}, \hyperlink{S44}{[S44]}, \hyperlink{S47}{[S47]}, \hyperlink{S49}{[S49]}, \hyperlink{S51}{[S51]}, \hyperlink{S55}{[S55]}, \hyperlink{S56}{[S56]}, \hyperlink{S59}{[S59]}, \hyperlink{S60}{[S60]}, \hyperlink{S61}{[S61]}, \hyperlink{S62}{[S62]}, \hyperlink{S63}{[S63]}, \hyperlink{S64}{[S64]}, \hyperlink{S66}{[S66]}, \hyperlink{S67}{[S67]}, \hyperlink{S68}{[S68]}, \hyperlink{S69}{[S69]}, \hyperlink{S70}{[S70]}, \hyperlink{S71}{[S71]}, \hyperlink{S72}{[S72]}, \hyperlink{S74}{[S74]}, \hyperlink{S76}{[S76]}, \hyperlink{S78}{[S78]}, \hyperlink{S81}{[S81]}, \hyperlink{S82}{[S82]}, \hyperlink{S84}{[S84]}, \hyperlink{S88}{[S88]}, \hyperlink{S94}{[S94]}, \hyperlink{S95}{[S95]}, \hyperlink{S97}{[S97]}, \hyperlink{S98}{[S98]}, \hyperlink{S99}{[S99]}, \hyperlink{S102}{[S102]}, \hyperlink{S104}{[S104]}, \hyperlink{S106}{[S106]}
 \\ \cline{2-3}
& Analysability                    & \hyperlink{S6}{[S6]}, \hyperlink{S67}{[S67]}, \hyperlink{S92}{[S92]}, \hyperlink{S94}{[S94]}, \hyperlink{S100}{[S100]}
 \\ \cline{2-3}
& Reusability                      & \hyperlink{S74}{[S74]}, \hyperlink{S99}{[S99]} \\ \hline
\multirow{2}{*}{\begin{minipage}[c][2\baselineskip][c]{3.0cm} \centering Reliability \end{minipage}}            & Fault Tolerance                 & \hyperlink{S5}{[S5]}, \hyperlink{S7}{[S7]}, \hyperlink{S8}{[S8]}, \hyperlink{S10}{[S10]}, \hyperlink{S30}{[S30]}, \hyperlink{S39}{[S39]}, \hyperlink{S40}{[S40]}, \hyperlink{S42}{[S42]}, \hyperlink{S47}{[S47]}, \hyperlink{S55}{[S55]}, \hyperlink{S57}{[S57]}, \hyperlink{S61}{[S61]}, \hyperlink{S63}{[S63]}, \hyperlink{S65}{[S65]}, \hyperlink{S70}{[S70]}, \hyperlink{S73}{[S73]}, \hyperlink{S81}{[S81]}, \hyperlink{S83}{[S83]}, \hyperlink{S87}{[S87]}, \hyperlink{S88}{[S88]}, \hyperlink{S92}{[S92]}, \hyperlink{S98}{[S98]}, \hyperlink{S100}{[S100]}, \hyperlink{S105}{[S105]}
\\ \cline{2-3}
& Faultlessness                    & \hyperlink{S1}{[S1]}, \hyperlink{S13}{[S13]}, \hyperlink{S24}{[S24]}, \hyperlink{S27}{[S27]}, \hyperlink{S48}{[S48]}, \hyperlink{S51}{[S51]}, \hyperlink{S52}{[S52]}, \hyperlink{S53}{[S53]}, \hyperlink{S62}{[S62]}, \hyperlink{S64}{[S64]}, \hyperlink{S77}{[S77]}, \hyperlink{S86}{[S86]}, \hyperlink{S94}{[S94]}, \hyperlink{S96}{[S96]}, \hyperlink{S104}{[S104]}, \hyperlink{S106}{[S106]}
 \\ \hline
\multirow{2}{*}{\begin{minipage}[c][2\baselineskip][c]{3.0cm} \centering Flexibility \end{minipage}}            & Adaptability                    & \hyperlink{S1}{[S1]}, \hyperlink{S2}{[S2]}, \hyperlink{S9}{[S9]}, \hyperlink{S12}{[S12]}, \hyperlink{S31}{[S31]}, \hyperlink{S33}{[S33]}, \hyperlink{S36}{[S36]}, \hyperlink{S52}{[S52]}, \hyperlink{S53}{[S53]}, \hyperlink{S57}{[S57]}, \hyperlink{S66}{[S66]}, \hyperlink{S75}{[S75]}, \hyperlink{S81}{[S81]}, \hyperlink{S82}{[S82]}, \hyperlink{S88}{[S88]}, \hyperlink{S91}{[S91]}, \hyperlink{S100}{[S100]}, \hyperlink{S101}{[S101]}, \hyperlink{S102}{[S102]}, \hyperlink{S104}{[S104]}, \hyperlink{S106}{[S106]}
 \\ \cline{2-3}
& Scalability                      & \hyperlink{S2}{[S2]}, \hyperlink{S13}{[S13]}, \hyperlink{S15}{[S15]}, \hyperlink{S17}{[S17]}, \hyperlink{S33}{[S33]}, \hyperlink{S37}{[S37]}, \hyperlink{S38}{[S38]}, \hyperlink{S52}{[S52]}, \hyperlink{S69}{[S69]}, \hyperlink{S82}{[S82]}, \hyperlink{S89}{[S89]}, \hyperlink{S94}{[S94]}, \hyperlink{S97}{[S97]}
\\ \hline
\multirow{2}{*}{\begin{minipage}[c][1\baselineskip][c]{3.0cm} \centering Compatibility \end{minipage}}          & Interoperability                & \hyperlink{S12}{[S12]}, \hyperlink{S23}{[S23]}, \hyperlink{S25}{[S25]}, \hyperlink{S46}{[S46]}, \hyperlink{S55}{[S55]}, \hyperlink{S76}{[S76]}, \hyperlink{S79}{[S79]}, \hyperlink{S84}{[S84]}, \hyperlink{S98}{[S98]}, \hyperlink{S101}{[S101]}
 \\ \cline{2-3}
& Co-Existence                     & \hyperlink{S2}{[S2]}, \hyperlink{S12}{[S12]}, \hyperlink{S23}{[S23]}, \hyperlink{S25}{[S25]}, \hyperlink{S76}{[S76]}, \hyperlink{S79}{[S79]}, \hyperlink{S80}{[S80]}
 \\ \hline
\multirow{2}{*}{\begin{minipage}[c][1\baselineskip][c]{3.0cm} \centering Security \end{minipage}}               & Confidentiality                 & \hyperlink{S3}{[S3]}, \hyperlink{S9}{[S9]}, \hyperlink{S21}{[S21]}, \hyperlink{S25}{[S25]}, \hyperlink{S44}{[S44]}, \hyperlink{S65}{[S65]}, \hyperlink{S77}{[S77]}, \hyperlink{S82}{[S82]}, \hyperlink{S85}{[S85]}
 \\ \cline{2-3}
& Integrity                        & \hyperlink{S64}{[S64]} \\ \hline
\multirow{2}{*}{\begin{minipage}[c][1\baselineskip][c]{3.0cm} \centering Interaction Capability \end{minipage}} & User Engagement                 & \hyperlink{S20}{[S20]}, \hyperlink{S21}{[S21]}, \hyperlink{S40}{[S40]}, \hyperlink{S74}{[S74]}, \hyperlink{S84}{[S84]}
 \\ \cline{2-3}
& Operability                      & \hyperlink{S40}{[S40]}, \hyperlink{S46}{[S46]}, \hyperlink{S49}{[S49]}, \hyperlink{S66}{[S66]}, \hyperlink{S76}{[S76]}
 \\ \hline
\end{tabular}
\end{table}
}

\fakesection[1em]{Functional Suitability} \rev{(101, 95.3\%)} is the capacity of an LLM-based MAS to implement functions that satisfy explicit user requirements and requirements derived from specified conditions (e.g., task constraints). There are three sub-Quality Attributes (sub-QAs) considered by the designers of LLM-based MASs under \textit{Functional Suitability}:

\begin{itemize}
    \item \textit{Functional Correctness} \rev{(97, 91.5\%)}: The capacity of an LLM-based MAS to deliver accurate results. 
    \item \textit{Functional Completeness} \rev{(26, 24.5\%)}: The capacity of an LLM-based MAS to yield full results when utilized by intended users. 
    \item \textit{Functional Appropriateness} \rev{(12, 11.3\%)}: The capacity of an LLM-based MAS to deliver functions that support the achievement of specified tasks and objectives.  
\end{itemize}

Ensuring a system satisfies specified functional requirements and produces correct outputs remains the primary concern for most designers of LLM-based MASs for SE tasks. \textit{Functional Suitability}, particularly \textit{Functional Correctness}, is a critical QA that needs to be considered in the design of LLM-based MASs for SE tasks.
For example, Zhang et al. introduced ``\textit{Deep Retrieval-Augmented Generation}'' in their LLM-based MAS to ``\textit{handle the complex inheritance relationships in exception types}'' of Java language; consequently, the MAS can ensure the correctness of exception handling~\hyperlink{S55}{[S55]}.

\fakesection[1em]{Maintainability} \rev{(55, 51.9\%)} is the capacity of an LLM-based MAS to be effectively and efficiently modified by designated maintainers, such as corrections, enhancements, and adaptations to environmental changes. There are three sub-QAs considered by the designers of LLM-based MASs under \textit{Maintainability}:

\begin{itemize}
    \item \textit{Modularity} \rev{(53, 50.0\%)}: The capacity of an LLM-based MAS to confine modifications to a single component, minimizing impact on other components, with the system structured into discrete modules and components exhibiting high cohesion and minimal coupling to other modules and components.
    \item \textit{Analysability} \rev{(5, 4.7\%)}: The capacity of an LLM-based MAS to be effectively and efficiently evaluated for the impact of proposed changes to one or more components, to diagnose deficiencies, failure causes, and to identify components requiring modification, incorporating mechanisms for self-analysis of faults and generation of reports prior to failures and other events.
    \item \textit{Reusability} \rev{(2, 1.9\%)}: The capacity of an LLM-based MAS and its constituent agents to be utilized as a component in multiple systems and in the development of other assets. 
\end{itemize}

Designers of LLM-based MASs are willing to build modular architectures to simplify maintenance, enable isolated updates, debugging, and evolution in these systems.
For example, Qin et al. introduced an MAS called ``\textit{AGENTFL}'' to ``\textit{decompose the (code) localization process into multiple phases}'' and ``\textit{specialize the LLMs}'' in various agents to complete fault localization. In this way, the modularity of the MAS can be improved~\hyperlink{S4}{[S4]}.

\fakesection[1em]{Performance Efficiency} \rev{(51, 48.1\%)} is the capacity of an LLM-based MAS to execute its functions within the designated time and performance parameters while maintaining efficient use of resources (e.g., CPU, memory, storage) under specified conditions. There are three sub-QAs considered by the designers of LLM-based MASs under \textit{Performance Efficiency}:

\begin{itemize}
    \item \textit{Resource Utilization} \rev{(29, 27.4\%)}: The capacity of an LLM-based MAS to perform designated functions under specified conditions while consuming no more than the allocated amount of resources.
    \item \textit{Time Behavior} \rev{(29, 27.4\%)}: The capacity of an LLM-based MAS to execute designated functions under specified conditions, ensuring that response time and throughput rates satisfy established requirements.
    \item \textit{Capacity} \rev{(11, 10.4\%)}: The capacity of an LLM-based MAS to fulfill requirements for the upper bound of system parameters (e.g., concurrent user support).  
\end{itemize}

Designers of LLM-based MASs prioritize optimizing computational resource usage and ensuring timely system responses to the users of MASs.
For example, Tao et al. introduced a memory mechanism in the MAS for software maintenance to reduce queries that ``\textit{contain the same code snippets as previous ones, leading to unnecessary  computational costs}''~\hyperlink{S28}{[S28]}.

\fakesection[1em]{Reliability} \rev{(40, 37.7\%)} is the capacity of an LLM-based MAS to execute designated functions under specified conditions for a defined duration without interruptions and failures. There are two sub-QAs considered by the designers of LLM-based MASs under \textit{Reliability}:

\begin{itemize}
    \item \textit{Fault Tolerance} \rev{(24, 22.6\%)}: The capability for an LLM-based MAS to function as intended even in the presence of hardware and software faults.
    \item \textit{Faultlessness} \rev{(16, 15.1\%)}: The capability of an LLM-based MAS to execute defined functions without faults during normal operation. In addition, the conception of \textit{Faultlessness} may also be extended to other QAs to reflect the extent to which they fulfill the required needs under typical operating conditions. 
\end{itemize}

Designers of LLM-based MASs place great importance on ensuring system robustness to reduce potential failures and unexpected errors. 
For example, Wu et al. introduced ``\textit{a novel interactive retrieval feature}'' in their MAS for code generation. If the agent fails to retrieve relevant context, it would ask the user to ``\textit{UPDATE CONTEXT}'', instead of ``\textit{terminating}'' the execution of the MAS~\hyperlink{S64}{[S64]}.

\fakesection[1em]{Flexibility} \rev{(30, 28.3\%)} is the capability of an LLM-based MAS to adjust in response to changes in requirements, usage contexts, and the system environment. Flexibility with respect to usage context involves two distinct aspects: the technical (e.g., software environment) and the non-technical (e.g., physical environment). There are two sub-QAs considered by the designers of LLM-based MASs under \textit{Flexibility}:

\begin{itemize}
    \item \textit{Adaptability} \rev{(21, 19.8\%)}: The capability of an LLM-based MAS to be effectively and efficiently modified and transferred for operation in different hardware, software, and operational environments.
    \item \textit{Scalability} \rev{(13, 12.3\%)}: The capability of an LLM-based MAS to manage increasing or decreasing workloads, and to adjust its capacity in response to fluctuations in requirements.  
\end{itemize}

Designers of LLM-based MASs focus on enabling systems to adjust to evolving requirements, dynamic runtime environments, and diverse application contexts.
For example, Hu et al. design an LLM-based MAS called \textit{EvoMAC} for code generation, which has ``\textit{ability to iteratively adapt both agents and their connections during test time for each task}''~\hyperlink{S53}{[S53]}.

\fakesection[1em]{Compatibility} \rev{(12, 11.3\%)} is the capability of an LLM-based MAS to exchange information with other systems and to carry out required functions while operating within a shared environment and utilizing common resources. There are two sub-QAs considered by the designers of LLM-based MASs under \textit{Compatibility}:

\begin{itemize}
    \item \textit{Interoperability} \rev{(10, 9.4\%)}: The capability of an LLM-based MAS to exchange information with other systems and to make mutual use of the exchanged information.
    \item \textit{Co-Existence} \rev{(7, 6.6\%)}: The capability of an LLM-based MAS to perform required functions efficiently within a shared environment and using common resources, without causing negative effects on the operation of other systems.  
\end{itemize}

Designers of LLM-based MASs place great importance on enabling seamless communication, interaction, and integration with other systems, platforms, and tools.
For example, Zhang et al. proposed an LLM-based MAS called \textit{CODEAGENT} for code generation that ``\textit{integrates five programming tools, enabling interaction with software artifacts for information retrieval, code implementation, and code testing}''~\hyperlink{S12}{[S12]}.

\fakesection[1em]{Security} \rev{(10, 9.4\%)} is the capability of an LLM-based MAS to safeguard information and data by ensuring that individuals or other systems access data according to their designated types and authorization levels, and by resisting attacks from malicious entities, which includes protection of both stored data and data in transmission. There are two sub-QAs considered by the designers of LLM-based MASs under \textit{Security}:

\begin{itemize}
    \item \textit{Confidentiality} \rev{(9, 8.5\%)}: The capability of an LLM-based MAS to ensure that data can be accessed exclusively by entities with appropriate authorization.
    \item \textit{Integrity} \rev{(1, 0.9\%)}: The capacity of an LLM-based MAS to protect the integrity of its system and data against unauthorized alteration and deletion, whether due to malicious intent or computational error.  
\end{itemize}

Protecting sensitive data and ensuring that information is accessible only to authorized entities cannot be ignored.
For example, Talebirad et al. used ``\textit{a stateless oracle agent, which can monitor each sensitive task and decide if it is indeed malicious or not}'' to promise the confidentiality of their LLM-based MAS for code generation~\hyperlink{S82}{[S82]}.

\fakesection[1em]{Interaction Capability} \rev{(9, 8.5\%)} is the capacity of an LLM-based MAS that enables specified users to exchange information through the user interface for achieving intended tasks. There are two sub-QAs considered by the designers of LLM-based MASs under \textit{Interaction Capability}:

\begin{itemize}
    \item \textit{Operability} \rev{(5, 4.7\%)}: The capability of an LLM-based MAS to provide users with functions and interface features that facilitate efficient monitoring, task control, and interaction with the agents, which is closely associated with controllability of agent behaviors, robustness of the MAS to user mistakes, and alignment with user expectations.
    \item \textit{User Engagement} \rev{(5, 4.7\%)}: The capacity of an LLM-based MAS to present functions and information in a manner that is appealing and motivating, thereby encouraging sustained user interaction, which encompasses system properties that enhance user pleasure and satisfaction, such as informative and user-friendly interface. 
\end{itemize}

Designers of LLM-based MASs prioritize ensuring ease of operation and effective human–system interaction to enhance the usability of MASs and their user experience. 
For example, Josifoski et al. proposed an LLM-based MAS called \textit{Flows} for code generation, which ``\textit{supports research in the design of interactions involving humans as computational building blocks in a way that maximizes the utility of the overall computation with minimal human effort}'', to promise that the whole system is user-friendly to human beings~\hyperlink{S74}{[S74]}.

\subsection{Category of Design Patterns Employed (RQ3)}\label{sec:RQ3_results}

{\tiny
\begin{table}[htbp]
\centering
\renewcommand{\arraystretch}{1.3}
\caption{\rev{Design Patterns used by Designers of LLM-based MASs for SE tasks}}
\label{Design Patterns of LLM-based MASs}
\begin{tabular}{>{\centering\arraybackslash}m{2.4cm}
                       >{\raggedright}m{4.3cm}
                       m{1.5cm}<{\centering}
                       >{\raggedright\arraybackslash}m{4.3cm}}
\hline
\textbf{Design Pattern}                                           & \textbf{Example} 
& \textbf{Count(\%)}       & \textbf{Studies}           \\ \hline
Role-Based Cooperation                                      & \textit{Specifically, INTERVENOR employs two LLM-based agents to play different roles in code repair.}~\hyperlink{S78}{[S78]}    
& \rev{49 (46.2\%)}               & \hyperlink{S2}{[S2]}, \hyperlink{S4}{[S4]}, \hyperlink{S5}{[S5]}, \hyperlink{S7}{[S7]}, \hyperlink{S11}{[S11]}, \hyperlink{S13}{[S13]}, \hyperlink{S14}{[S14]}, \hyperlink{S16}{[S16]}, \hyperlink{S18}{[S18]}, \hyperlink{S19}{[S19]}, \hyperlink{S21}{[S21]}, \hyperlink{S26}{[S26]}, \hyperlink{S28}{[S28]}, \hyperlink{S29}{[S29]}, \hyperlink{S30}{[S30]}, \hyperlink{S31}{[S31]}, \hyperlink{S32}{[S32]}, \hyperlink{S35}{[S35]}, \hyperlink{S36}{[S36]}, \hyperlink{S37}{[S37]}, \hyperlink{S38}{[S38]}, \hyperlink{S42}{[S42]}, \hyperlink{S44}{[S44]}, \hyperlink{S46}{[S46]}, \hyperlink{S47}{[S47]}, \hyperlink{S49}{[S49]}, \hyperlink{S51}{[S51]}, \hyperlink{S55}{[S55]}, \hyperlink{S56}{[S56]}, \hyperlink{S57}{[S57]}, \hyperlink{S58}{[S58]}, \hyperlink{S59}{[S59]}, \hyperlink{S60}{[S60]}, \hyperlink{S61}{[S61]}, \hyperlink{S63}{[S63]}, \hyperlink{S64}{[S64]}, \hyperlink{S67}{[S67]}, \hyperlink{S69}{[S69]}, \hyperlink{S71}{[S71]}, \hyperlink{S78}{[S78]}, \hyperlink{S81}{[S81]}, \hyperlink{S86}{[S86]}, \hyperlink{S93}{[S93]}, \hyperlink{S94}{[S94]}, \hyperlink{S95}{[S95]}, \hyperlink{S96}{[S96]}, \hyperlink{S97}{[S97]}, \hyperlink{S99}{[S99]}, \hyperlink{S103}{[S103]}
  \\ \hline
Self-Reflection                                             & \textit{..., in order to reduce hallucinations and inefficient planning, we apply a self-reflection mechanism.}~\hyperlink{S43}{[S43]}        
& \rev{36 (34.0\%)}               & \hyperlink{S1}{[S1]}, \hyperlink{S2}{[S2]}, \hyperlink{S3}{[S3]}, \hyperlink{S11}{[S11]}, \hyperlink{S19}{[S19]}, \hyperlink{S22}{[S22]}, \hyperlink{S24}{[S24]}, \hyperlink{S27}{[S27]}, \hyperlink{S31}{[S31]}, \hyperlink{S36}{[S36]}, \hyperlink{S37}{[S37]}, \hyperlink{S38}{[S38]}, \hyperlink{S39}{[S39]}, \hyperlink{S42}{[S42]}, \hyperlink{S43}{[S43]}, \hyperlink{S45}{[S45]}, \hyperlink{S47}{[S47]}, \hyperlink{S49}{[S49]}, \hyperlink{S53}{[S53]}, \hyperlink{S57}{[S57]}, \hyperlink{S58}{[S58]}, \hyperlink{S60}{[S60]}, \hyperlink{S62}{[S62]}, \hyperlink{S63}{[S63]}, \hyperlink{S72}{[S72]}, \hyperlink{S73}{[S73]}, \hyperlink{S74}{[S74]}, \hyperlink{S75}{[S75]}, \hyperlink{S79}{[S79]}, \hyperlink{S81}{[S81]}, \hyperlink{S87}{[S87]}, \hyperlink{S90}{[S90]}, \hyperlink{S91}{[S91]}, \hyperlink{S92}{[S92]}, \hyperlink{S100}{[S100]}, \hyperlink{S102}{[S102]}
 \\ \hline
Tool-Agent Registry                                         & \textit{Across five tasks with Python and WikiSearch API as tools, Lemur-70B-Chat outperforms both Llama-2-70B-Chat and CodeLlama-34B-INST by large margins.}~\hyperlink{S79}{[S79]}                          
& \rev{16 (15.1\%)}               & \hyperlink{S2}{[S2]}, \hyperlink{S8}{[S8]}, \hyperlink{S9}{[S9]}, \hyperlink{S12}{[S12]}, \hyperlink{S25}{[S25]}, \hyperlink{S27}{[S27]}, \hyperlink{S75}{[S75]}, \hyperlink{S76}{[S76]}, \hyperlink{S79}{[S79]}, \hyperlink{S80}{[S80]}, \hyperlink{S82}{[S82]}, \hyperlink{S85}{[S85]}, \hyperlink{S88}{[S88]}, \hyperlink{S94}{[S94]}, \hyperlink{S101}{[S101]}, \hyperlink{S104}{[S104]}
 \\ \hline
Cross-Reflection                                            & \textit{Meanwhile, this Stderr information is provided to the questioner, who will generate a natural language description based on the Stderr. This natural language description is also appended to the dialogue messages. Next, both the natural language description and the Stderr are provided as new questions to the programmer, who will continue to modify the code.}~\hyperlink{S7}{[S7]}     
& \rev{14 (13.2\%)}               & \hyperlink{S7}{[S7]}, \hyperlink{S13}{[S13]}, \hyperlink{S26}{[S26]}, \hyperlink{S41}{[S41]}, \hyperlink{S50}{[S50]}, \hyperlink{S61}{[S61]}, \hyperlink{S65}{[S65]}, \hyperlink{S67}{[S67]}, \hyperlink{S69}{[S69]}, \hyperlink{S71}{[S71]}, \hyperlink{S74}{[S74]}, \hyperlink{S78}{[S78]}, \hyperlink{S98}{[S98]}, \hyperlink{S105}{[S105]}
 \\ \hline
Retrieval-Augmented Generation (RAG)                        & \textit{Furthermore, we draw inspiration from the RAG (Retrieval-Augmented Generation) approach, where we match similar  content from a memory pool as additional information.}~\hyperlink{S36}{[S36]}   
& \rev{12 (11.3\%)}               & \hyperlink{S8}{[S8]}, \hyperlink{S23}{[S23]}, \hyperlink{S36}{[S36]}, \hyperlink{S45}{[S45]}, \hyperlink{S51}{[S51]}, \hyperlink{S54}{[S54]}, \hyperlink{S55}{[S55]}, \hyperlink{S64}{[S64]}, \hyperlink{S89}{[S89]}, \hyperlink{S94}{[S94]}, \hyperlink{S101}{[S101]}, \hyperlink{S104}{[S104]}
 \\ \hline
Incremental Model Querying                                  & \textit{These patches might be from running a single SWE agent multiple times or running multiple SWE agents.}~\hyperlink{S17}{[S17]} 
& \rev{10 (9.4\%)}                & \hyperlink{S4}{[S4]}, \hyperlink{S15}{[S15]}, \hyperlink{S17}{[S17]}, \hyperlink{S98}{[S98]}, \hyperlink{S100}{[S100]}, \hyperlink{S102}{[S102]}, \hyperlink{S103}{[S103]}, \hyperlink{S104}{[S104]}, \hyperlink{S105}{[S105]}, \hyperlink{S106}{[S106]}
 \\ \hline
Agent Adapter                                               & \textit{We give the agent access to tools, including access to: ...}~\hyperlink{S25}{[S25]}   
& \rev{8 (7.5\%)}                 & \hyperlink{S6}{[S6]}, \hyperlink{S9}{[S9]}, \hyperlink{S22}{[S22]}, \hyperlink{S25}{[S25]}, \hyperlink{S27}{[S27]}, \hyperlink{S94}{[S94]}, \hyperlink{S97}{[S97]}, \hyperlink{S101}{[S101]}
 \\ \hline
Single-Path Plan Generator                                  & \textit{AXNav consists of three main components that are used to prepare for, execute, and export test results: ...}~\hyperlink{S66}{[S66]}    
& \rev{7 (6.6\%)}                & \hyperlink{S10}{[S10]}, \hyperlink{S15}{[S15]}, \hyperlink{S48}{[S48]}, \hyperlink{S66}{[S66]}, \hyperlink{S68}{[S68]}, \hyperlink{S97}{[S97]}, \hyperlink{S99}{[S99]}
 \\ \hline
Human-Reflection                                            & \textit{AISD is designed to keep the user in the loop, i.e., by repeatedly taking user feedback on use cases, high-level system designs, and prototype implementations through system testing.}~\hyperlink{S20}{[S20]}
& \rev{5 (4.7\%)}                & \hyperlink{S20}{[S20]}, \hyperlink{S27}{[S27]}, \hyperlink{S40}{[S40]}, \hyperlink{S84}{[S84]}, \hyperlink{S94}{[S94]}
 \\ \hline
Debate-Based Cooperation                                    & \textit{To address this, we implemented a Multi-Agent Debate (MAD) mechanism to establish a loop between generator and validator.}~\hyperlink{S3}{[S3]}
& \rev{5 (4.7\%)}                & \hyperlink{S3}{[S3]}, \hyperlink{S35}{[S35]}, \hyperlink{S73}{[S73]}, \hyperlink{S83}{[S83]}, \hyperlink{S98}{[S98]}
 \\ \hline
Hierarchical Cooperation                                   & \textit{Our method employs a novel hierarchical coordination paradigm, inspired by a cognitive debugging model, to efficiently manage cognitive steps with minimal communication and dynamically adjust to bug complexity through its three-level architecture.}~\hyperlink{S2}{[S2]} 
& \rev{5 (4.7\%)}                & \hyperlink{S2}{[S2]}, \hyperlink{S34}{[S34]}, \hyperlink{S76}{[S76]}, \hyperlink{S104}{[S104]}, \hyperlink{S106}{[S106]}
 \\ \hline
Prompt/Response Optimiser                                   & \textit{The Requirements Engineering agent is an integral part of our  engine, leveraging AI capabilities to automate the generation and  prioritization of software requirements.}~\hyperlink{S21}{[S21]}     
& \rev{4 (3.8\%)}                & \hyperlink{S4}{[S4]}, \hyperlink{S9}{[S9]}, \hyperlink{S21}{[S21]}, \hyperlink{S33}{[S33]}
 \\ \hline
Layered-Based Cooperation                                   & \textit{..., we propose a layered approach for implementing capabilities in LLM-based applications by mapping them to the layers and components with corresponding attributes.}~\hyperlink{S45}{[S45]} 
& \rev{4 (3.8\%)}                & \hyperlink{S15}{[S15]}, \hyperlink{S45}{[S45]}, \hyperlink{S52}{[S52]}, \hyperlink{S77}{[S77]}
 \\ \hline
Agent Evaluator                                             & \textit{To verify that generated fix suggestions can plausibly fix each issue, FixAlly evaluates each code modification  in its Suggestion Assessment module.}~\hyperlink{S60}{[S60]} 
& \rev{4 (3.8\%)}                & \hyperlink{S60}{[S60]}, \hyperlink{S61}{[S61]}, \hyperlink{S70}{[S70]}, \hyperlink{S95}{[S95]}
 \\ \hline
Multi-Path Plan Generator                                   & \textit{It divides the Vulnerability Identification phase into 40 targeted scenarios, each focused on a specific vulnerability type (as listed in our repository).}~\hyperlink{S26}{[S26]} 
& \rev{4 (3.8\%)}                & \hyperlink{S14}{[S14]}, \hyperlink{S23}{[S23]}, \hyperlink{S26}{[S26]}, \hyperlink{S96}{[S96]}
 \\ \hline
Voting-Based Cooperation                                    & \textit{Finally, the model merges and votes on all candidate solutions, selecting the highest-voted one as the final repair solution.}~\hyperlink{S31}{[S31]} 
& \rev{3 (2.8\%)}                & \hyperlink{S17}{[S17]}, \hyperlink{S31}{[S31]}, \hyperlink{S33}{[S33]}
 \\ \hline
\end{tabular}
\end{table}
}

As mentioned in Section \ref{sec:RQs}, design patterns have been employed to guide the construction and implementation of LLM-based MASs. In this section, we report the results of RQ3 using the architecture patterns for foundation model based agents identified by Liu et al. \citep{liu2024agent} as a starting point, together with the additional design patterns identified in our study. Table~\ref{Design Patterns of LLM-based MASs} presents the taxonomy of the design patterns extracted from the included studies. Results show that \textit{Role-Based Cooperation} \rev{(46.2\%)} is the most frequently used design pattern by the designers of LLM-based MASs. \textit{Self-Reflection} \rev{(34.0\%)} is also widely used in designing LLM-based MASs for SE tasks. The remaining design patterns employed are \textit{Tool-Agent Registry} \rev{(15.1\%)}, \textit{Cross-Reflection} \rev{(13.2\%)}, \textit{Retrieval-Augmented Generation (RAG)} \rev{(11.3\%)}, \textit{Incremental Model Querying} \rev{(9.4\%)}, \textit{Agent Adapter} \rev{(7.5\%)}, \textit{Single-Path Plan Generator} \rev{(6.6\%)}, \textit{Human-Reflection} \rev{(4.7\%)}, \textit{Debate-Based Cooperation} \rev{(4.7\%)}, \textit{Hierarchical Cooperation} \rev{(4.7\%)}, \textit{Prompt/Response Optimiser} \rev{(3.8\%)}, \textit{Layered-Based Cooperation} \rev{(3.8\%)}, \textit{Agent Evaluator} \rev{(3.8\%)}, \textit{Multi-Path Plan Generator} \rev{(3.8\%)}, and \textit{Voting-Based Cooperation} \rev{(2.8\%)}. Since one MAS may also use multiple design patterns, the sum of the percentages of the design patterns exceeds 100\%.

\fakesection[1em]{Role-Based Cooperation} \rev{(49, 46.2\%)} is a collaborative pattern in which each LLM-based agent adopts a distinct role that specifies functional responsibilities and interaction protocols within the system. These roles aim to improve overall agent performance by leveraging specialized capabilities and ensuring efficient task allocation and coordination. Applying \textit{Role-Based Cooperation} in MAS design enables clear allocation of responsibility to agents, enhances the scalability and adaptability of the MAS, and supports flexible coordination strategies between agents. For example, Dong et al. proposed an LLM-based MAS for code generation, and assigned the agents to three roles: \textit{Analyst}, \textit{Coder}, \textit{Tester}~\hyperlink{S86}{[S86]}.

\fakesection[1em]{Self-Reflection} \rev{(36, 34.0\%)} denotes a pattern in which certain agents use LLM capabilities to perform introspective analysis of its own actions, decisions, and performance. Through this process, the agents with the reflection mechanism can identify improvement opportunities, adapt to dynamic environments, and optimize contributions to collaborative tasks, thereby increasing the effectiveness of cooperation between agents. This pattern indicates that designers of LLM-based MASs enable agents to carry out introspection and to iteratively refine outputs, which reduces errors and accommodates changing task requirements. For instance, Hong et al. proposed an LLM-based MAS called \textit{MetaGPT}  for software development, and its code agent can ``\textit{improve code using its own historical execution and debugging memory}'' with the help of \textit{Self-Reflection}~\hyperlink{S81}{[S81]}.

\fakesection[1em]{Tool-Agent Registry} \rev{(16, 15.1\%)} refers to a centralized repository that maintains a comprehensive and unified resource, readily accessible for assigning appropriate tools to agents according to task requirements, thereby enabling effective and reliable task execution. The registry serves as an integral mechanism for agent collaboration by enabling dynamic discovery, selection, and utilization of tools. Through this mechanism, it improves output quality, reduces model workload and latency, thereby enhancing the flexibility of the whole MAS. For example, Fan et al. proposed an LLM-based MAS called \textit{ICAA} for code analysis, and set up a special ``\textit{agent integrating a toolbox that includes context-aware splitting, code retrieval, documentation retrieval, Web search, and static code analysis tools}''. This agent incorporates a ``\textit{thinking–decision–action loop}'' to select and schedule tools for different ``\textit{sub-agents}''~\hyperlink{S88}{[S88]}.

\fakesection[1em]{Cross-Reflection} \rev{(14, 13.2\%)} denotes a collaborative pattern in which each agent uses LLMs to observe, analyze, and interpret the behaviors, decisions, and outputs of other agents in the MAS. This process allows agents to gain insights into the strategies and performance of their peers, enabling adaptive learning, strategic alignment, and improved coordination between agents. It also supports error detection and correction of artifacts generated by peer agents, reduces individual biases of agents, and fosters collective calibration and consensus between agents. For example, Wang et al. proposed an LLM-based MAS named \textit{INTERVENOR} for code generation, which uses a method ``\textit{Chain of Repair}'' to introduce a ``\textit{Code Teacher}'' agent that provides feedback for ``\textit{Code Learner}'' to reflect upon and revise the generated code~\hyperlink{S78}{[S78]}.

\fakesection[1em]{Retrieval-Augmented Generation (RAG)} \rev{(12, 11.3\%)} refers to a collaborative pattern that integrates retrieval-based methods with generative capabilities to enhance the problem-solving and decision-making processes of agents. This pattern allows agents to retrieve relevant information from external knowledge sources and databases, and integrate the information seamlessly into their generated outputs and actions. \textit{RAG} enables agents to access up-to-date and contextually appropriate data, enhancing the accuracy, relevance, and effectiveness of their contributions to collaborative tasks. For instance, Wu et al. proposed an LLM-based MAS called \textit{AutoGen} that uses ``\textit{RAG}'' to enhance the quality of code generation and question answering~\hyperlink{S64}{[S64]}.

\fakesection[1em]{Incremental Model Querying} \rev{(10, 9.4\%)} is a collaborative pattern in which agents decompose complex tasks into smaller, sequential sub-queries and iteratively query LLMs, with each query building on the outputs of prior ones to refine understanding, integrate intermediate results, and gradually converge toward a comprehensive and high-quality solution. For example, Qin et al. proposed an LLM-based MAS called \textit{AGENTFL}, which adopts ``\textit{a Multi-Round Dialogue strategy in the Method Review task}'' to improve the efficiency of fault localization~\hyperlink{S4}{[S4]}.

\fakesection[1em]{Agent Adapter} \rev{(8, 7.5\%)} is a pattern that enables smooth interaction and collaboration between various agents and  external tools by standardizing communication protocols, translating data formats, and managing interfaces within the MAS. This pattern allows agents to exchange information and coordinate actions effectively, thereby improving the overall performance and adaptability of the MAS. For example, Tufano et al. proposed an LLM-based MAS for code generation, using a component called ``\textit{Conversation Manager}'', which can ``\textit{parse the commands and invoke the Tools Library}'', to help the agents operate the external tools~\hyperlink{S9}{[S9]}.

\fakesection[1em]{Single-Path Plan Generator} \rev{(7, 6.6\%)} refers to a collaborative pattern tasked with designing a linear and sequential plan to accomplish a specific goal and task. Initially, a \textit{Planner} agent (or a team of cooperating \textit{Planner} agents) constructs a cohesive, linear, and executable plan to achieve a specified objective. The resulting plan consists of a single, non-branching pathway of actions and decisions, determined by analyzing the input data, the operational capabilities of the involved agents, and the environmental constraints. The plan is decomposed into constituent sub-tasks which are allocated to other agents for cooperative execution. This pattern provides a clear global plan that reduces coordination ambiguity and conflict among agents, and enables efficient task decomposition and deterministic assignment of sub-tasks which lowers negotiation overhead and improves throughput. For example, Taeb et al. proposed an LLM-based MAS named \textit{AXNav} for software testing, which can ``\textit{formulate a step-by-step plan that accomplishes the goal}'' provided by the users of the MAS~\hyperlink{S66}{[S66]}.

\fakesection[1em]{Human-Reflection} \rev{(5, 4.7\%)} refers to a collaborative pattern that human users observe, evaluate, and provide reflective feedback on the actions, decisions, and outputs of agents. This feedback is then utilized by the agents to adjust their strategies, improve their performance, and enhance their collaborative capability, thereby ensuring better alignment with human goals and ethical considerations. For instance, Fakih et al. proposed an LLM-based MAS for code generation that ``\textit{integrates user feedback loops}'' into the pipeline for agents to reflect upon~\hyperlink{S27}{[S27]}.

\fakesection[1em]{Debate-Based Cooperation} \rev{(5, 4.7\%)} denotes a collaborative pattern by which multiple agents, leveraging the reasoning and argumentative capabilities of LLMs, engage in structured debates to evaluate various perspectives, challenge assumptions and converge on optimal solutions for a given task. Each agent adopts a distinct viewpoint, presenting arguments supported by evidence and logic, while critically assessing the arguments of others. This pattern enhances MAS design by structuring adversarial argumentation among agents to surface diverse perspectives, reveal and correct errors and weak reasoning. For example, Zhang et al. proposed an LLM-based MAS named \textit{ACFIX} for software maintenance, which introduced ``\textit{Multi-Agent Debate (MAD) mechanism}'' between the different agent roles~\hyperlink{S3}{[S3]}.

\fakesection[1em]{Hierarchical Cooperation} \rev{(5, 4.7\%)} refers to a dynamic and scalable collaborative pattern that employs a tiered approach to resolve the given SE tasks. Initially, a minimal set of agents that leverage the capabilities of LLMs collaborate through simple coordination strategies to address a given SE task. If the task remains unsolved, the MAS can gradually expand the collaboration framework by introducing additional agents and tools, thereby enhancing its capacity with increased complexity. This iterative process continues until the problem is successfully solved or the system reaches its predefined complexity limit. For example, Lee et al. proposed an LLM-based MAS called \textit{FixAgent} under the guidance of \textit{Hierarchical Cooperation} in which simple software maintenance tasks are handled by ``\textit{the simple L1} agents'', while more complex maintenance tasks trigger the incremental incorporation of additional agents and external tools to support deeper analysis and more sophisticated task solving~\hyperlink{S2}{[S2]}.

\fakesection[1em]{Prompt/Response Optimiser} \rev{(4, 3.8\%)} is a specialized pattern that dynamically refines and enhances the prompts provided to LLMs and optimizes the generated responses to improve their relevance, accuracy, and alignment with the objectives of the agent system. This pattern employs iterative feedback loops, context analysis, and performance metrics to fine-tune input prompts and evaluate output quality, ensuring effective communication and coordination among agents. For instance, Tufano et al. proposed an LLM-based MAS named \textit{AutoDev} for code generation, which can refine the prompts provided by the users ``\textit{in a predefined format}''~\hyperlink{S9}{[S9]}.

\fakesection[1em]{Layered-Based Cooperation} \rev{(4, 3.8\%)} refers to a structured collaborative pattern wherein agents are organized into hierarchical layers, each layer assigned specific agent roles, tasks, and levels of abstraction based on their capabilities. Agents within each layer leverage the generative and reasoning capacities of LLMs to process inputs, execute tasks, and produce outputs that are passed to subsequent layers for further refinement and integration. This pattern separates concerns across hierarchical abstraction levels, enabling role specialization, modularity, and scalable coordination. For instance, Zan et al. proposed an LLM-based MAS named \textit{CODES}, which is formulated in a ``\textit{Multi-Layer Sketch}'' to accomplish the repository-level code generation~\hyperlink{S15}{[S15]}.

\fakesection[1em]{Agent Evaluator} \rev{(4, 3.8\%)} refers to a specialized pattern, which introduces an independent, specialized agent that systematically conducts metric-based evaluation and governance of the performance, decisions, and outputs of other agents in the MAS, thereby supporting targeted remediation, quality assurance, and continuous improvement of the MAS. For example, Huang et al. proposed an LLM-based MAS named \textit{AgentCoder}, which implements a ``\textit{test agent}'' to evaluate the results of code generation. ``\textit{If all test cases are passed, the agent returns the code to the human developer}''~\hyperlink{S61}{[S61]}.

\fakesection[1em]{Multi-Path Plan Generator} \rev{(4, 3.8\%)} denotes a specialized pattern that generates multiple alternative paths and action options of agents at each intermediate step, then assembles them into the final task execution plan. Each plan represents a unique sequence of actions and decisions that account for various scenarios, agent capabilities, and environmental variables. This pattern increases the robustness and adaptability of an LLM-based MAS by generating multiple plans that support the selection of optimal or complementary strategies across varying agent capabilities and environmental conditions. For example, Ma et al. proposed an LLM-based MAS for code generation, which can generate multiple plans to complete the code generation tasks given by users, and the MAS ``\textit{iteratively narrows down the search space and guides the agents to focus on the most relevant area by simulating multiple workflows and evaluating their reward score}''~\hyperlink{S23}{[S23]}.

\fakesection[1em]{Voting-Based Cooperation} \rev{(3, 2.8\%)} is a collaborative pattern in which multiple agents, using the reasoning and evaluation capabilities of LLMs, make collective decisions by voting on proposed actions, strategies, and solutions. Each agent contributes its perspective based on its role, expertise, and task analysis, and the final decision is determined through a predefined voting protocol, such as majority rule or weighted voting. This pattern enables democratic participation, reduces individual bias, and promotes consensus-based outcomes, thereby enhancing the robustness and fairness of collaborative task execution. For instance, Li et al. proposed an LLM-based MAS that uses voting between agents to ``\textit{determine the final answer}'' to the code generation task given by users~\hyperlink{S33}{[S33]}.

\subsection{Category of Design Rationale (RQ4)}\label{sec:RQ4_results}
As mentioned in Section \ref{sec:RQs}, LLM-based MASs are designed according to the rationale that guides their construction to support specific SE tasks. In this section, we report the results of RQ4. Table~\ref{Design Rationale of LLM-based MASs} presents the taxonomy of the design rationale of LLM-based MASs for SE tasks extracted from the included studies. Results show that \textit{Improving the Quality of Generated Code} \rev{(48.1\%)} is the most commonly used design rationale. Meanwhile, \textit{Simulating Human Processes of Solving SE Tasks} \rev{(29.2\%)} is also used by designers of LLM-based MASs. The remaining design rationale includes: \textit{Optimizing Software Resource Management} \rev{(27.4\%)}, \textit{Improving the Efficiency of Generating Software Artifacts} \rev{(23.6\%)}, \textit{Reducing the Difficulty of Task Resolution} \rev{(23.6\%)}, \textit{Improving the Adaptability of Agent Systems} \rev{(18.9\%)}, \textit{Enhancing the Diversity of Generated Software Artifacts} \rev{(13.2\%)}, and \textit{Ensuring Software Security} \rev{(2.8\%)}.

\fakesection[1em]{Improving the Quality of Generated Code} \rev{(51, 48.1\%)} represents the most frequently adopted design rationale. 
Code is a key artifact generated within an LLM-based MAS, and the quality of the generated code has a direct impact on the final product produced by LLM-based MASs for SE tasks. 
Defects in generated code can propagate to subsequent artifacts and undermine the reliability, maintainability, and overall quality of the final software products. Therefore, designers place great emphasis on improving code quality and adopt a variety of strategies to achieve this goal. For example, Lei et al. proposed an LLM-based MAS for code generation, which introduces ``\textit{an interaction system comprising two agents}'' (\textit{questioner} and \textit{programmer}) to simulate the process of programmers writing code according to project requirements and conducting unit tests. The MAS ``\textit{uses multiple rounds of execution feedback to check the generated code}'' for improving the code quality~\hyperlink{S7}{[S7]}.

{\tiny
\begin{table}[htbp]
\centering
\renewcommand{\arraystretch}{1.3}
\caption{\rev{Design Rationale of LLM-based MASs for SE Tasks}}
\label{Design Rationale of LLM-based MASs}
\begin{tabular}{>{\centering\arraybackslash}m{2.4cm}
                       >{\raggedright}m{4.3cm}
                       m{1.5cm}<{\centering}
                       >{\raggedright\arraybackslash}m{4.3cm}}
\hline
\textbf{Design Rationale}                                           & \textbf{Example} 
& \textbf{Count(\%)}   & \textbf{Studies}           \\ \hline
Improving the Quality of Generated Code                     & \textit{We instructed each agent to iterate with the others at least three times to refine and update the code based on feedback from the previous agents, ensuring a thorough and collaborative development process.}~\hyperlink{S13}{[S13]}    
& \rev{51 (48.1\%)}               & \hyperlink{S1}{[S1]}, \hyperlink{S5}{[S5]}, \hyperlink{S6}{[S6]}, \hyperlink{S7}{[S7]}, \hyperlink{S11}{[S11]}, \hyperlink{S12}{[S12]}, \hyperlink{S13}{[S13]}, \hyperlink{S24}{[S24]}, \hyperlink{S27}{[S27]}, \hyperlink{S29}{[S29]}, \hyperlink{S31}{[S31]}, \hyperlink{S33}{[S33]}, \hyperlink{S36}{[S36]}, \hyperlink{S40}{[S40]}, \hyperlink{S41}{[S41]}, \hyperlink{S42}{[S42]}, \hyperlink{S45}{[S45]}, \hyperlink{S46}{[S46]}, \hyperlink{S47}{[S47]}, \hyperlink{S49}{[S49]}, \hyperlink{S50}{[S50]}, \hyperlink{S51}{[S51]}, \hyperlink{S52}{[S52]}, \hyperlink{S54}{[S54]}, \hyperlink{S57}{[S57]}, \hyperlink{S58}{[S58]}, \hyperlink{S59}{[S59]}, \hyperlink{S60}{[S60]}, \hyperlink{S61}{[S61]}, \hyperlink{S62}{[S62]}, \hyperlink{S64}{[S64]}, \hyperlink{S65}{[S65]}, \hyperlink{S66}{[S66]}, \hyperlink{S70}{[S70]}, \hyperlink{S73}{[S73]}, \hyperlink{S79}{[S79]}, \hyperlink{S81}{[S81]}, \hyperlink{S83}{[S83]}, \hyperlink{S90}{[S90]}, \hyperlink{S91}{[S91]}, \hyperlink{S92}{[S92]}, \hyperlink{S93}{[S93]}, \hyperlink{S95}{[S95]}, \hyperlink{S96}{[S96]}, \hyperlink{S97}{[S97]}, \hyperlink{S98}{[S98]}, \hyperlink{S99}{[S99]}, \hyperlink{S100}{[S100]}, \hyperlink{S101}{[S101]}, \hyperlink{S103}{[S103]}, \hyperlink{S105}{[S105]}
 \\ \hline
Simulating Human Processes of Solving SE Tasks              & \textit{Specifically, iAudit is inspired by the observation that expert human auditors first perceive what could be wrong and then perform a detailed analysis of the code to identify the cause.}~\hyperlink{S16}{[S16]}        
& \rev{31 (29.2\%)}               & \hyperlink{S2}{[S2]}, \hyperlink{S5}{[S5]}, \hyperlink{S7}{[S7]}, \hyperlink{S10}{[S10]}, \hyperlink{S11}{[S11]}, \hyperlink{S12}{[S12]}, \hyperlink{S14}{[S14]}, \hyperlink{S16}{[S16]}, \hyperlink{S19}{[S19]}, \hyperlink{S26}{[S26]}, \hyperlink{S28}{[S28]}, \hyperlink{S29}{[S29]}, \hyperlink{S30}{[S30]}, \hyperlink{S35}{[S35]}, \hyperlink{S42}{[S42]}, \hyperlink{S44}{[S44]}, \hyperlink{S53}{[S53]}, \hyperlink{S55}{[S55]}, \hyperlink{S58}{[S58]}, \hyperlink{S59}{[S59]}, \hyperlink{S64}{[S64]}, \hyperlink{S67}{[S67]}, \hyperlink{S71}{[S71]}, \hyperlink{S72}{[S72]}, \hyperlink{S78}{[S78]}, \hyperlink{S81}{[S81]}, \hyperlink{S86}{[S86]}, \hyperlink{S87}{[S87]}, \hyperlink{S95}{[S95]}, \hyperlink{S97}{[S97]}, \hyperlink{S99}{[S99]}
 \\ \hline
Optimizing Software Resource Management                     & \textit{We posit that our approach will considerably augment the field of software quality assurance, rendering it more efficient, precise, and cost-effective.}~\hyperlink{S88}{[S88]}                          
& \rev{29 (27.4\%)}               & \hyperlink{S2}{[S2]}, \hyperlink{S4}{[S4]}, \hyperlink{S6}{[S6]}, \hyperlink{S8}{[S8]}, \hyperlink{S14}{[S14]}, \hyperlink{S18}{[S18]}, \hyperlink{S19}{[S19]}, \hyperlink{S20}{[S20]}, \hyperlink{S25}{[S25]}, \hyperlink{S28}{[S28]}, \hyperlink{S32}{[S32]}, \hyperlink{S34}{[S34]}, \hyperlink{S35}{[S35]}, \hyperlink{S36}{[S36]}, \hyperlink{S37}{[S37]}, \hyperlink{S40}{[S40]}, \hyperlink{S45}{[S45]}, \hyperlink{S46}{[S46]}, \hyperlink{S47}{[S47]}, \hyperlink{S48}{[S48]}, \hyperlink{S49}{[S49]}, \hyperlink{S58}{[S58]}, \hyperlink{S63}{[S63]}, \hyperlink{S68}{[S68]}, \hyperlink{S81}{[S81]}, \hyperlink{S87}{[S87]}, \hyperlink{S88}{[S88]}, \hyperlink{S104}{[S104]}, \hyperlink{S105}{[S105]}
 \\ \hline
Improving the Efficiency of Generating Software Artifacts   & \textit{ Moreover, our  analysis of agent interactions within AGENTVERSE reveals the emergence of specific collaborative behaviors, contributing to heightened group efficiency.}~\hyperlink{S62}{[S62]}                          
& \rev{25 (23.6\%)}               &  \hyperlink{S2}{[S2]}, \hyperlink{S3}{[S3]}, \hyperlink{S4}{[S4]}, \hyperlink{S9}{[S9]}, \hyperlink{S13}{[S13]}, \hyperlink{S17}{[S17]}, \hyperlink{S19}{[S19]}, \hyperlink{S21}{[S21]}, \hyperlink{S27}{[S27]}, \hyperlink{S39}{[S39]}, \hyperlink{S40}{[S40]}, \hyperlink{S45}{[S45]}, \hyperlink{S49}{[S49]}, \hyperlink{S61}{[S61]}, \hyperlink{S62}{[S62]}, \hyperlink{S63}{[S63]}, \hyperlink{S64}{[S64]}, \hyperlink{S73}{[S73]}, \hyperlink{S75}{[S75]}, \hyperlink{S87}{[S87]}, \hyperlink{S88}{[S88]}, \hyperlink{S90}{[S90]}, \hyperlink{S91}{[S91]}, \hyperlink{S104}{[S104]}, \hyperlink{S106}{[S106]}
 \\  \hline
Reducing the Difficulty of Task Resolution                  & \textit{Specifically, by decomposing the localization process into several steps and achieving FL through the collaboration of multiple LLM-driven agents (i.e., intelligent entities that can perceive the environment, make decisions, and perform actions), the advantages we expect are twofold.}~\hyperlink{S4}{[S4]}     
& \rev{25 (23.6\%)}               & \hyperlink{S1}{[S1]}, \hyperlink{S2}{[S2]}, \hyperlink{S4}{[S4]}, \hyperlink{S13}{[S13]}, \hyperlink{S14}{[S14]}, \hyperlink{S15}{[S15]}, \hyperlink{S20}{[S20]},
\hyperlink{S22}{[S22]},
\hyperlink{S32}{[S32]}, \hyperlink{S43}{[S43]}, \hyperlink{S53}{[S53]}, \hyperlink{S56}{[S56]}, \hyperlink{S62}{[S62]}, \hyperlink{S64}{[S64]}, \hyperlink{S69}{[S69]}, \hyperlink{S72}{[S72]}, \hyperlink{S74}{[S74]}, \hyperlink{S84}{[S84]}, \hyperlink{S86}{[S86]}, \hyperlink{S93}{[S93]}, \hyperlink{S100}{[S100]}, \hyperlink{S101}{[S101]}, \hyperlink{S102}{[S102]}, \hyperlink{S104}{[S104]}, \hyperlink{S106}{[S106]}
 \\ \hline
Improving the Adaptability of Agent Systems                  & \textit{Our proposal for dynamic process generation aims to accommodate this variability, enabling a wider array of diverse instances to emerge and guide the development process accordingly.}~\hyperlink{S57}{[S57]}   
& \rev{20 (18.9\%)}               &  \hyperlink{S1}{[S1]}, \hyperlink{S2}{[S2]}, \hyperlink{S9}{[S9]}, \hyperlink{S12}{[S12]}, \hyperlink{S31}{[S31]}, \hyperlink{S33}{[S33]}, \hyperlink{S36}{[S36]}, \hyperlink{S37}{[S37]}, \hyperlink{S38}{[S38]}, \hyperlink{S47}{[S47]}, \hyperlink{S52}{[S52]}, \hyperlink{S53}{[S53]}, \hyperlink{S57}{[S57]}, \hyperlink{S75}{[S75]}, \hyperlink{S76}{[S76]}, \hyperlink{S89}{[S89]}, \hyperlink{S91}{[S91]}, \hyperlink{S94}{[S94]}, \hyperlink{S100}{[S100]}, \hyperlink{S101}{[S101]}
 \\  \hline
Enhancing the Diversity of Generated Software Artifacts     & \textit{The parallel generation method with KMeans filtering helps improve the diversity than using parallel generation only, albeit not significantly.}~\hyperlink{S18}{[S18]}   
& \rev{14 (13.2\%)}               & \hyperlink{S18}{[S18]}, \hyperlink{S21}{[S21]}, \hyperlink{S23}{[S23]}, \hyperlink{S26}{[S26]}, \hyperlink{S38}{[S38]}, \hyperlink{S45}{[S45]}, \hyperlink{S48}{[S48]}, \hyperlink{S57}{[S57]}, \hyperlink{S59}{[S59]}, \hyperlink{S80}{[S80]}, \hyperlink{S85}{[S85]}, \hyperlink{S89}{[S89]}, \hyperlink{S96}{[S96]}, \hyperlink{S103}{[S103]}
 \\ \hline
Ensuring Software Security                                  & \textit{The system should be designed with robust security measures in place to prevent unauthorized access or misuse.}~\hyperlink{S82}{[S82]}             
& \rev{3 (2.8\%)}                &  \hyperlink{S77}{[S77]}, \hyperlink{S82}{[S82]}, \hyperlink{S83}{[S83]}
 \\ \hline
\end{tabular}
\end{table}
}

\fakesection[1em]{Simulating Human Processes of Solving SE Tasks} \rev{(31, 29.2\%)} is another design rationale frequently employed by designers. LLM-based MASs are inspired by human approaches to solving SE tasks, and their designs and implementations are naturally built on this foundation. For example, Zhang et al. proposed an LLM-based MAS named \textit{Self-Edit}, which ``\textit{is inspired by the problem-solving process of human programmers}'' to assign roles and tasks to agents~\hyperlink{S87}{[S87]}.

\fakesection[1em]{Optimizing Software Resource Management} \rev{(29, 27.4\%)} is also a widely considered design rationale. Designers of LLM-based MASs focus on the efficient allocation, utilization, and coordination of computational and memory assets among autonomous agents. For example, Qin et al. proposed an LLM-based MAS named \textit{AGENTFL}, which decomposes the whole fault localization process into three stages, and implemented the ``\textit{Document-Guided Search strategy}'' and ``\textit{retained only the Top-N classes that possess relatively high method-level coverage}'' to significantly reduce the input scale given to agents, thereby lowering costs~\hyperlink{S4}{[S4]}.

\fakesection[1em]{Improving the Efficiency of Generating Software Artifacts} \rev{(25, 23.6\%)} is a major design rationale used to support decisions. Accelerating the generation of artifacts and enhancing their success rate is an essential concern when designing LLM-based MASs for SE tasks. For example, Yang et al. proposed an LLM-based MAS for code generation, and the MAS introduced ``\textit{well-designed Agent-Computer Interfaces (ACIs)}'' for each agent. The ACIs provide agents with a structured view of the runtime environment and apply strict history management and context pruning to supply only the minimal information relevant to the current task, which helps agents ``\textit{understand the state of the application given previous changes}'' and ``\textit{avoid unnecessary context from prior observations}'', thereby improving the efficiency of generating code~\hyperlink{S40}{[S40]}.

\fakesection[1em]{Reducing the Difficulty of Task Resolution} \rev{(25, 23.6\%)} is design rationale considered by designers. Simplifying the resolution of SE tasks can facilitate the design and implementation of LLM-based MASs, enhance the quality of produced software artifacts, increase production efficiency, and reduce the costs associated with completing SE tasks. For instance, Qian et al. proposed an LLM-based MAS called \textit{ChatDev} for software development, which ``\textit{uses a chat chain to divide each phase into smaller sub-tasks further}'' and assigns simpler sub-tasks to related agents~\hyperlink{S93}{[S93]}.

\fakesection[1em]{Improving the Adaptability of Agent Systems} \rev{(20, 18.9\%)} is a commonly utilized design rationale. Designers of LLM-based MASs for SE tasks aim to enable these systems to operate in diverse runtime environments, which in turn places emphasis on enhancing the adaptability of the MASs. For instance, Xu et al. proposed an LLM-based MAS called \textit{Gentopia.AI} for code generation, which allows agents to be built in a ``\textit{hierarchical architecture}'' or a ``\textit{non-hierarchical architecture}''. This strategy enables agents to easily call external tools, integrate multiple LLMs, and invoke other ``\textit{sub-agents as  plugins}'', which improves the adaptability of the MAS~\hyperlink{S76}{[S76]}.

\fakesection[1em]{Enhancing the Diversity of Generated Software Artifacts} \rev{(14, 13.2\%)} is considered by designers. Enhancing the diversity of produced artifacts allows for the selection of artifacts from a broader range, facilitating the identification of higher-quality artifacts and increasing the likelihood of finding optimal solutions to the SE task. For example, Ataei et al. proposed an LLM-based MAS called \textit{Elicitron} for requirements engineering, which automatically creates multiple \textit{interviewer} agents. Each agent is configured with a distinct interviewing style and perspective, and the dialogue context of each agent is maintained independently, to ``\textit{identify a diverse set of user needs}''~\hyperlink{S18}{[S18]}.

\fakesection[1em]{Ensuring Software Security} \rev{(3, 2.8\%)} cannot be ignored. The security of LLM-based MASs for SE tasks is of paramount importance. Designers of LLM-based MASs have incorporated specialized measures to ensure system security. For example, Talebirad et al. proposed an LLM-based MAS for code generation, which ``\textit{addresses limitations and challenges such as security risks}''. Since the MAS has the ``\textit{ability to interact with files and execute code}'', which introduces potential security risks. This MAS uses ``\textit{a state-less oracle agent, which can monitor each sensitive task and decide if it is indeed malicious or not}''~\hyperlink{S82}{[S82]}.

\section{Discussions}\label{sec:Discussions}

\subsection{Interpretations of the Results}\label{sec:interpretation_of_results}
\begin{figure*}[htbp]
    \centering
    \includegraphics[width=\linewidth]{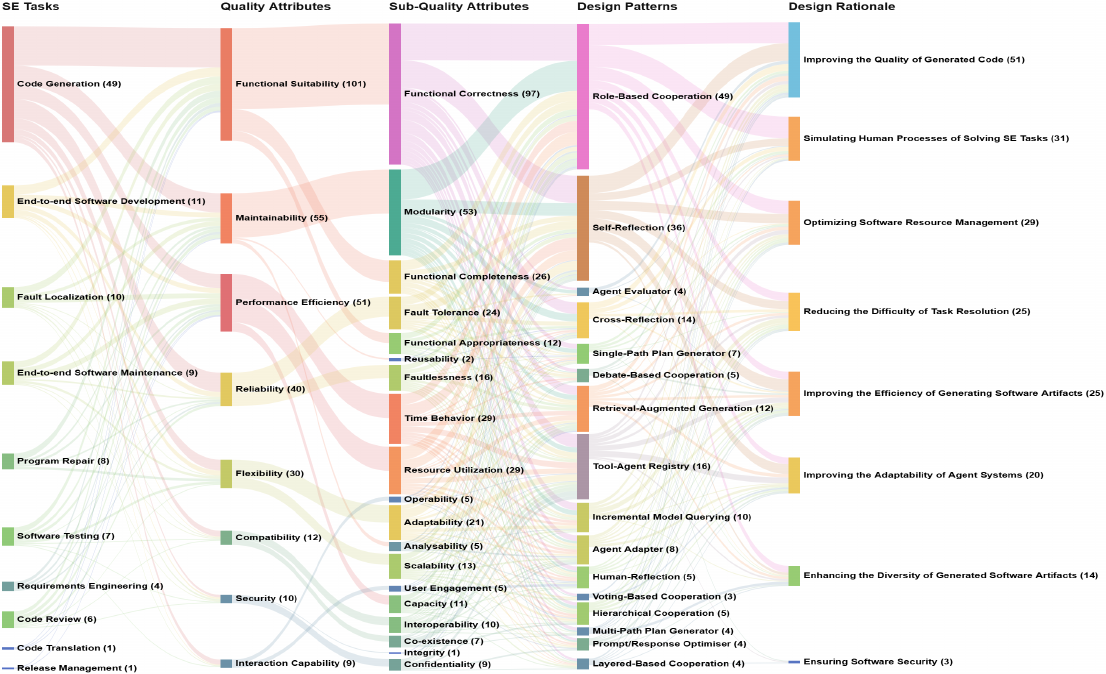}
    \caption{\rev{Pairwise Mappings between SE Tasks, Quality Attributes, Sub-Quality Attributes, Design Patterns, and Design Rationale}}
    \label{fig:mapping between SET_DP_DR}
\end{figure*}

In this section, we discuss and explain the relationship between the results of the four RQs: \textit{SE Tasks}, \textit{Quality Attributes}, \textit{Design Patterns}, and \textit{Design Rationale}. \rev{Figure~\ref{fig:mapping between SET_DP_DR} provides an overview of the pairwise mappings examined in this section across five dimensions: SE tasks, QAs, sub-QAs, design patterns, and design rationale. Each node represents a category within one dimension, and the number beside it indicates the number of studies in that category. Each band connects two categories in adjacent dimensions, and its width represents the number of studies belonging to both categories. For example, the band between \textit{Functional Correctness} and \textit{Role-Based Cooperation} represents the 44 studies that consider the former and adopt the latter. A study may consider multiple QAs and sub-QAs, adopt multiple patterns, and report multiple categories of design rationale, so it may be counted in more than one category within these dimensions. A continuous visual path across multiple dimensions does not necessarily represent the same set of studies throughout that path.}

\subsubsection{\rev{Task-Specific Quality Concerns}}\rev{This subsection examines how the quality concerns reported in the included studies vary across SE tasks. We interpret the distributions in Table~\ref{tab:QA-SET} in relation to the outputs produced, the activities performed, and the dependencies within each task workflow. This interpretation helps clarify why particular quality concerns may be relevant when designing LLM-based MASs for different SE tasks.}
\rev{Table~\ref{tab:QA-SET} maps quality concerns to SE task categories, grouped by their primary activity or workflow scope. \textit{Code Review}, \textit{Software Testing}, \textit{Fault Localization}, and \textit{Program Repair} may recur during development and maintenance, whereas \textit{End-to-End Software Development} and \textit{End-to-End Software Maintenance} span multiple activities. Each study receives one task label; systems integrating closely related activities are assigned the corresponding composite task label.} \rev{To compare categories with different study counts, we calculate the within-task prevalence~\citep{Agresti2013} of each QA or sub-QA:}

\begin{equation}
\rev{
P(q \mid t)=\frac{N(q,t)}{N(t)},
}
\label{eq:qa_task_prevalence}
\end{equation}

\rev{where $N(q,t)$ is the number of studies categorized under task $t$ that consider QA or sub-QA $q$, and $N(t)$ is the total number of studies in that task category. These proportions describe reported quality concerns. A study may consider multiple QAs or sub-QAs, therefore, the percentages within each task category do not necessarily sum to 100\%.}

\begin{table*}[h]
\centering
\caption{Mapping between Quality Attributes (vertical) and SE Tasks (horizontal)}
\label{tab:QA-SET}
\begin{adjustbox}{width=\textwidth,center}
\begin{tabular}{clcccccccccc}
\toprule
& & \multicolumn{1}{c}{\textbf{Requirements}}
  & \multicolumn{2}{c}{\textbf{Implementation}}
  & \multicolumn{4}{c}{\textbf{Quality Assurance}}
  & \multicolumn{1}{c}{\textbf{Release}}
  & \multicolumn{2}{c}{\textbf{\shortstack{End-to-End\\Workflows}}} \\
\arrayrulecolor{black}
\midrule
\textbf{Quality Attribute} & \textbf{Sub-QA}
& \textbf{RE}
& \textbf{CG}
& \textbf{CT}
& \textbf{CR}
& \textbf{ST}
& \textbf{FL}
& \textbf{PR}
& \textbf{RM}
& \textbf{ETESD}
& \textbf{ETESM} \\
\midrule


\multirow{3}{*}{\textbf{{\normalsize Functional Suitability}}}
& \cellcolor[HTML]{E0F7E0}Functional Correctness
& \cellcolor[HTML]{D9E2F3}3
& \cellcolor[HTML]{4472C4}47
& \cellcolor[HTML]{D9E2F3}1
& \cellcolor[HTML]{D9E2F3}4
& \cellcolor[HTML]{D9E2F3}7
& \cellcolor[HTML]{D9E2F3}9
& \cellcolor[HTML]{D9E2F3}8
& \cellcolor[HTML]{D9E2F3}1
& \cellcolor[HTML]{D9E2F3}9
& \cellcolor[HTML]{D9E2F3}8 \\

& \cellcolor[HTML]{E0F7E0}Functional Completeness
& \cellcolor[HTML]{D9E2F3}2
& \cellcolor[HTML]{8EAADB}13
& \cellcolor[HTML]{F2F2F2}0
& \cellcolor[HTML]{D9E2F3}3
& \cellcolor[HTML]{D9E2F3}4
& \cellcolor[HTML]{F2F2F2}0
& \cellcolor[HTML]{F2F2F2}0
& \cellcolor[HTML]{F2F2F2}0
& \cellcolor[HTML]{D9E2F3}4
& \cellcolor[HTML]{F2F2F2}0 \\

& \cellcolor[HTML]{E0F7E0}Functional Appropriateness
& \cellcolor[HTML]{F2F2F2}0
& \cellcolor[HTML]{D9E2F3}6
& \cellcolor[HTML]{F2F2F2}0
& \cellcolor[HTML]{D9E2F3}1
& \cellcolor[HTML]{F2F2F2}0
& \cellcolor[HTML]{D9E2F3}1
& \cellcolor[HTML]{D9E2F3}1
& \cellcolor[HTML]{D9E2F3}1
& \cellcolor[HTML]{D9E2F3}2
& \cellcolor[HTML]{F2F2F2}0 \\

\multirow{3}{*}{\textbf{{\normalsize Performance Efficiency}}}
& \cellcolor[HTML]{FFFFE0}Resource Utilization
& \cellcolor[HTML]{D9E2F3}2
& \cellcolor[HTML]{8EAADB}13
& \cellcolor[HTML]{F2F2F2}0
& \cellcolor[HTML]{D9E2F3}2
& \cellcolor[HTML]{D9E2F3}1
& \cellcolor[HTML]{D9E2F3}5
& \cellcolor[HTML]{D9E2F3}1
& \cellcolor[HTML]{F2F2F2}0
& \cellcolor[HTML]{D9E2F3}1
& \cellcolor[HTML]{D9E2F3}4 \\

& \cellcolor[HTML]{FFFFE0}Time Behavior
& \cellcolor[HTML]{F2F2F2}0
& \cellcolor[HTML]{D9E2F3}8
& \cellcolor[HTML]{D9E2F3}1
& \cellcolor[HTML]{D9E2F3}2
& \cellcolor[HTML]{D9E2F3}3
& \cellcolor[HTML]{D9E2F3}2
& \cellcolor[HTML]{D9E2F3}2
& \cellcolor[HTML]{D9E2F3}1
& \cellcolor[HTML]{D9E2F3}5
& \cellcolor[HTML]{D9E2F3}5 \\

& \cellcolor[HTML]{FFFFE0}Capacity
& \cellcolor[HTML]{D9E2F3}2
& \cellcolor[HTML]{D9E2F3}3
& \cellcolor[HTML]{F2F2F2}0
& \cellcolor[HTML]{D9E2F3}1
& \cellcolor[HTML]{D9E2F3}1
& \cellcolor[HTML]{D9E2F3}1
& \cellcolor[HTML]{D9E2F3}1
& \cellcolor[HTML]{F2F2F2}0
& \cellcolor[HTML]{D9E2F3}1
& \cellcolor[HTML]{D9E2F3}1 \\

\multirow{3}{*}{\textbf{{\normalsize Maintainability}}}
& \cellcolor[HTML]{E0F7E0}Modularity
& \cellcolor[HTML]{D9E2F3}3
& \cellcolor[HTML]{4472C4}21
& \cellcolor[HTML]{D9E2F3}1
& \cellcolor[HTML]{D9E2F3}5
& \cellcolor[HTML]{D9E2F3}4
& \cellcolor[HTML]{D9E2F3}4
& \cellcolor[HTML]{D9E2F3}3
& \cellcolor[HTML]{F2F2F2}0
& \cellcolor[HTML]{D9E2F3}6
& \cellcolor[HTML]{D9E2F3}6 \\

& \cellcolor[HTML]{E0F7E0}Analysability
& \cellcolor[HTML]{F2F2F2}0
& \cellcolor[HTML]{D9E2F3}4
& \cellcolor[HTML]{F2F2F2}0
& \cellcolor[HTML]{D9E2F3}1
& \cellcolor[HTML]{F2F2F2}0
& \cellcolor[HTML]{F2F2F2}0
& \cellcolor[HTML]{F2F2F2}0
& \cellcolor[HTML]{F2F2F2}0
& \cellcolor[HTML]{F2F2F2}0
& \cellcolor[HTML]{F2F2F2}0 \\

& \cellcolor[HTML]{E0F7E0}Reusability
& \cellcolor[HTML]{F2F2F2}0
& \cellcolor[HTML]{D9E2F3}1
& \cellcolor[HTML]{F2F2F2}0
& \cellcolor[HTML]{F2F2F2}0
& \cellcolor[HTML]{F2F2F2}0
& \cellcolor[HTML]{F2F2F2}0
& \cellcolor[HTML]{F2F2F2}0
& \cellcolor[HTML]{F2F2F2}0
& \cellcolor[HTML]{D9E2F3}1
& \cellcolor[HTML]{F2F2F2}0 \\

\multirow{2}{*}{\textbf{{\normalsize Reliability}}}
& \cellcolor[HTML]{FFFFE0}Fault Tolerance
& \cellcolor[HTML]{D9E2F3}1
& \cellcolor[HTML]{8EAADB}12
& \cellcolor[HTML]{F2F2F2}0
& \cellcolor[HTML]{D9E2F3}3
& \cellcolor[HTML]{D9E2F3}1
& \cellcolor[HTML]{F2F2F2}0
& \cellcolor[HTML]{D9E2F3}2
& \cellcolor[HTML]{F2F2F2}0
& \cellcolor[HTML]{D9E2F3}4
& \cellcolor[HTML]{D9E2F3}1 \\

& \cellcolor[HTML]{FFFFE0}Faultlessness
& \cellcolor[HTML]{D9E2F3}1
& \cellcolor[HTML]{8EAADB}11
& \cellcolor[HTML]{F2F2F2}0
& \cellcolor[HTML]{F2F2F2}0
& \cellcolor[HTML]{F2F2F2}0
& \cellcolor[HTML]{D9E2F3}1
& \cellcolor[HTML]{F2F2F2}0
& \cellcolor[HTML]{F2F2F2}0
& \cellcolor[HTML]{D9E2F3}2
& \cellcolor[HTML]{D9E2F3}1 \\

\multirow{2}{*}{\textbf{{\normalsize Flexibility}}}
& \cellcolor[HTML]{E0F7E0}Adaptability
& \cellcolor[HTML]{F2F2F2}0
& \cellcolor[HTML]{D9E2F3}9
& \cellcolor[HTML]{F2F2F2}0
& \cellcolor[HTML]{D9E2F3}1
& \cellcolor[HTML]{D9E2F3}3
& \cellcolor[HTML]{D9E2F3}1
& \cellcolor[HTML]{F2F2F2}0
& \cellcolor[HTML]{F2F2F2}0
& \cellcolor[HTML]{D9E2F3}4
& \cellcolor[HTML]{D9E2F3}3 \\

& \cellcolor[HTML]{E0F7E0}Scalability
& \cellcolor[HTML]{F2F2F2}0
& \cellcolor[HTML]{D9E2F3}7
& \cellcolor[HTML]{F2F2F2}0
& \cellcolor[HTML]{F2F2F2}0
& \cellcolor[HTML]{F2F2F2}0
& \cellcolor[HTML]{F2F2F2}0
& \cellcolor[HTML]{D9E2F3}3
& \cellcolor[HTML]{F2F2F2}0
& \cellcolor[HTML]{D9E2F3}2
& \cellcolor[HTML]{D9E2F3}1 \\

\multirow{2}{*}{\textbf{{\normalsize Compatibility}}}
& \cellcolor[HTML]{FFFFE0}Interoperability
& \cellcolor[HTML]{F2F2F2}0
& \cellcolor[HTML]{D9E2F3}7
& \cellcolor[HTML]{F2F2F2}0
& \cellcolor[HTML]{D9E2F3}1
& \cellcolor[HTML]{F2F2F2}0
& \cellcolor[HTML]{D9E2F3}1
& \cellcolor[HTML]{F2F2F2}0
& \cellcolor[HTML]{F2F2F2}0
& \cellcolor[HTML]{D9E2F3}1
& \cellcolor[HTML]{F2F2F2}0 \\

& \cellcolor[HTML]{FFFFE0}Co-Existence
& \cellcolor[HTML]{F2F2F2}0
& \cellcolor[HTML]{D9E2F3}4
& \cellcolor[HTML]{F2F2F2}0
& \cellcolor[HTML]{F2F2F2}0
& \cellcolor[HTML]{D9E2F3}1
& \cellcolor[HTML]{D9E2F3}1
& \cellcolor[HTML]{F2F2F2}0
& \cellcolor[HTML]{F2F2F2}0
& \cellcolor[HTML]{F2F2F2}0
& \cellcolor[HTML]{D9E2F3}1 \\

\multirow{2}{*}{\textbf{{\normalsize Security}}}
& \cellcolor[HTML]{E0F7E0}Confidentiality
& \cellcolor[HTML]{F2F2F2}0
& \cellcolor[HTML]{D9E2F3}2
& \cellcolor[HTML]{F2F2F2}0
& \cellcolor[HTML]{D9E2F3}1
& \cellcolor[HTML]{D9E2F3}1
& \cellcolor[HTML]{D9E2F3}2
& \cellcolor[HTML]{F2F2F2}0
& \cellcolor[HTML]{F2F2F2}0
& \cellcolor[HTML]{D9E2F3}2
& \cellcolor[HTML]{D9E2F3}1 \\

& \cellcolor[HTML]{E0F7E0}Integrity
& \cellcolor[HTML]{F2F2F2}0
& \cellcolor[HTML]{D9E2F3}1
& \cellcolor[HTML]{F2F2F2}0
& \cellcolor[HTML]{F2F2F2}0
& \cellcolor[HTML]{F2F2F2}0
& \cellcolor[HTML]{F2F2F2}0
& \cellcolor[HTML]{F2F2F2}0
& \cellcolor[HTML]{F2F2F2}0
& \cellcolor[HTML]{F2F2F2}0
& \cellcolor[HTML]{F2F2F2}0 \\

\multirow{2}{*}{\textbf{{\normalsize Interaction Capability}}}
& \cellcolor[HTML]{FFFFE0}User Engagement
& \cellcolor[HTML]{F2F2F2}0
& \cellcolor[HTML]{D9E2F3}4
& \cellcolor[HTML]{F2F2F2}0
& \cellcolor[HTML]{F2F2F2}0
& \cellcolor[HTML]{F2F2F2}0
& \cellcolor[HTML]{F2F2F2}0
& \cellcolor[HTML]{F2F2F2}0
& \cellcolor[HTML]{F2F2F2}0
& \cellcolor[HTML]{D9E2F3}1
& \cellcolor[HTML]{F2F2F2}0 \\

& \cellcolor[HTML]{FFFFE0}Operability
& \cellcolor[HTML]{F2F2F2}0
& \cellcolor[HTML]{D9E2F3}3
& \cellcolor[HTML]{F2F2F2}0
& \cellcolor[HTML]{F2F2F2}0
& \cellcolor[HTML]{D9E2F3}1
& \cellcolor[HTML]{D9E2F3}1
& \cellcolor[HTML]{F2F2F2}0
& \cellcolor[HTML]{F2F2F2}0
& \cellcolor[HTML]{F2F2F2}0
& \cellcolor[HTML]{F2F2F2}0 \\

\bottomrule
\end{tabular}
\end{adjustbox}

\begin{minipage}{\textwidth}
\vspace{0.2cm}
\scriptsize
\textbf{Full names of each SE task:}
RE: \textit{Requirements Engineering};
CG: \textit{Code Generation};
CT: \textit{Code Translation};
CR: \textit{Code Review};
ST: \textit{Software Testing};
FL: \textit{Fault Localization};
PR: \textit{Program Repair};
RM: \textit{Release Management};
ETESD: \textit{End-to-End Software Development};
ETESM: \textit{End-to-End Software Maintenance}.
\rev{The background color of each numerical cell represents its occurrence frequency, with darker shades of blue indicating higher frequencies.}
\end{minipage}
\end{table*}

\rev{\textbf{Functional Correctness} is considered in 47 of the 49 studies on \textit{Code Generation} (95.9\%) and 9 of the 10 studies on \textit{Fault Localization} (90.0\%). CodeAgent~\hyperlink{S12}{[S12]}, an MAS for \textit{Code Generation}, evaluates a generated implementation against a task-specific test suite that combines repository tests with additional tests written by study participants. AgentFL~\hyperlink{S4}{[S4]}, an MAS for \textit{Fault Localization}, evaluates whether a known faulty method appears among the first one, three, or five candidates returned by the MAS. In both tasks, a completed run can produce a plausible but unusable answer: a function can execute and return the wrong value, and a list of suspicious methods can direct a developer away from the fault. So that designers can have a reason to assess success against a task-specific reference.}

\rev{\textbf{Resource Utilization} is considered in a larger proportion of studies on \textit{Fault Localization} (5/10, 50.0\%) and \textit{End-to-End Software Maintenance} (4/9, 44.4\%) than on \textit{Code Generation} (13/49, 26.5\%). A failing test or an issue report may describe a symptom without identifying the code responsible. AgentFL~\hyperlink{S4}{[S4]}, an MAS for \textit{Fault Localization}, uses method coverage to narrow the set of classes that the LLM inspects. MAGIS~\hyperlink{S28}{[S28]}, an MAS for \textit{End-to-End Software Maintenance}, reuses a summary when a candidate file is unchanged and can update an existing summary from the changed lines when the file has been modified. By comparison, a standalone programming problem of the kind addressed by MapCoder~\hyperlink{S29}{[S29]}, an MAS for \textit{Code Generation}, can provide enough information to begin implementation without searching a repository. When the relevant code must first be located, the breadth of the search and the repeated inspection of candidate files can both increase token consumption. Designers of localization and maintenance systems therefore have a reason to control how much code agents inspect and how much previously inspected code agents must process again.}

\rev{\textbf{Time Behavior} is considered in a higher proportion of studies on \textit{End-to-End Software Development} (5/11, 45.5\%) and \textit{End-to-End Software Maintenance} (5/9, 55.6\%) than on \textit{Code Generation} (8/49, 16.3\%). In MegaAgent's game development experiment~\hyperlink{S106}{[S106]}, an MAS for \textit{End-to-End Software Development}, seven agents completed the task in 800 seconds when task groups could run in parallel. With the same number of agents, sequential execution took 4,505 seconds. Some dependencies in maintenance leave less scope for parallel execution. FixAlly~\hyperlink{S60}{[S60]}, an MAS for \textit{End-to-End Software Maintenance}, applies a proposed fix and rebuilds the app before rerunning a GUI test to reach the affected screen. An accessibility scanner then checks whether the issue persists. End-to-end tasks therefore give designers a reason to consider both scheduling and validation when assessing completion time. Overlapping independent work can shorten the wait for a finished system, while a required build or check can keep the user waiting after code generation has finished.}

\rev{\textbf{Modularity} is considered in 21 of the 49 studies on \textit{Code Generation} (42.9\%), 6 of the 11 on \textit{End-to-End Software Development} (54.5\%), and 6 of the 9 on \textit{End-to-End Software Maintenance} (66.7\%). In MetaGPT~\hyperlink{S81}{[S81]}, an MAS for \textit{End-to-End Software Development}, agents publish structured messages to a shared pool, and each role subscribes to the messages needed for its work. In MASAI~\hyperlink{S32}{[S32]}, an MAS for \textit{End-to-End Software Maintenance}, the test template generator supplies the issue reproducer with a test template and the command for running the test. Developers can change how the generator finds a suitable template while preserving the inputs the reproducer expects. One explanation for the higher proportions in the end-to-end SE tasks is that more activities depend on artifacts produced within the MAS. A change to the format or meaning of a shared artifact can require changes in every agent that consumes it. Designers therefore have a reason to keep interfaces between activities stable so that revising one agent's procedure does not require corresponding revisions throughout the workflow.}

\subsubsection{\rev{Design Pattern Usage: Quality Attributes and Design Rationale}}

\rev{We examine which design patterns are adopted in studies that consider each MAS sub-QA. For each design pattern, we also examine the categories of design rationale reported in studies that adopt the design pattern (Figure~\ref{fig:mapping between SET_DP_DR}). Table~\ref{tab:DP-QA} gives the number of studies that both consider sub-QA $q$ and adopt design pattern $p$. Among the studies that consider $q$, the proportion that adopt $p$ is}

\begin{equation}
\rev{
P(p \mid q)=\frac{N(q,p)}{N(q)},
}
\label{eq:pattern_qa_prevalence}
\end{equation}

\rev{where $N(q)$ is the number of studies that consider $q$, and $N(q,p)$ is the number that also adopt $p$. A study may adopt several patterns; therefore, the percentages for a given sub-QA may sum to more than 100\%.}

\begin{table*}[h]
\caption{Mapping between Design Patterns (horizontal) and Quality Attributes (vertical)}
\label{tab:DP-QA}
\begin{adjustbox}{width=\textwidth,center}
\begin{tabular}{clcccccccccccccccc}
                                  & & \multicolumn{1}{l}{\textbf{RBC}}     & \multicolumn{1}{l}{\textbf{SR}} & \multicolumn{1}{l}{\textbf{TAR}} & \multicolumn{1}{l}{\textbf{CR}} & \multicolumn{1}{l}{\textbf{RAG}} & \multicolumn{1}{l}{\textbf{IMQ}}& \multicolumn{1}{l}{\textbf{AA}} & \multicolumn{1}{l}{\textbf{SPPG}} & \multicolumn{1}{l}{\textbf{HR}} & \multicolumn{1}{l}{\textbf{DBC}} & \multicolumn{1}{l}{\textbf{HC}} & \multicolumn{1}{l}{\textbf{PRO}} & \multicolumn{1}{l}{\textbf{LBC}} & \multicolumn{1}{l}{\textbf{AE}} & \multicolumn{1}{l}{\textbf{MPPG}} & \multicolumn{1}{l}{\textbf{VBC}} \\
\multirow{3}{*}{\textbf{{\normalsize Functional Suitability}}} &\cellcolor[HTML]{E0F7E0}Functional Correctness       
&\cellcolor[HTML]{4472C4}44 &\cellcolor[HTML]{4472C4}33        &\cellcolor[HTML]{8EAADB}16          &\cellcolor[HTML]{8EAADB}14          &\cellcolor[HTML]{8EAADB}11 &\cellcolor[HTML]{D9E2F3}9  &\cellcolor[HTML]{D9E2F3}7     &\cellcolor[HTML]{D9E2F3}6          &\cellcolor[HTML]{D9E2F3}5               &\cellcolor[HTML]{D9E2F3}4               &\cellcolor[HTML]{D9E2F3}4                &\cellcolor[HTML]{D9E2F3}3                &\cellcolor[HTML]{D9E2F3}3               &\cellcolor[HTML]{D9E2F3}3               &\cellcolor[HTML]{D9E2F3}4               &\cellcolor[HTML]{D9E2F3}3\\
&\cellcolor[HTML]{E0F7E0}Functional Completeness            
&\cellcolor[HTML]{8EAADB}12 &\cellcolor[HTML]{D9E2F3}6        & \cellcolor[HTML]{D9E2F3}6          & \cellcolor[HTML]{F2F2F2}0          & \cellcolor[HTML]{D9E2F3}4   &\cellcolor[HTML]{F2F2F2}0 &\cellcolor[HTML]{D9E2F3}3 &\cellcolor[HTML]{D9E2F3}2       & \cellcolor[HTML]{D9E2F3}2          & \cellcolor[HTML]{D9E2F3}2 &\cellcolor[HTML]{D9E2F3}2                &\cellcolor[HTML]{D9E2F3}1                &\cellcolor[HTML]{F2F2F2}0              &\cellcolor[HTML]{F2F2F2}0               &\cellcolor[HTML]{F2F2F2}0               &\cellcolor[HTML]{F2F2F2}0\\
&\cellcolor[HTML]{E0F7E0}Functional Appropriateness       
& \cellcolor[HTML]{D9E2F3}7 & \cellcolor[HTML]{D9E2F3}4        & \cellcolor[HTML]{D9E2F3}1          & \cellcolor[HTML]{D9E2F3}3          & \cellcolor[HTML]{D9E2F3}3    &\cellcolor[HTML]{D9E2F3}1 &\cellcolor[HTML]{D9E2F3}2  &\cellcolor[HTML]{F2F2F2}0          & \cellcolor[HTML]{D9E2F3}1         & \cellcolor[HTML]{D9E2F3}1  &\cellcolor[HTML]{D9E2F3}2                &\cellcolor[HTML]{F2F2F2}0                &\cellcolor[HTML]{F2F2F2}0               &\cellcolor[HTML]{F2F2F2}0              &\cellcolor[HTML]{F2F2F2}0               &\cellcolor[HTML]{F2F2F2}0\\
\multirow{3}{*}{\textbf{{\normalsize Performance Efficiency}}} &\cellcolor[HTML]{FFFFE0}Resource Utilization                     
& \cellcolor[HTML]{8EAADB}16 & \cellcolor[HTML]{8EAADB}12        & \cellcolor[HTML]{D9E2F3}6          & \cellcolor[HTML]{D9E2F3}1    &\cellcolor[HTML]{D9E2F3}6 &\cellcolor[HTML]{D9E2F3}3  & \cellcolor[HTML]{D9E2F3}3          & \cellcolor[HTML]{D9E2F3}2     & \cellcolor[HTML]{D9E2F3}3          & \cellcolor[HTML]{D9E2F3}1  &\cellcolor[HTML]{D9E2F3}2                &\cellcolor[HTML]{D9E2F3}1                &\cellcolor[HTML]{D9E2F3}1               &\cellcolor[HTML]{F2F2F2}0               &\cellcolor[HTML]{F2F2F2}0               &\cellcolor[HTML]{F2F2F2}0\\
& \cellcolor[HTML]{FFFFE0}Time Behavior                    
& \cellcolor[HTML]{8EAADB}13 & \cellcolor[HTML]{8EAADB}15        & \cellcolor[HTML]{D9E2F3}7          & \cellcolor[HTML]{D9E2F3}2     &\cellcolor[HTML]{D9E2F3}5 &\cellcolor[HTML]{D9E2F3}3  &\cellcolor[HTML]{D9E2F3}3          & \cellcolor[HTML]{D9E2F3}1     & \cellcolor[HTML]{D9E2F3}2          & \cellcolor[HTML]{D9E2F3}2  &\cellcolor[HTML]{D9E2F3}3                &\cellcolor[HTML]{D9E2F3}2                &\cellcolor[HTML]{D9E2F3}1               &\cellcolor[HTML]{D9E2F3}1               &\cellcolor[HTML]{F2F2F2}0               &\cellcolor[HTML]{F2F2F2}0\\
& \cellcolor[HTML]{FFFFE0}Capacity   
& \cellcolor[HTML]{D9E2F3}6       & \cellcolor[HTML]{D9E2F3}3          & \cellcolor[HTML]{D9E2F3}3          & \cellcolor[HTML]{D9E2F3}1    &\cellcolor[HTML]{D9E2F3}3 &\cellcolor[HTML]{F2F2F2}0  &\cellcolor[HTML]{D9E2F3}1          & \cellcolor[HTML]{D9E2F3}1     & \cellcolor[HTML]{D9E2F3}1          & \cellcolor[HTML]{F2F2F2}0  &\cellcolor[HTML]{F2F2F2}0                &\cellcolor[HTML]{F2F2F2}0                &\cellcolor[HTML]{D9E2F3}1               &\cellcolor[HTML]{F2F2F2}0               &\cellcolor[HTML]{D9E2F3}1               &\cellcolor[HTML]{F2F2F2}0\\
\multirow{3}{*}{\textbf{{\normalsize Maintainability}}} &\cellcolor[HTML]{E0F7E0}Modularity                  
& \cellcolor[HTML]{4472C4}36 & \cellcolor[HTML]{8EAADB}15        & \cellcolor[HTML]{D9E2F3}6          & \cellcolor[HTML]{D9E2F3}9          & \cellcolor[HTML]{D9E2F3}6     &\cellcolor[HTML]{D9E2F3}6 &\cellcolor[HTML]{D9E2F3}3 & \cellcolor[HTML]{D9E2F3}5          & \cellcolor[HTML]{D9E2F3}2     & \cellcolor[HTML]{D9E2F3}1 &\cellcolor[HTML]{D9E2F3}4                &\cellcolor[HTML]{D9E2F3}2                &\cellcolor[HTML]{D9E2F3}1               &\cellcolor[HTML]{D9E2F3}4               &\cellcolor[HTML]{D9E2F3}2               &\cellcolor[HTML]{D9E2F3}1\\
& \cellcolor[HTML]{E0F7E0}Analysability         
& \cellcolor[HTML]{D9E2F3}2 & \cellcolor[HTML]{D9E2F3}2        & \cellcolor[HTML]{D9E2F3}1          & \cellcolor[HTML]{D9E2F3}1          & \cellcolor[HTML]{D9E2F3}1    &\cellcolor[HTML]{D9E2F3}1 &\cellcolor[HTML]{D9E2F3}2  & \cellcolor[HTML]{F2F2F2}0          & \cellcolor[HTML]{D9E2F3}1     & \cellcolor[HTML]{F2F2F2}0   &\cellcolor[HTML]{F2F2F2}0                &\cellcolor[HTML]{F2F2F2}0               &\cellcolor[HTML]{F2F2F2}0               &\cellcolor[HTML]{F2F2F2}0               &\cellcolor[HTML]{F2F2F2}0              &\cellcolor[HTML]{F2F2F2}0\\
& \cellcolor[HTML]{E0F7E0}Reusability             
& \cellcolor[HTML]{D9E2F3}1        & \cellcolor[HTML]{D9E2F3}1          & \cellcolor[HTML]{F2F2F2}0          & \cellcolor[HTML]{D9E2F3}1    &\cellcolor[HTML]{F2F2F2}0 &\cellcolor[HTML]{F2F2F2}0  & \cellcolor[HTML]{F2F2F2}0          & \cellcolor[HTML]{D9E2F3}1     & \cellcolor[HTML]{F2F2F2}0          & \cellcolor[HTML]{F2F2F2}0 &\cellcolor[HTML]{F2F2F2}0                &\cellcolor[HTML]{F2F2F2}0                &\cellcolor[HTML]{F2F2F2}0               &\cellcolor[HTML]{F2F2F2}0               &\cellcolor[HTML]{F2F2F2}0               &\cellcolor[HTML]{F2F2F2}0\\
\multirow{2}{*}{\textbf{{\normalsize Reliability}}} &\cellcolor[HTML]{FFFFE0}Fault Tolerance                 
& \cellcolor[HTML]{D9E2F3}10        & \cellcolor[HTML]{D9E2F3}10          & \cellcolor[HTML]{D9E2F3}2          & \cellcolor[HTML]{D9E2F3}5    &\cellcolor[HTML]{D9E2F3}2 &\cellcolor[HTML]{D9E2F3}3  &\cellcolor[HTML]{F2F2F2}0          & \cellcolor[HTML]{D9E2F3}1     & \cellcolor[HTML]{D9E2F3}1          & \cellcolor[HTML]{D9E2F3}3  &\cellcolor[HTML]{F2F2F2}0                &\cellcolor[HTML]{F2F2F2}0                &\cellcolor[HTML]{F2F2F2}0               &\cellcolor[HTML]{D9E2F3}2               &\cellcolor[HTML]{F2F2F2}0               &\cellcolor[HTML]{F2F2F2}0\\
& \cellcolor[HTML]{FFFFE0}Faultlessness                  
& \cellcolor[HTML]{D9E2F3}6 & \cellcolor[HTML]{D9E2F3}5        & \cellcolor[HTML]{D9E2F3}3               & \cellcolor[HTML]{D9E2F3}1    &\cellcolor[HTML]{D9E2F3}4 &\cellcolor[HTML]{D9E2F3}2  &\cellcolor[HTML]{D9E2F3}2         & \cellcolor[HTML]{D9E2F3}1     & \cellcolor[HTML]{D9E2F3}2          & \cellcolor[HTML]{F2F2F2}0  &\cellcolor[HTML]{D9E2F3}2                &\cellcolor[HTML]{F2F2F2}0                &\cellcolor[HTML]{D9E2F3}2               &\cellcolor[HTML]{F2F2F2}0               &\cellcolor[HTML]{D9E2F3}1               &\cellcolor[HTML]{F2F2F2}0\\
\multirow{2}{*}{\textbf{{\normalsize Flexibility}}} & \cellcolor[HTML]{E0F7E0}Adaptability
& \cellcolor[HTML]{D9E2F3}5 & \cellcolor[HTML]{8EAADB}11
& \cellcolor[HTML]{D9E2F3}8 & \cellcolor[HTML]{F2F2F2}0
& \cellcolor[HTML]{D9E2F3}3 & \cellcolor[HTML]{D9E2F3}4
& \cellcolor[HTML]{D9E2F3}2 & \cellcolor[HTML]{D9E2F3}1
& \cellcolor[HTML]{F2F2F2}0 & \cellcolor[HTML]{F2F2F2}0
& \cellcolor[HTML]{D9E2F3}3 & \cellcolor[HTML]{D9E2F3}2
& \cellcolor[HTML]{D9E2F3}1 & \cellcolor[HTML]{F2F2F2}0
& \cellcolor[HTML]{F2F2F2}0 & \cellcolor[HTML]{D9E2F3}2\\
& \cellcolor[HTML]{E0F7E0}Scalability          
& \cellcolor[HTML]{D9E2F3}7 & \cellcolor[HTML]{D9E2F3}3        & \cellcolor[HTML]{D9E2F3}3          & \cellcolor[HTML]{D9E2F3}2         & \cellcolor[HTML]{D9E2F3}2    &\cellcolor[HTML]{D9E2F3}2 &\cellcolor[HTML]{D9E2F3}2  & \cellcolor[HTML]{D9E2F3}2          & \cellcolor[HTML]{D9E2F3}1              & \cellcolor[HTML]{F2F2F2}0  &\cellcolor[HTML]{D9E2F3}1                &\cellcolor[HTML]{D9E2F3}1                &\cellcolor[HTML]{D9E2F3}2               &\cellcolor[HTML]{F2F2F2}0               &\cellcolor[HTML]{F2F2F2}0               &\cellcolor[HTML]{D9E2F3}2\\
\multirow{2}{*}{\textbf{{\normalsize Compatibility}}} &\cellcolor[HTML]{FFFFE0}Interoperability           
& \cellcolor[HTML]{D9E2F3}2 & \cellcolor[HTML]{D9E2F3}1        & \cellcolor[HTML]{D9E2F3}5                   & \cellcolor[HTML]{D9E2F3}1    &\cellcolor[HTML]{D9E2F3}3 &\cellcolor[HTML]{D9E2F3}1  & \cellcolor[HTML]{D9E2F3}2          & \cellcolor[HTML]{F2F2F2}0     & \cellcolor[HTML]{D9E2F3}1          & \cellcolor[HTML]{D9E2F3}1   &\cellcolor[HTML]{D9E2F3}1                &\cellcolor[HTML]{F2F2F2}0                &\cellcolor[HTML]{F2F2F2}0               &\cellcolor[HTML]{F2F2F2}0              &\cellcolor[HTML]{D9E2F3}1                &\cellcolor[HTML]{F2F2F2}0\\
& \cellcolor[HTML]{FFFFE0}Co-Existence           
& \cellcolor[HTML]{D9E2F3}1 & \cellcolor[HTML]{D9E2F3}2        & \cellcolor[HTML]{D9E2F3}6          & \cellcolor[HTML]{F2F2F2}0          & \cellcolor[HTML]{D9E2F3}1    &\cellcolor[HTML]{F2F2F2}0 &\cellcolor[HTML]{D9E2F3}1  &\cellcolor[HTML]{F2F2F2}0          & \cellcolor[HTML]{F2F2F2}0               & \cellcolor[HTML]{F2F2F2}0 &\cellcolor[HTML]{D9E2F3}2                &\cellcolor[HTML]{F2F2F2}0                &\cellcolor[HTML]{F2F2F2}0               &\cellcolor[HTML]{F2F2F2}0               &\cellcolor[HTML]{D9E2F3}1               &\cellcolor[HTML]{F2F2F2}0\\
\multirow{2}{*}{\textbf{{\normalsize Security}}} & \cellcolor[HTML]{E0F7E0}Confidentiality          
& \cellcolor[HTML]{D9E2F3}2 & \cellcolor[HTML]{D9E2F3}1        & \cellcolor[HTML]{D9E2F3}4          & \cellcolor[HTML]{D9E2F3}1          & \cellcolor[HTML]{F2F2F2}0     &\cellcolor[HTML]{F2F2F2}0 &\cellcolor[HTML]{D9E2F3}2 & \cellcolor[HTML]{F2F2F2}0          & \cellcolor[HTML]{F2F2F2}0          & \cellcolor[HTML]{D9E2F3}1  &\cellcolor[HTML]{F2F2F2}0                &\cellcolor[HTML]{D9E2F3}2                &\cellcolor[HTML]{D9E2F3}1               &\cellcolor[HTML]{F2F2F2}0               &\cellcolor[HTML]{F2F2F2}0               &\cellcolor[HTML]{F2F2F2}0\\
& \cellcolor[HTML]{E0F7E0}Integrity          
& \cellcolor[HTML]{D9E2F3}1 & \cellcolor[HTML]{F2F2F2}0        & \cellcolor[HTML]{F2F2F2}0          & \cellcolor[HTML]{F2F2F2}0         & \cellcolor[HTML]{D9E2F3}1   &\cellcolor[HTML]{F2F2F2}0 &\cellcolor[HTML]{F2F2F2}0  & \cellcolor[HTML]{F2F2F2}0          & \cellcolor[HTML]{F2F2F2}0             & \cellcolor[HTML]{F2F2F2}0  &\cellcolor[HTML]{F2F2F2}0                &\cellcolor[HTML]{F2F2F2}0                &\cellcolor[HTML]{F2F2F2}0               &\cellcolor[HTML]{F2F2F2}0               &\cellcolor[HTML]{F2F2F2}0               &\cellcolor[HTML]{F2F2F2}0\\
\multirow{2}{*}{\textbf{{\normalsize Interaction Capability}}} &\cellcolor[HTML]{FFFFE0}User Engagement           
& \cellcolor[HTML]{D9E2F3}1 & \cellcolor[HTML]{D9E2F3}1        & \cellcolor[HTML]{F2F2F2}0                   & \cellcolor[HTML]{D9E2F3}1    &\cellcolor[HTML]{F2F2F2}0 &\cellcolor[HTML]{F2F2F2}0  & \cellcolor[HTML]{F2F2F2}0          & \cellcolor[HTML]{F2F2F2}0     & \cellcolor[HTML]{D9E2F3}3         & \cellcolor[HTML]{F2F2F2}0    &\cellcolor[HTML]{F2F2F2}0                &\cellcolor[HTML]{D9E2F3}1                &\cellcolor[HTML]{F2F2F2}0               &\cellcolor[HTML]{F2F2F2}0               &\cellcolor[HTML]{F2F2F2}0               &\cellcolor[HTML]{F2F2F2}0\\
& \cellcolor[HTML]{FFFFE0}Operability          
& \cellcolor[HTML]{D9E2F3}2 & \cellcolor[HTML]{D9E2F3}1        & \cellcolor[HTML]{D9E2F3}1          & \cellcolor[HTML]{F2F2F2}0          & \cellcolor[HTML]{F2F2F2}0    &\cellcolor[HTML]{F2F2F2}0 &\cellcolor[HTML]{F2F2F2}0  & \cellcolor[HTML]{D9E2F3}1          & \cellcolor[HTML]{D9E2F3}1               & \cellcolor[HTML]{F2F2F2}0  &\cellcolor[HTML]{D9E2F3}1                &\cellcolor[HTML]{F2F2F2}0               &\cellcolor[HTML]{F2F2F2}0               &\cellcolor[HTML]{F2F2F2}0               &\cellcolor[HTML]{F2F2F2}0               &\cellcolor[HTML]{F2F2F2}0\\
\end{tabular}
\end{adjustbox}
\begin{minipage}{13.5cm} 
\vspace{0.1cm}
\vspace{0.1cm}
\scriptsize  \textbf{Full names of each Design Pattern:} RBC: \textit{Role-Based Cooperation}; SR: \textit{Self-Reflection}; TAR: \textit{Tool-Agent Registry}; CR: \textit{Cross-Reflection}; RAG: \textit{Retrieval-Augmented Generation}; IMQ: \textit{Incremental Model Querying}; AA: \textit{Agent Adapter}; SPPG: \textit{Single-Path Plan Generator}; HR: \textit{Human-Reflection}; DBC: \textit{Debate-Based Cooperation}; HC: \textit{Hierarchical Cooperation}; PRO: \textit{Prompt/Response Optimiser}; LBC: \textit{Layered-Based Cooperation}; AE: \textit{Agent Evaluator}; MPPG: \textit{Multi-Path Plan Generator}; VBC: \textit{Voting-Based Cooperation}. 
\rev{The background color of numerical cells indicates occurrence frequency, where darker blue represents higher frequency.}
\end{minipage}
\end{table*}

\rev{\textbf{Functional Suitability.} Among the 97 studies that consider \textit{Functional Correctness}, 44 (45.4\%) adopt \textit{Role-Based Cooperation} and 33 (34.0\%) adopt \textit{Self-Reflection}. The corresponding adoption rates across all 106 studies are 46.2\% and 34.0\%. Of the 33 studies adopting \textit{Self-Reflection} and considering \textit{Functional Correctness}, 19 also report \textit{Improving the Quality of Generated Code} as design rationale. In MetaGPT~\hyperlink{S81}{[S81]}, an MAS for \textit{End-to-End Software Development}, the architect defines component interfaces, and the engineer implements the components according to the requirements and design. After early code reviews missed errors, the authors enabled the engineer to write and run unit tests and revise the implementation when a test failed. Designers can use the two patterns together when checking correctness requires both a specification of expected behavior and evidence of actual behavior. Assigning design and implementation to separate roles gives the engineer a reference for checking the code. Self-reflection allows the engineer to use a discrepancy revealed by testing to guide the next revision.}

\rev{\textbf{Performance Efficiency.} Of the 29 studies that consider \textit{Resource Utilization}, 16 (55.2\%) adopt \textit{Role-Based Cooperation}. Fourteen of the 16 also report \textit{Optimizing Software Resource Management} as design rationale. In MAGIS~\hyperlink{S28}{[S28]}, the repository custodian filters candidate files for relevance to the issue before the manager assigns implementation tasks. Assigning file selection to a dedicated agent can save resources when filtering once costs less than having several downstream agents process the same irrelevant code. Of the 29 studies that consider \textit{Time Behavior}, 15 (51.7\%) adopt \textit{Self-Reflection}. Ten of the 15 also report \textit{Improving the Efficiency of Generating Software Artifacts} as design rationale. In an MAS proposed in~\hyperlink{S90}{[S90]} for \textit{Program Repair}, the model used execution feedback to revise failed programs and achieved comparable accuracy with fewer independently generated candidates. For designers concerned with completion time, revising a failed candidate is attractive when feedback points to a small correction and generating a replacement would require starting from scratch. The time advantage depends on the available parallelism: successive revisions must wait for earlier results, while independent candidates can be generated and tested concurrently.}

\rev{\textbf{Maintainability.} Among the 53 studies that consider \textit{Modularity}, 36 (67.9\%) adopt \textit{Role-Based Cooperation}, compared with 13 of the remaining 53 studies (24.5\%). The authors~\hyperlink{S64}{[S64]} of AutoGen, an MAS for \textit{Code Generation} identify the ability to develop, test, and maintain agents separately as a benefit of assigning distinct responsibilities. In AutoGen, an assistant agent can generate code and send the code to a user proxy agent for execution. Developers can then change the execution environment within the proxy, provided the proxy continues to accept the generated code and return results in a form the assistant can interpret. The ability to contain a change within one agent offers a plausible explanation for the higher adoption rate among studies concerned with modularity. Designers have a reason to separate responsibilities that are likely to change independently and to keep the interface between the responsible agents stable.}

\rev{\textbf{Reliability.} Among the 24 studies that consider \textit{Fault Tolerance}, 10 (41.7\%) adopt \textit{Role-Based Cooperation} and 10 (41.7\%) adopt \textit{Self-Reflection}. Five studies adopt both patterns. Assigning observation to a separate role gives the acting agent evidence about what actually happened, and self-reflection helps the acting agent choose a different action so testing can continue. For \textit{Faultlessness}, designers can act earlier when the information needed to detect an error is already available. Infant Agent~\hyperlink{S104}{[S104]}, an MAS for \textit{End-to-End Software Maintenance} for example, checks whether the first and last lines of a proposed edit match the expected text. Checking the edit location before changing the file allows the MAS to request a correction while the original code is still intact.}

\rev{\textbf{Flexibility.} Among the 21 studies that consider \textit{Adaptability}, 11 (52.4\%) adopt \textit{Self-Reflection} and 8 (38.1\%) adopt \textit{Tool-Agent Registry}. Nine of the 11 studies adopting \textit{Self-Reflection} and five of the eight adopting \textit{Tool-Agent Registry} report \textit{Improving the Adaptability of Agent Systems} as design rationale. Consider a project that changes its test report format. An agent may extract no failures even though the report explicitly marks a test as failed. Comparing the extraction result with the failed entry gives the agent a basis for revising its extraction rule. Designers can therefore use self-reflection when an environmental change invalidates an assumption in the agent's procedure. A registry is useful when an available tool already supports the changed input and the agent needs to find a compatible parser. For \textit{Scalability}, 7 of 13 studies (53.8\%) adopt \textit{Role-Based Cooperation}. In SoA~\hyperlink{S38}{[S38]}, an MAS for \textit{Code Generation}, a coordinating agent divides a complex function into subfunctions and delegates implementation to additional agents. Separating coordination from implementation can increase capacity when each additional agent can implement a subfunction from a local specification without having to process the entire codebase.}

\rev{\textbf{Compatibility.} Five of the 10 studies that consider \textit{Interoperability} (50.0\%) adopt \textit{Tool-Agent Registry}. A registry can describe the input a tool accepts and the information the tool returns. In CodeAgent~\hyperlink{S12}{[S12]}, an MAS for \textit{Code Generation}, for example, an agent supplies a class name to a symbol search tool and uses the returned class definition to implement a subclass. A registry also appears in 6 of the 7 studies that consider \textit{Co-Existence} (85.7\%). Four of the six registry studies also consider \textit{Interoperability}. The need to exchange information with external tools therefore offers a plausible reason for registry adoption in several studies concerned with co-existence. Designers can use a registry to help agents construct valid calls and interpret tool results. When running a tool could impair other applications on the same host, designers also need to control execution, for example by limiting concurrent calls to memory-intensive tools.}

\rev{\textbf{Security.} Of the nine studies that consider \textit{Confidentiality}, four (44.4\%) adopt \textit{Tool-Agent Registry}. One of the four studies~\hyperlink{S82}{[S82]}, an MAS for \textit{Code Generation}, reports \textit{Ensuring Software Security} as design rationale and proposes a separate agent to review sensitive operations. A registry can identify tools capable of reading repository files. Because a single tool may provide access to files with different restrictions, the MAS can check the requesting agent's permission for the specific file before releasing its contents. The only study, an MAS for \textit{Code Generation}, that considers \textit{Integrity}~\hyperlink{S64}{[S64]} separates code generation, review, and execution in the AutoGen implementation of OptiGuide. The Writer generates code, the Safeguard reviews the proposed code, and the Commander executes the code only after clearance. Across 100 coding tasks, separate writing and review roles yielded higher F1 scores for detecting unsafe code than a single agent performing both tasks. Designers have a reason to separate review from execution when generated code can modify protected data: even if the Writer is authorized to produce code, the Commander can withhold execution when the Safeguard identifies a potentially unauthorized change.}

\rev{\textbf{Interaction Capability.} Three of the five studies that consider \textit{User Engagement} adopt \textit{Human-Reflection} (60.0\%). In AISD~\hyperlink{S20}{[S20]}, an MAS for \textit{Code Generation}, a draft use case specified that classification results would be saved to a JSON file. The user requested a graphical display, and AISD used the revised use case to generate the system design. When a requirement leaves a user preference unresolved, designers can use human reflection to give users a reason to participate: reviewing a concrete proposal allows users to influence what the MAS builds. Two of the five studies that consider \textit{Operability} adopt \textit{Role-Based Cooperation} (40.0\%). In AgileGen~\hyperlink{S46}{[S46]}, an MAS for \textit{Code Generation}, the scenario design component asks users to confirm the expected behavior. The prototyping component builds the application from the confirmed scenarios and revises the code in response to feedback on the prototype. Role-based cooperation can make the MAS easier to operate when each user decision has a clear consequence for the workflow. Users can approve a scenario or request a change to the prototype, while the MAS coordinates the implementation work required by each decision.}

\subsubsection{\rev{Design Pattern Usage and Design Rationale across SE Tasks}}

\rev{Within each SE task category, we count how many studies adopt each design pattern and, separately, how many report each design rationale. We divide each count by the total number of studies in that task category. A study may adopt several patterns and design rationale, so the percentages in either distribution can sum to more than 100\%. Attributing a design pattern choice to specific design rationale requires evidence linking the pattern to design rationale in the primary study.}

\rev{\textbf{Role-Based Cooperation} is more prevalent in studies of \textit{End-to-End Software Development} (8/11, 72.7\%) than in studies of \textit{Code Generation} (19/49, 38.8\%). In MetaGPT~\hyperlink{S81}{[S81]}, the architect agent defines component interfaces, and the engineer agents use the interface definitions to implement the components. Two of the six \textit{Code Review} studies and two of the four \textit{Requirements Engineering} studies both adopt the pattern and report \textit{Simulating Human Processes of Solving SE Tasks} as design rationale. In MARE~\hyperlink{S30}{[S30]}, an MAS for \textit{Requirements Engineering}, a collector agent interviews a stakeholder agent and uses the responses to draft requirements. One reason for the higher adoption in end-to-end software development may be that a coding task can supply a function's interface and expected behavior, while an end-to-end software development task may require the MAS to define component interfaces before implementation can begin. Designers can use complementary roles to manage dependencies between activities by specifying which agent must produce a decision or artifact and which agent must use or review the result.}

\rev{\textbf{Self-Reflection} is used in 19 of 49 \textit{Code Generation} studies (38.8\%) and 4 of 7 \textit{Software Testing} studies (57.1\%). Of the 19 code generation studies adopting the pattern, 13 report \textit{Improving the Quality of Generated Code} as design rationale. In Reflexion~\hyperlink{S100}{[S100]}, an MAS for \textit{Code Generation},the coding agent examines a failed unit test, explains a possible error in the previous implementation, and uses the explanation to revise the function. In XUAT-Copilot~\hyperlink{S43}{[S43]}, an MAS for \textit{Software Testing}, an unsupported operation prompts the operation agent to review the proposed test action and select an operation available in the test environment. In a stepwise acceptance test, an invalid action can prevent the application from reaching the state required for the next step. Designers have a reason to use self-reflection when execution feedback identifies a specific failure and the agent responsible for the function or test action can use the feedback to correct its own work before proceeding.}

\rev{\textbf{Cross-Reflection} appears in similar proportions of studies on \textit{End-to-End Software Development} (2/11, 18.2\%) and \textit{Code Generation} (9/49, 18.4\%). Among the studies adopting this design pattern, both end-to-end development studies and five of the nine code generation studies report \textit{Improving the Quality of Generated Code} as their design rationale. In CodePori~\hyperlink{S13}{[S13]}, an MAS for \textit{End-to-End Software Development}, one developer agent write draft code, a second developer agent revises the code, and the two agents exchange revisions before verification. In AutoCoder~\hyperlink{S7}{[S7]}, an MAS for \textit{Code Generation}, the questioner agent uses the interpreter's error output to describe what went wrong. The programmer agent then uses the description and the error output to revise the code. The resulting dialogue forms one example in the dataset. These cases illustrate two ways to applying cross-reflection in MAS design for SE tasks. Before conducting the final verification, designers can ask a second agent to review and revise the draft artifacts produced by the first agent. When something goes wrong in the final output of the MAS, designers can ask another agent to interpret the error output so that the agent generating the error artifacts can use the explanation to revise the output artifacts.}

\subsection{Implications}
\rev{Building on the results presented in Section~\ref{sec:Results} and the interpretation of how the findings across the four RQs relate to one another in Section~\ref{sec:interpretation_of_results}, we derive six implications for designing LLM-based MASs for SE tasks. These recommendations draw on the observed co-occurrences between SE tasks, QAs, sub-QAs, design patterns, and design rationale, as well as the distributions of QAs, sub-QAs, design patterns, and design rationale within each SE task category.}
 
\begin{tcolorbox}[arc=0mm,width=\columnwidth,
                  top=1mm,left=1mm,right=1mm,bottom=1mm,
                  boxrule=.2pt]
\textbf{Implication 1}. \rev{Designers should incorporate reflection mechanisms with explicit review criteria and revision procedures to support the \textit{Functional Correctness} of LLM-based MASs.}
\end{tcolorbox}

\rev{As shown in Table~\ref{tab:DP-QA}, \textit{Self-Reflection}, \textit{Cross-Reflection}, and \textit{Human-Reflection} co-occur with \textit{Functional Correctness} in 33, 14, and five studies, respectively. These patterns differ in who reviews an agent's output: the agent that produced the output, another agent, or a human reviewer. Reviewing intermediate artifacts and final results provides opportunities to identify discrepancies between the outputs and the task requirements. Feedback describing these discrepancies can then guide the revision of the affected outputs.}

\rev{Designers should select reflection mechanisms according to the knowledge and evidence required to assess an output. \textit{Self-Reflection} allows the generating agent to review its own output using the task requirements and available validation evidence. For example, \textit{MetaGPT}~\hyperlink{S81}{[S81]} enables its Engineer agent to use execution results and debugging history to revise its own code. \textit{Cross-Reflection} can be used when another agent can contribute relevant expertise, additional evidence, or a different perspective on the output. In \textit{INTERVENOR}~\hyperlink{S78}{[S78]}, a Code Teacher examines the code produced by a Code Learner and uses compiler error feedback to provide repair guidance to the learner. \textit{Human-Reflection} can be incorporated when requirements need to be clarified by users or automated checks cannot adequately assess whether an output meets their expectations. For example, \textit{AISD}~\hyperlink{S20}{[S20]} involves users in reviewing generated use cases and validating prototype implementations, allowing them to clarify requirements and provide feedback for subsequent revisions.}

\rev{Building on these examples, designers should specify the output to be inspected, the task requirements and correctness criteria to apply, and the agent responsible for revising the output. Feedback should identify the part of the output considered incorrect and explain the discrepancy with reference to the relevant requirement or validation result. When an output contains executable code, execution results and test failure information can help substantiate the feedback. Complementary evidence comes from CRITIC, which uses interpreter error messages and execution results to guide correction in mathematical program synthesis. Its ablation experiments show that removing interpreter feedback can limit correction performance and make it unstable~\citep{gou2024critic}. The workflow should pass the feedback to the agent responsible for revision and require the revised output to be checked against the relevant task requirements. Applicable validation steps should also be repeated after revision; for example, \textit{MetaGPT}~\hyperlink{S81}{[S81]} repeats testing and debugging until the tests pass or the retry limit is reached. Designers should use these checks to assess whether the identified errors have been addressed and whether the revision has introduced additional incorrect behavior.}

\begin{tcolorbox}[arc=0mm,width=\columnwidth,
                  top=1mm,left=1mm,right=1mm,bottom=1mm,
                  boxrule=.2pt]
\textbf{Implication 2}. \rev{Designers should clearly define the task responsibilities and functional boundaries of individual agents to support the \textit{Maintainability} of LLM-based MASs.}
\end{tcolorbox}

\rev{Table~\ref{tab:DP-QA} shows that 36 of the 53 studies considering \textit{Modularity} adopt the design pattern \textit{Role-Based Cooperation} (67.9\%), which provides a starting point for separating responsibilities within an MAS. To support modularity, the assigned responsibilities should correspond to components with defined interfaces and explicit dependencies. Such boundaries can help confine modifications to the components responsible for the affected functionality.}

\rev{For each agent, designers should specify its assigned tasks, expected inputs and outputs, permitted tool operations, and conditions for transferring work to another agent. These specifications should be reflected in the agent's prompts, tool interfaces, and coordination logic. For example, a software maintenance workflow can distinguish diagnosis, patch generation, and validation responsibilities and define the artifacts exchanged between them. Where responsibilities overlap, the workflow should specify how the agents' outputs are reconciled. Qin et al.~\hyperlink{S4}{[S4]} illustrated role specialization in \textit{AGENTFL}, which assigns distinct responsibilities to four agents with customized instructions and external components. To assess whether the specified agent responsibilities and functional boundaries support \textit{Maintainability}, designers can examine how modifications to one agent affect other agents in the LLM-based MAS. For example, in the software maintenance workflow described above, designers can revise the patch-generation prompt of the agent or repair strategy while preserving its input and output contracts, and check whether corresponding modifications are required in the diagnosis or validation agent. If such modifications are needed, designers should examine whether they arise from overlapping responsibilities or dependencies outside the defined interfaces, and refine the allocation of responsibilities or the interfaces where appropriate. Keeping changes to an agent's internal implementation localized can reduce the number of MAS components that require modification and revalidation during maintenance.}
 
\begin{tcolorbox}[arc=0mm,width=\columnwidth,
                  top=1mm,left=1mm,right=1mm,bottom=1mm,
                  boxrule=.2pt]
\textbf{Implication 3}. \rev{When adapting human SE practices to agent workflows, designers should define explicit task handoffs, verification steps, and escalation conditions.}
\end{tcolorbox}

\rev{\textit{Simulating Human Processes of Solving SE Tasks} is reported in 31 of the 106 studies (29.2\%), including four of the six studies on \textit{Code Review} and two of the four studies on \textit{Requirements Engineering}. This reported rationale identifies human SE practices as a source of design ideas for agent workflows. Designers drawing on these practices should select and adapt the relevant activities and decision rules according to the target SE task, the capabilities of the agents, and the communication and execution costs involved. For code review, for example, designers can specify how a reviewing agent supports a finding with code evidence, how the finding is verified, and how conflicting assessments are resolved.}

\rev{For each stage included in the agent workflow, designers should define the required inputs, expected outputs, and checks performed before transferring the outputs to another stage. The workflow should also specify the conditions for revising an output, invoking additional analysis, requesting human input, or terminating execution. Lee et al.~\hyperlink{S2}{[S2]} provided an example in \textit{FixAgent}, whose hierarchical workflow is informed by a cognitive debugging model. The system begins with bug localization and patch generation and activates additional analysis when an earlier level fails to produce a patch that passes the available tests. This example illustrates how a human problem-solving strategy can inform conditional transitions between stages in an agent workflow. Designers adopting similar workflows should check whether each handoff provides the receiving agent with the information needed for its task. They should also compare the adapted workflow with a simpler alternative under the same execution budget to assess whether additional stages improve task outcomes sufficiently to justify their communication and execution costs.}

\rev{Software architecture and design also deserve careful consideration throughout \textit{End-to-End Software Development} workflows. In our task classification, software design activities embedded in these workflows are represented under \textit{End-to-End Software Development} rather than treated as separate SE tasks. Consequently, our SE task-level mappings do not separately characterize the QAs or design patterns associated with software architecture and design activities embedded within these software development workflows. For workflows involving architecture and design activities, designers can apply the handoff and verification principles described above by specifying the design artifacts required by implementation agents, including component responsibilities and interface specifications, and validating these artifacts against the relevant requirements before implementation begins. Evaluating these intermediate artifacts separately would help assess the design stage independently of the outcome of the complete development workflow.}

\begin{tcolorbox}[arc=0mm,width=\columnwidth,
                  top=1mm,left=1mm,right=1mm,bottom=1mm,
                  boxrule=.2pt]
\textbf{Implication 4}. \rev{For code generation with executable tests, designers should separate test design from code generation; for project-level fault localization, they should assess context-selection strategies against both localization accuracy and resource use.}
\end{tcolorbox}

\rev{For \textit{Code Generation}, Table~\ref{tab:QA-SET} shows that 47 of the 49 studies (95.9\%) consider \textit{Functional Correctness}. One concrete design option is to assign test execution and feedback reporting to an \textit{Agent Evaluator}, while defining test-design responsibilities separately from code-generation responsibilities. For example, \textit{AgentCoder}~\hyperlink{S61}{[S61]} uses separate programmer, test designer, and test executor agents. Its test designer independently generates tests from the coding requirements, and its test executor returns execution results to the programmer for refinement. Designers adopting this arrangement can provide the test designer with the task requirements and interface specification without the candidate implementation. They should check the expected test outputs against those requirements, return failing inputs and execution results to the code-generation agent, and rerun the tests after revision. Checking the generated tests is necessary because an incorrect expected output can direct the programmer to modify otherwise correct code.}

\rev{For \textit{Fault Localization}, nine of the 10 studies (90.0\%) consider \textit{Functional Correctness}, and five (50.0\%) consider \textit{Resource Utilization}. For systems that inspect large repositories, designers can use \textit{Role-Based Cooperation} to assign failure analysis, repository navigation, and candidate assessment to agents with defined responsibilities. \textit{AGENTFL}~\hyperlink{S4}{[S4]} illustrates this approach through specialized agents, document-guided search, and coverage-based filtering of candidate classes before detailed method inspection. When failure logs or coverage information are available, designers can use this evidence to select the initial code context and define when additional inspection is needed. They should evaluate the selection strategy on representative bugs by recording whether the retained candidates contain the known faulty locations, together with final localization accuracy and token consumption. These checks help determine whether reduced context-processing costs are accompanied by the exclusion of evidence needed for accurate localization.}

\begin{tcolorbox}[arc=0mm,width=\columnwidth,
                  top=1mm,left=1mm,right=1mm,bottom=1mm,
                  boxrule=.2pt]
\textbf{Implication 5}. \rev{Designers should set task-specific execution budgets and reduce unnecessary processing while checking that task outcomes meet the required quality criteria.}
\end{tcolorbox}

\rev{\textit{Optimizing Software Resource Management} and \textit{Improving the Efficiency of Generating Software Artifacts} are reported in 29 (27.4\%) and 25 (23.6\%) studies, respectively. \textit{Resource Utilization} and \textit{Time Behavior} are each considered in 29 studies. These findings highlight execution costs as a design concern, particularly for workflows involving repeated repository inspection, artifact generation, and verification. Designers can measure end-to-end latency, token consumption, LLM calls, and tool execution costs on representative tasks to identify the stages responsible for substantial or repeated processing. These measurements can inform limits on execution time, resource consumption, and refinement rounds, together with explicit stopping conditions.}

\rev{Possible changes include supplying each agent with relevant context, reusing analysis results while the inputs and program state on which they depend remain unchanged, and limiting repeated review or refinement. For workflows with stable task dependencies, task assignments and transitions can also be represented explicitly. For example, \textit{CODER}, proposed by Chen et al.~\hyperlink{S14}{[S14]}, uses JSON-formatted task graphs to specify participating agents, their subtasks, and downstream transitions. Such representations make execution paths available for inspection and adjustment. Designers should calculate the execution cost per successfully completed task by dividing the total execution cost across all attempts, including unsuccessful attempts and retries, by the number of successfully completed tasks. Comparing this measure with the corresponding baseline value, while checking that task success and artifact quality meet the required criteria, can help assess whether the modified workflow reduces execution costs without compromising the required quality.}

\begin{tcolorbox}[arc=0mm,width=\columnwidth,
                  top=1mm,left=1mm,right=1mm,bottom=1mm,
                  boxrule=.2pt]
\textbf{Implication 6}. \rev{To jointly address \textit{Performance Efficiency} and \textit{Maintainability}, designers should compare the execution cost and modification effort of alternative agent configurations under anticipated system changes and deployment constraints.}
\end{tcolorbox}

\rev{Among the 106 studies, 51 (48.1\%) consider \textit{Performance Efficiency}, 55 (51.9\%) consider \textit{Maintainability}, and 23 (21.7\%) consider both. The reviewed studies provide concrete design practices that designers can examine when addressing these concerns. For example, \textit{MASAI}~\hyperlink{S32}{[S32]} specifies each sub-agent through its inputs, problem-solving strategy, and outputs, and connects sub-agents by passing their outputs to subsequent sub-agents. \textit{AutoGen}~\hyperlink{S64}{[S64]} provides unified agent interaction interfaces and supports configurable tool execution, termination conditions, and limits on automatic replies. Complementary evidence comes from Wang et al.~\citep{Wang2025AgentPractices}, whose analysis of developer discussions across ten LLM-based agent frameworks identifies both resource-management limitations and maintenance difficulties arising from API changes and dependency conflicts. These findings reinforce the need to consider both execution efficiency and modification effort when selecting agent configurations.}

\rev{Building on these practices, designers can define explicit input and output contracts when tools or agent strategies are expected to change. They should then check whether replacing a tool or revising a strategy requires modifications outside the responsible agent. When latency budgets or API call quotas constrain execution, designers can make stopping conditions and interaction limits configurable, allowing budget adjustments without rewriting each agent's task instructions. For each intended model backend and tool execution environment, designers should assess whether these limits control execution cost while preserving the required level of task success.}

\rev{To evaluate both QAs together, designers should first specify the system modifications they expect to make during maintenance or further development, such as replacing an analysis tool, revising an agent's problem-solving strategy, or adding a verification step. Each specified modification should be implemented separately in every candidate configuration. Designers should evaluate all configurations using a fixed set of SE tasks that the MAS is intended to solve. For a program-repair MAS, these SE tasks could consist of bug reports paired with their corresponding repository versions. The same SE tasks should be run repeatedly before and after each modification in the intended execution environments. Those settings unrelated to the modification should remain unchanged across these runs. The comparison should record task success, latency, token consumption, and model and tool call counts, together with the time required to implement each modification, the agents and coordination rules modified, and the results of regression checks. These measurements support selecting a configuration that satisfies task-success and execution requirements while requiring acceptable implementation effort for the specified system modifications.}

\section{Threats on Validity}\label{sec:Threats}
In this section, we outline the potential threats to the validity of our study. We identify the threats encountered during the research and clarify the measures we employed to mitigate them. Internal validity is not discussed, as our study does not involve any experimental manipulation of variables and thus does not support causal inference.

\textbf{Construct validity}: Since data collection and data extraction in our study were conducted manually, there is a potential risk of individual bias in the data extraction results. To mitigate this risk, a pilot data extraction was conducted before the formal data extraction, which partially alleviated the threats to the construct validity of the study. Furthermore, following both the pilot and formal data extraction, the first author \rev{engaged in} discussions with the second and third authors to ensure that all of the data items could be extracted from our dataset and \rev{that} the data extraction results were aligned with \rev{the} four RQs. \rev{Another potential threat concerns the operationalization of SE tasks in RQ1. SE task categories are not naturally mutually exclusive and collectively exhaustive at the same level of abstraction, as some tasks exhibit hierarchical or functional overlap. For example, \textit{Fault Localization} and \textit{Program Repair} may be investigated independently, while they can also constitute successive activities within \textit{End-to-End Software Maintenance}. To ensure consistent quantitative analysis and avoid double-counting, each study was assigned a single \textit{SE Task} label. Systems covering multiple closely related tasks were categorized under an appropriate higher-level composite task. For studies assigned to a composite task category, the constituent tasks were not counted separately in the distribution of \textit{SE Tasks}.}

\textbf{External validity}: In this study, external validity primarily concerns the selection of data sources. To ensure the credibility of our data, our data collection is based on three recent literature surveys on LLM-based agent systems for SE tasks by \rev{He et al. \citep{He2025survey},} Liu et al. \citep{Liu2024survey}, and Wang et al. \citep{Wang2024agent}. To further enhance the comprehensiveness of the dataset, we additionally included arXiv \citep{arXiv} as a data source, which is an open-access preprint platform maintained by Cornell University and is widely recognized within the academic community. Besides, arXiv features a dedicated ``Software Engineering'' category, which facilitates the identification of papers relevant to our RQs. \rev{A potential threat to external validity arises from the rapid evolution of LLM-based MASs for SE. New studies, design approaches, and design patterns may have emerged since the end of the data collection period. Of the three literature surveys used as data sources in this study, the most recent, conducted by He et al. \citep{He2025survey}, covered studies published up to November 14, 2024. Accordingly, we adopted this date as the cutoff for our literature search and further examined relevant studies from both the literature surveys and arXiv. As a result, the identified design patterns and related findings may not fully represent approaches published after this cutoff date. Although our dataset includes studies addressing diverse SE tasks and adopting different MAS designs, studies published after the cutoff date may introduce additional design patterns or variants. The identified patterns therefore characterize the design practices reported in the included studies, and their applicability to later studies requires further investigation.}

\textbf{Reliability}: To mitigate potential uncertainties associated with the adopted research methodology, we implemented several measures to enhance the reliability of the study. Throughout the processes of data extraction and data analysis, extensive discussions were held among the first, second, and third authors to resolve any internal inconsistencies and to ensure the consistency and accuracy of the results. Moreover, we have made our dataset \citep{dataset} publicly available to enable other researchers to replicate the study and validate our findings.

\section{Conclusions and Future Work}\label{sec:Conclusion}
In this study, we focused on the SE tasks addressed by the specially designed LLM-based MASs, as well as the \textit{quality attributes} considered by the designers of LLM-based MASs to address the SE tasks, the \textit{design patterns} employed to build LLM-based MASs for SE tasks, and the \textit{design rationale} supporting the construction of LLM-based MASs to facilitate SE tasks. We collected 106 papers that met our criteria from three recent surveys on LLM-based agent systems for SE tasks by He et al. \citep{He2025survey}, Liu et al. \citep{Liu2024survey} and Wang et al. \citep{Wang2024agent}, and the SE category of arXiv~\citep{arXiv}. The study results show that: \textit{Code Generation} is the most common SE task addressed by LLM-based MASs, \textit{Functional Suitability} is the QA that is most commonly considered by designers of LLM-based MASs for SE tasks, \textit{Role-Based Cooperation} is the most frequently used design pattern to develop LLM-based MASs to support SE tasks, and \textit{Improving the Quality of Generated Code} is the most common rationale behind the design of LLM-based MASs for SE tasks.

Based on our study results, we provide implications for designing LLM-based MASs for SE tasks. \rev{These implications emphasize defining clear agent responsibilities and functional boundaries, incorporating reflection mechanisms with explicit criteria for checking outputs, and translating human SE practices into agent workflows with defined task handoffs, verification steps, and escalation conditions. The implications also recommend selecting and combining design patterns according to the target SE task and its quality requirements, setting task-specific execution budgets while checking that task outcomes meet the required quality criteria, and assessing \textit{Performance Efficiency} and \textit{Maintainability} of LLM-based MASs jointly. In future work, we will empirically investigate how these design choices influence the quality of LLM-based MASs.}

\section*{Data Availability}
The dataset for this work has been made available at \cite{dataset}.

\begin{acks}
This work has been partially supported by the National Natural Science Foundation of China (NSFC) with Grant No. 92582203 and 62402348.
\end{acks}

\bibliographystyle{ACM-Reference-Format}
\bibliography{ref}

\appendix
\section{Included Studies}
\begin{center} 
\scriptsize
\begin{longtable}{m{0.8cm}m{12.6cm}}
  \caption{List of the included studies in this work\label{tab:included-studies}}\\
  \hline
  \textbf{ID} & \textbf{Author, Publication Title, and Venue} \\
  \hline
  \endfirsthead
  \textbf{ID} & \textbf{Author, Publication Title, and Venue} \\
  \hline
  \endhead
\hypertarget{S1}{[S1]}                     & Sarah Fakhoury, Markus Kuppe, Shuvendu K. Lahiri, Tahina Ramananandro, Nikhil Swamy. \textbf{3DGen: AI-Assisted Generation of Provably Correct Binary Format Parsers}. arXiv preprint arXiv:2404.10362 (2024).  
\\ \hline
\hypertarget{S2}{[S2]}                     & Cheryl Lee, Chunqiu Steven Xia, Longji Yang, Jen-tse Huang, Zhouruixin Zhu, Lingming Zhang, Michael R. Lyu. \textbf{FixAgent: Hierarchical Multi-Agent Framework for Unified Software  Debugging }. arXiv preprint arXiv:2404.17153 (2024).
\\ \hline
\hypertarget{S3}{[S3]}                     & Lyuye Zhang, Kaixuan Li, Kairan Sun, Daoyuan Wu, Ye Liu, Haoye Tian, Yang Liu. \textbf{ACFIX: Guiding LLMs with Mined Common RBAC Practices for Context-Aware Repair of Access Control Vulnerabilities in Smart Contracts}. arXiv preprint arXiv:2403.06838 (2024).
\\ \hline
\hypertarget{S4}{[S4]}                     & Yihao Qin, Shangwen Wang, Yiling Lou, Jinhao Dong, Kaixin Wang, Xiaoling Li, Xiaoguang Mao. \textbf{AGENTFL: Scaling LLM-based Fault Localization to Project-Level Context}. arXiv preprint arXiv:2403.16362 (2024).
\\ \hline
\hypertarget{S5}{[S5]}                     & Minh Huynh Nguyen, Thang Phan Chau, Phong X. Nguyen, Nghi D. Q. Bui. \textbf{AgileCoder: Dynamic Collaborative Agents for Software Development based on Agile Methodology}. arXiv preprint arXiv:2406.11912 (2024).
\\ \hline
\hypertarget{S6}{[S6]}                     & Zeeshan Rasheed, Malik Abdul Sami, Muhammad Waseem, Kai-Kristian Kemell, Xiaofeng Wang, Anh Nguyen, Kari Systä, Pekka Abrahamsson. \textbf{AI-powered Code Review with LLMs: Early Results}. arXiv preprint arXiv:2404.18496 (2024).
\\ \hline
\hypertarget{S7}{[S7]}                     & Bin Lei, Yuchen Li, Qiuwu Chen. \textbf{AutoCoder: Enhancing Code Large Language Model with AIEV-INSTRUCT}. arXiv preprint arXiv:2405.14906 (2024).
\\ \hline
\hypertarget{S8}{[S8]}                     & Yuntong Zhang, Haifeng Ruan, Zhiyu Fan, Abhik Roychoudhury. 2024. \textbf{AutoCodeRover: Autonomous Program Improvement}. In Proceedings of the 33rd ACM SIGSOFT International Symposium on Software Testing and Analysis (ISSTA), ACM, 1592-1604.
\\ \hline
\hypertarget{S9}{[S9]}                     & Michele Tufano, Anisha Agarwal, Jinu Jang, Roshanak Zilouchian Moghaddam, Neel Sundaresan. \textbf{AutoDev: Automated AI-Driven Development}. arXiv preprint arXiv:2403.08299 (2024).
\\ \hline
\hypertarget{S10}{[S10]}                    & Chenyuan Yang, Xuheng Li, Md Rakib Hossain Misu, Jianan Yao, Weidong Cui, Yeyun Gong, Chris Hawblitzel, Shuvendu Lahiri, Jacob R. Lorch, Shuai Lu, Fan Yang, Ziqiao Zhou, Shan Lu. \textbf{AutoVerus: Automated Proof Generation for Rust Code}. arXiv preprint arXiv:2409.13082 (2024).
\\ \hline
\hypertarget{S11}{[S11]}                    &Sanjiban Choudhury, Paloma Sodhi. \textit{Better than Your Teacher: LLM Agents that learn from Privileged AI Feedback}. arXiv preprint arXiv:2410.05434 (2024).
\\ \hline
\hypertarget{S12}{[S12]}                    &Kechi Zhang, Jia Li, Ge Li, Xianjie Shi, Zhi Jin. 2024. \textbf{CodeAgent: Enhancing Code Generation with Tool-Integrated Agent Systems for Real-World Repo-level Coding Challenges}. In Proceedings of the 62nd Annual Meeting of the Association for Computational Linguistics (ACL), ACL, 13643-13658
\\ \hline
\hypertarget{S13}{[S13]}                    &Zeeshan Rasheed, Malik Abdul Sami, Kai-Kristian Kemell, Muhammad Waseem, Mika Saari, Kari Systä, Pekka Abrahamsson. \textbf{CodePori: Large-Scale System for Autonomous Software Development Using Multi-Agent Technology}. arXiv preprint arXiv:2402.01411v2 (2024).
\\ \hline
\hypertarget{S14}{[S14]}                    &Dong Chen, Shaoxin Lin, Muhan Zeng, Daoguang Zan, Jian-Gang Wang, Anton Cheshkov, Jun Sun, Hao Yu, Guoliang Dong, Artem Aliev, Jie Wang, Xiao Cheng, Guangtai Liang, Yuchi Ma, Pan Bian, Tao Xie, Qianxiang Wang. \textbf{CODER: ISSUE RESOLVING WITH MULTI-AGENT AND TASK GRAPHS}. arXiv preprint arXiv:2406.01304 (2024).
\\ \hline
\hypertarget{S15}{[S15]}                    &Daoguang Zan, Ailun Yu, Wei Liu, Dong Chen, Bo Shen, Wei Li, Yafen Yao, Yongshun Gong, Xiaolin Chen, Bei Guan, Zhiguang Yang, Yongji Wang, Qianxiang Wang, Lizhen Cui. \textbf{CodeS: Natural Language to Code Repository via Multi-Layer Sketch}. arXiv preprint arXiv:2403.16443 (2024).
\\ \hline
\hypertarget{S16}{[S16]}                    &Wei Ma, Daoyuan Wu, Yuqiang Sun, Tianwen Wang, Shangqing Liu, Jian Zhang, Yue Xue, Yang Liu. \textbf{Combining Fine-tuning and LLM-based Agents for Intuitive Smart Contract Auditing with Justifications}. arXiv preprint arXiv:2403.16073 (2024).
\\ \hline
\hypertarget{S17}{[S17]}                    &Kexun Zhang, Weiran Yao, Zuxin Liu, Yihao Feng, Zhiwei Liu, Rithesh Murthy, Tian Lan, Lei Li, Renze Lou, Jiacheng Xu, Bo Pang, Yingbo Zhou, Shelby Heinecke, Silvio Savarese, Huan Wang, Caiming Xiong. \textbf{Diversity Empowers Intelligence: Integrating Expertise of Software Engineering Agents}. arXiv preprint arXiv:2408.07060 (2024).
\\ \hline
\hypertarget{S18}{[S18]}                    &Mohammadmehdi Ataei, Hyunmin Cheong, Daniele Grandi, Ye Wang, Nigel Morris, Alexander Tessier. \textbf{Elicitron: An LLM Agent-Based Simulation Framework for Design Requirements Elicitation}. arXiv preprint arXiv:2404.16045 (2024).
\\ \hline
\hypertarget{S19}{[S19]}                    &Md Nakhla Rafi, Dong Jae Kim, Tse-Hsun Chen, Shaowei Wang. \textbf{Enhancing Fault Localization Through Ordered Code Analysis with LLM Agents and Self-Reflection}. arXiv preprint arXiv:2409.13642 (2024).
\\ \hline
\hypertarget{S20}{[S20]}                    &Simiao Zhang, Jiaping Wang, Guoliang Dong, Jun Sun, Yueling Zhang, Geguang Pu. \textbf{Experimenting a New Programming Practice with LLMs}. arXiv preprint arXiv:2401.01062 (2024).
\\ \hline
\hypertarget{S21}{[S21]}                    &Malik Abdul Sami, Muhammad Waseem, Zeeshan Rasheed, Mika Saari, Kari Systä, Pekka Abrahamsson. \textbf{Experimenting with Multi-Agent Software Development: Towards a Unified Platform}. arXiv preprint arXiv:2406.05381 (2024).
\\ \hline
\hypertarget{S22}{[S22]}                    &Arsham Gholamzadeh Khoee, Yinan Yu, Robert Feldt, Andris Freimanis, Patrick Andersson Rhodin, Dhasarathy Parthasarathy. \textbf{GoNoGo: An Efficient LLM-based Multi-Agent System for Streamlining Automotive Software Release Decision-Making}. arXiv preprint arXiv:2408.09785 (2024).
\\ \hline
\hypertarget{S23}{[S23]}                    &Yingwei Ma, Qingping Yang, Rongyu Cao, Binhua Li, Fei Huang, Yongbin Li. \textbf{How to Understand Whole Software Repository?}. arXiv preprint arXiv:2406.01422 (2024).
\\ \hline
\hypertarget{S24}{[S24]}                    &Chen Qian, Jiahao Li, Yufan Dang, Wei Liu, YiFei Wang, Zihao Xie, Weize Chen, Cheng Yang, Yingli Zhang, Zhiyuan Liu, Maosong Sun. \textbf{Iterative Experience Refinement of Software-Developing Agents}. arXiv preprint arXiv:2405.04219 (2024).
\\ \hline
\hypertarget{S25}{[S25]}                    &Richard Fang, Rohan Bindu, Akul Gupta, Daniel Kang. \textbf{LLM Agents can Autonomously Exploit One-day Vulnerabilities}. arXiv preprint arXiv:2404.08144 (2024).
\\ \hline
\hypertarget{S26}{[S26]}                    &Zhiyuan Wei, Jing Sun, Zijiang Zhang, Xianhao Zhang. \textbf{LLM-SmartAudit: Advanced Smart Contract Vulnerability Detection}. arXiv preprint arXiv:2410.09381 (2024).
\\ \hline
\hypertarget{S27}{[S27]}                    &Mohamad Fakih, Rahul Dharmaji, Yasamin Moghaddas, Gustavo Quiros Araya, Oluwatosin Ogundare, Mohammad Abdullah Al Faruque. 2024. \textbf{LLM4PLC: Harnessing Large Language Models for Verifiable Programming of PLCs in Industrial Control Systems}. In Proceedings of the 46th International Conference on Software Engineering: Software Engineering in Practice (ICSE-SEIP), 192-203.
\\ \hline
\hypertarget{S28}{[S28]}                    &Wei Tao, Yucheng Zhou, Yanlin Wang, Wenqiang Zhang, Hongyu Zhang, Yu Cheng. \textbf{MAGIS: LLM-Based Multi-Agent Framework for GitHub Issue ReSolution}. arXiv preprint arXiv:2403.17927 (2024).
\\ \hline
\hypertarget{S29}{[S29]}                    &Md. Ashraful Islam, Mohammed Eunus Ali, Md Rizwan Parvez. \textbf{MapCoder: Multi-Agent Code Generation for Competitive Problem Solving}. 2024. In Proceedings of the 62nd Annual Meeting of the Association for Computational Linguistics (ACL), 4912-4944.
\\ \hline
\hypertarget{S30}{[S30]}                    &Dongming Jin, Zhi Jin, Xiaohong Chen, Chunhui Wang. \textbf{MARE: Multi-Agents Collaboration Framework for Requirements Engineering}. arXiv preprint arXiv:2405.03256 (2024).
\\ \hline
\hypertarget{S31}{[S31]}                    &Yizhou Liu, Pengfei Gao, Xinchen Wang, Jie Liu, Yexuan Shi, Zhao Zhang, Chao Peng. \textbf{MarsCode Agent: AI-native Automated Bug Fixing}. arXiv preprint arXiv:2409.00899 (2024).
\\ \hline
\hypertarget{S32}{[S32]}                    &Daman Arora, Atharv Sonwane, Nalin Wadhwa, Abhav Mehrotra, Saiteja Utpala, Ramakrishna Bairi, Aditya Kanade, Nagarajan Natarajan. \textbf{MASAI: Modular Architecture for Software-engineering AI Agents}. arXiv preprint arXiv:2406.11638 (2024).
\\ \hline
\hypertarget{S33}{[S33]}                    &Junyou Li, Qin Zhang, Yangbin Yu, Qiang Fu, Deheng Ye. \textbf{More Agents Is All You Need}. arXiv preprint arXiv:2402.05120 (2024).
\\ \hline
\hypertarget{S34}{[S34]}                    &Zhuoyun Du, Chen Qian, Wei Liu, Zihao Xie, Yifei Wang, Yufan Dang, Weize Chen, Cheng Yang. \textbf{Multi-Agent Software Development through Cross-Team Collaboration}. arXiv preprint arXiv:2406.08979 (2024).
\\ \hline
\hypertarget{S35}{[S35]}                    &Zhenyu Mao, Jialong Li, Dongming Jin, Munan Li, Kenji Tei. \textbf{Multi-role Consensus through LLMs Discussions for Vulnerability Detection}. arXiv preprint arXiv:2403.14274 (2024).
\\ \hline
\hypertarget{S36}{[S36]}                    &Haolin Jin, Zechao Sun, Huaming Chen. \textbf{RGD: Multi-LLM Based Agent Debugger via Refinement and Generation Guidance}. arXiv preprint arXiv:2410.01242 (2024).
\\ \hline
\hypertarget{S37}{[S37]}                    &Chen Qian, Zihao Xie, Yifei Wang, Wei Liu, Yufan Dang, Zhuoyun Du, Weize Chen, Cheng Yang, Zhiyuan Liu, Maosong Sun. \textbf{Scaling Large-Language-Model-based Multi-Agent Collaboration}. arXiv preprint arXiv:2406.07155 (2024).
\\ \hline
\hypertarget{S38}{[S38]}                    &Yoichi Ishibashi, Yoshimasa Nishimura. \textbf{Self-Organized Agents: A LLM Multi-Agent Framework toward Ultra Large-Scale Code Generation and Optimization}. arXiv preprint arXiv:2404.02183 (2024).
\\ \hline
\hypertarget{S39}{[S39]}                    &Haifeng Ruan, Yuntong Zhang, Abhik Roychoudhury. \textbf{SpecRover: Code Intent Extraction via LLMs}. arXiv preprint arXiv:2408.02232 (2024).
\\ \hline
\hypertarget{S40}{[S40]}                    &John Yang, Carlos E. Jimenez, Alexander Wettig, Kilian Lieret, Shunyu Yao, Karthik Narasimhan, Ofir Press. \textbf{SWE-agent: Agent-Computer Interfaces Enable Automated Software Engineering}. arXiv preprint arXiv:2405.15793 (2024).
\\ \hline
\hypertarget{S41}{[S41]}                    &Noble Saji Mathews, Meiyappan Nagappan. \textbf{Test-Driven Development for Code Generation}. arXiv preprint arXiv:2402.13521 (2024).
\\ \hline
\hypertarget{S42}{[S42]}                    &Shubham Gandhi, Manasi Patwardhan, Lovekesh Vig, Gautam Shroff. \textbf{BudgetMLAgent: A Cost-Effective LLM Multi-Agent system for Automating Machine Learning Tasks}. 2025. In Proceedings of the 4th International Conference on AI-ML Systems (AIMLSystems’24). 1-9.
\\ \hline
\hypertarget{S43}{[S43]}                    &Zhitao Wang, Wei Wang, Zirao Li, Long Wang, Can Yi, Xinjie Xu, Luyang Cao, Hanjing Su, Shouzhi Chen, Jun Zhou. \textbf{XUAT-Copilot: Multi-Agent Collaborative System for Automated User Acceptance Testing with Large Language Model}. arXiv preprint arXiv:2401.02705 (2024).
\\ \hline
\hypertarget{S44}{[S44]}                    &Xunzhu Tang, Kisub Kim, Yewei Song, Cedric Lothritz, Bei Li, Saad Ezzini, Haoye Tian, Jacques Klein, Tegawende F. Bissyande. 2024. \textbf{CodeAgent: Autonomous Communicative Agents for Code Review}. In Proceedings of the 2024 Conference on Empirical Methods in Natural Language Processing (EMNLP), 11279-11313.
\\ \hline
\hypertarget{S45}{[S45]}                    &Dawen Zhang, Xiwei Xu, Chen Wang, Zhenchang Xing, Robert Mao. \textbf{A Layered Architecture for Developing and Enhancing Capabilities in Large Language Model-based Software Systems}. arXiv preprint arXiv:2411.12357 (2024).
\\ \hline
\hypertarget{S46}{[S46]}                    &Sai Zhang, Zhenchang Xing, Ronghui Guo, Fangzhou Xu, Lei Chen, Zhaoyuan Zhang, Xiaowang Zhang, Zhiyong Feng, Zhiqiang Zhuang. 2025. \textbf{Empowering Agile-Based Generative Software Development through Human-AI Teamwork}. ACM Transactions on Software Engineering and Methodology, Volume 34, Issue 6, 1-46.
\\ \hline
\hypertarget{S47}{[S47]}                    &Juyeon Yoon; Robert Feldt; Shin Yoo. \textbf{Intent-Driven Mobile GUI Testing with Autonomous Large Language Model Agents}. 2024. In Proceedings of the 17th IEEE International Conference on Software Testing, Verification \& Validation (ICST), 129-139.
\\ \hline
\hypertarget{S48}{[S48]}                    &Lu Shi, Bin Qi, Jiarui Luo, Yang Zhang, Zhanzhao Liang, Zhaowei Gao, Wenke Deng, Lin Sun. \textbf{Aegis:An Advanced LLM-Based Multi-Agent for Intelligent Functional Safety Engineering}. 2024. In Proceedings of the 2024 Conference on Empirical Methods in Natural Language Processing (EMNLP), 1571-1583.
\\ \hline
\hypertarget{S49}{[S49]}                    &Zhe Liu, Cheng Li, Chunyang Chen, Junjie Wang, Boyu Wu, Yawen Wang, Jun Hu, Qing Wang. \textbf{Vision-driven Automated Mobile GUI Testing via Multimodal Large Language Model}. arXiv preprint arXiv:2407.03037 (2024).
\\ \hline
\hypertarget{S50}{[S50]}                    &Jiahong Xiang, Xiaoyang Xu, Fanchu Kong, Mingyuan Wu, Zizheng Zhang, Haotian Zhang, Yuqun Zhang. \textbf{How Far Can We Go with Practical Function-Level Program Repair?}. arXiv preprint arXiv:2404.12833 (2024).
\\ \hline
\hypertarget{S51}{[S51]}                    &Zixiao Zhao, Jing Sun, Zhiyuan Wei, Cheng-Hao Cai, Zhe Hou, Jin Song Dong. \textbf{VisionCoder: Empowering Multi-Agent Auto-Programming for Image Processing with Hybrid LLMs}. arXiv preprint arXiv:2410.19245 (2024).
\\ \hline
\hypertarget{S52}{[S52]}                    &Ahmed R. Sadik, Sebastian Brulin, Markus Olhofer, Antonello Ceravola, Frank Joublin. \textbf{LLM as a code generator in Agile Model Driven Development}. arXiv preprint arXiv:2410.18489 (2024).
\\ \hline
\hypertarget{S53}{[S53]}                    &Yue Hu, Yuzhu Cai, Yaxin Du, Xinyu Zhu, Xiangrui Liu, Zijie Yu, Yuchen Hou, Shuo Tang, Siheng Chen. \textbf{Self-Evolving Multi-Agent Collaboration Networks for Software Development}. arXiv preprint arXiv:2410.16946 (2024).
\\ \hline
\hypertarget{S54}{[S54]}                    &Zihan Liu, Ruinan Zeng, Dongxia Wang, Gengyun Peng, Jingyi Wang, Qiang Liu, Peiyu Liu, Wenhai Wang. \textbf{Agents4PLC: Automating Closed-loop PLC Code Generation and Verification in Industrial Control Systems using LLM-based Agents}. arXiv preprint arXiv:2410.14209 (2024).
\\ \hline
\hypertarget{S55}{[S55]}                    &Xuanming Zhang, Yuxuan Chen, Yuan Yuan, Minlie Huang. \textbf{Seeker: Enhancing Exception Handling in Code with LLM-based Multi-Agent Approach}. arXiv preprint arXiv:2410.06949 (2024).
\\ \hline
\hypertarget{S56}{[S56]}                    &Zhiqiang Yuan, Weitong Chen, Hanlin Wang, Kai Yu, Xin Peng, Yiling Lou. \textbf{TRANSAGENT: An LLM-Based Multi-Agent System for Code Translation}. arXiv preprint arXiv:2409.19894 (2024).
\\ \hline
\hypertarget{S57}{[S57]}                    &Leilei Lin, Yingming Zhou, Wenlong Chen, Chen Qian. \textbf{Think-on-Process: Dynamic Process Generation for Collaborative Development of Multi-Agent System}. arXiv preprint arXiv:2409.06568 (2024).
\\ \hline
\hypertarget{S58}{[S58]}                    &Huan Zhang, Wei Cheng, Yuhan Wu, Wei Hu. \textbf{A Pair Programming Framework for Code Generation via Multi-Plan Exploration and Feedback-Driven Refinement}. 2024. In Proceedings of the 39th IEEE/ACM International Conference on Automated Software Engineering (ASE), 1319-1331.
\\ \hline
\hypertarget{S59}{[S59]}                    &Weiqing Yang, Hanbin Wang, Zhenghao Liu, Xinze Li, Yukun Yan, Shuo Wang, Yu Gu, Minghe Yu, Zhiyuan Liu, Ge Yu. \textbf{Enhancing the Code Debugging Ability of LLMs via Communicative Agent Based Data Refinement}. arXiv preprint arXiv:2408.05006 (2024).
\\ \hline
\hypertarget{S60}{[S60]}                    &Forough Mehralian, Titus Barik, Jeff Nichols, Amanda Swearngin. \textbf{Automated Code Fix Suggestions for Accessibility Issues in Mobile Apps}. arXiv preprint arXiv:2408.03827 (2024). 
\\ \hline
\hypertarget{S61}{[S61]}                    &Dong Huang, Jie M.Zhang, Michael Luck, Qingwen Bu, Yuhao Qing, Heming Cui. \textbf{AgentCoder: Multi-Agent Code Generation with Effective Testing and Self-optimisation}. arXiv preprint arXiv:2312.13010 (2023).
\\ \hline
\hypertarget{S62}{[S62]}                    &Weize Chen, Yusheng Su, Jingwei Zuo, Cheng Yang, Chenfei Yuan, Chi-Min Chan, Heyang Yu, Yaxi Lu, Yi-Hsin Hung, Chen Qian, Yujia Qin, Xin Cong, Ruobing Xie, Zhiyuan Liu, Maosong Sun, Jie Zhou. 2024. \textbf{AgentVerse: Facilitating Multi-Agent Collaboration and Exploring Emergent Behaviors}. In Proceedings of 12th International Conference on Learning Representations (ICLR), 1-43.
\\ \hline
\hypertarget{S63}{[S63]}                    &Guangyao Chen, Siwei Dong, Yu Shu, Ge Zhang, Jaward Sesay, Börje F. Karlsson, Jie Fu, Yemin Shi. \textbf{AutoAgents: A Framework for Automatic Agent Generation}. arXiv preprint arXiv:2309.17288 (2023).
\\ \hline
\hypertarget{S64}{[S64]}                    &Qingyun Wu, Gagan Bansal, Jieyu Zhang, Yiran Wu, Beibin Li, Erkang Zhu, Li Jiang, Xiaoyun Zhang, Shaokun Zhang, Jiale Liu, Ahmed Hassan Awadallah, Ryen W White, Doug Burger, Chi Wang. \textbf{AutoGen: Enabling Next-Gen LLM Applications via Multi-Agent Conversation}. arXiv preprint arXiv:2308.08155 (2023).
\\ \hline
\hypertarget{S65}{[S65]}                    &Zeeshan Rasheed, Muhammad Waseem, Kai-Kristian Kemell, Wang Xiaofeng, Anh Nguyen Duc, Kari Systä, Pekka Abrahamsson. \textbf{Autonomous Agents in Software Development: A Vision Paper}. arXiv preprint arXiv:2311.18440 (2023).
\\ \hline
\hypertarget{S66}{[S66]}                    &Maryam Taeb, Amanda Swearngin, Eldon Schoop, Ruijia Cheng, Yue Jiang, Jeffrey Nichols. 2024. \textbf{AXNav: Replaying Accessibility Tests from Natural Language}. In Proceedings of the 2024 CHI Conference on Human Factors in Computing Systems (CHI), 1-16.
\\ \hline
\hypertarget{S67}{[S67]}                    & Guohao Li, Hasan Abed Al Kader Hammoud, Hani Itani, Dmitrii Khizbullin, Bernard Ghanem. 2023. \textbf{CAMEL: communicative agents for ``mind'' exploration of large language model society}. In Proceedings of the 37th International Conference on Neural Information Processing Systems (NeurIPS), 51991-52008.
\\ \hline
\hypertarget{S68}{[S68]}                    &Guang Yang, Yu Zhou, Xiang Chen, Xiangyu Zhang, Terry Yue Zhuo, Taolue Chen. 2024. \textbf{Chain-of-Thought in Neural Code Generation: From and For Lightweight Language Models}. IEEE Transactions on Software Engineering, Volume 50, Issue 9, 2437-2457.
\\ \hline
\hypertarget{S69}{[S69]}                    &Seungjun Moon, Hyungjoo Chae, Yongho Song, Taeyoon Kwon, Dongjin Kang, Kai Tzu-iunn Ong, Seung-won Hwang, Jinyoung Yeo. \textbf{Coffee: Boost Your Code LLMs by Fixing Bugs with Feedback}. arXiv preprint arXiv:2311.07215 (2023).
\\ \hline
\hypertarget{S70}{[S70]}                    &Nalin Wadhwa, Jui Pradhan, Atharv Sonwane, Surya Prakash Sahu, Nagarajan Natarajan, Aditya Kanade, Suresh Parthasarathy, Sriram Rajamani. 2023. \textbf{CORE: Resolving Code Quality Issues using LLMs}. In Proceedings of the 31th ACM on Software Engineering (FSE), 789-811.
\\ \hline
\hypertarget{S71}{[S71]}                    &Zijun Liu, Yanzhe Zhang, Peng Li, Yang Liu, Diyi Yang. \textbf{A Dynamic LLM-Agent Network: An LLM-agent Collaboration Framework with Agent Team Optimization}. arXiv preprint arXiv:2310.02170 (2023).
\\ \hline
\hypertarget{S72}{[S72]}                    &Yu Hao, Weiteng Chen, Ziqiao Zhou, Weidong Cui. \textbf{E\&V: Prompting Large Language Models to Perform Static Analysis by Pseudo-code Execution and Verification}. arXiv preprint arXiv:2312.08477 (2023).
\\ \hline
\hypertarget{S73}{[S73]}                    &Chen Qian, Yufan Dang, Jiahao Li, Wei Liu, Zihao Xie, Yifei Wang, Weize Chen, Cheng Yang, Xin Cong, Xiaoyin Che, Zhiyuan Liu, Maosong Sun. 2024. \textbf{Experiential Co-Learning of Software-Developing Agents}. In Proceedings of the 62nd Annual Meeting of the Association for Computational Linguistics (ACL), 5628-5640.
\\ \hline
\hypertarget{S74}{[S74]}                    &Martin Josifoski, Lars Klein, Maxime Peyrard, Nicolas Baldwin, Yifei Li, Saibo Geng, Julian Paul Schnitzler, Yuxing Yao, Jiheng Wei, Debjit Paul, Robert West. \textbf{Flows: Building Blocks of Reasoning and Collaborating AI}. arXiv preprint arXiv:2308.01285 (2023).
\\ \hline
\hypertarget{S75}{[S75]}                    &Chunqiu Steven Xia, Matteo Paltenghi, Jia Le Tian, Michael Pradel, Lingming Zhang. 2024. \textbf{Fuzz4All: Universal Fuzzing with Large Language Models}. In Proceedings of the IEEE/ACM 46th International Conference on Software Engineering (ICSE), 1-13.
\\ \hline
\hypertarget{S76}{[S76]}                    &Binfeng Xu, Xukun Liu, Hua Shen, Zeyu Han, Yuhan Li, Murong Yue, Zhiyuan Peng, Yuchen Liu, Ziyu Yao, Dongkuan Xu. 2023. \textbf{Gentopia.AI: A Collaborative Platform for Tool-Augmented LLMs}. 
In Proceedings of the 2023 Conference on Empirical Methods in Natural Language Processing: System Demonstrations (EMNLP), 237-245.
\\ \hline
\hypertarget{S77}{[S77]}                    &Zhou Yang, Zhipeng Zhao, Chenyu Wang, Jieke Shi, Dongsum Kim, Donggyun Han, David Lo. \textbf{Gotcha! This Model Uses My Code! Evaluating Membership Leakage Risks in Code Models}. arXiv preprint arXiv:2310.01166 (2023).
\\ \hline
\hypertarget{S78}{[S78]}                    &Hanbin Wang, Zhenghao Liu, Shuo Wang, Ganqu Cui, Ning Ding, Zhiyuan Liu, Ge Yu. 2024. \textbf{INTERVENOR: Prompting the Coding Ability of Large Language Models with the Interactive Chain of Repair}. In Proceedings of the 62nd Findings of the Association for Computational Linguistics (ACL), 2081-2107.
\\ \hline
\hypertarget{S79}{[S79]}                    &Yiheng Xu, Hongjin Su, Chen Xing, Boyu Mi, Qian Liu, Weijia Shi, Binyuan Hui, Fan Zhou, Yitao Liu, Tianbao Xie, Zhoujun Cheng, Siheng Zhao, Lingpeng Kong, Bailin Wang, Caiming Xiong, Tao Yu. 2024. \textbf{Lemur: Harmonizing Natural Language and Code for Language Agents}. In Proceedings of the 12th International Conference on Representation Learning (ICLR), 1-21.
\\ \hline
\hypertarget{S80}{[S80]}                    &Zhe Liu, Chunyang Chen, Junjie Wang, Mengzhuo Chen, Boyu Wu, Xing Che, Dandan Wang, Qing Wang. 2024. \textbf{Make LLM a Testing Expert: Bringing Human-like Interaction to Mobile GUI Testing via Functionality-aware Decisions}. In Proceedings of the IEEE/ACM 46th International Conference on Software Engineering (ICSE), 1-13.
\\ \hline
\hypertarget{S81}{[S81]}                    &Sirui Hong, Mingchen Zhuge, Jonathan Chen, Xiawu Zheng, Yuheng Cheng, Ceyao Zhang, Jinlin Wang, Zili Wang, Steven Ka Shing Yau, Zijuan Lin, Liyang Zhou, Chenyu Ran, Lingfeng Xiao, Chenglin Wu, Jürgen Schmidhuber. \textbf{MetaGPT: Meta Programming for A Multi-Agent Collaborative Framework}. arXiv preprint arXiv:2308.00352 (2023).
\\ \hline
\hypertarget{S82}{[S82]}                    &Yashar Talebirad, Amirhossein Nadiri. \textbf{Multi-Agent Collaboration: Harnessing the Power of Intelligent LLM Agents}. arXiv preprint arXiv:2306.03314 (2023).
\\ \hline
\hypertarget{S83}{[S83]}                    &Eric Zelikman, Qian Huang, Gabriel Poesia, Noah D. Goodman, Nick Haber. 2023. \textbf{Parsel: Algorithmic Reasoning with Language Models by Composing Decompositions}. In Proceedings of 37th Conference on Neural Information Processing Systems (NeurIPS).
\\ \hline
\hypertarget{S84}{[S84]}                    &Zhenchang Xing, Qing Huang, Yu Cheng, Liming Zhu, Qinghua Lu, Xiwei Xu. \textbf{Prompt Sapper: LLM-Empowered Software Engineering Infrastructure for AI-Native Services}. arXiv preprint arXiv:2306.02230 (2023).
\\ \hline
\hypertarget{S85}{[S85]}                    &Zefan Wang, Zichuan Liu, Yingying Zhang, Aoxiao Zhong, Jihong Wang, Fengbin Yin, Lunting Fan, Lingfei Wu, Qingsong Wen. \textbf{RCAgent: Cloud Root Cause Analysis by Autonomous Agents with Tool-Augmented Large Language Models}. arXiv preprint arXiv:2310.16340 (2023).
\\ \hline
\hypertarget{S86}{[S86]}                    &Yihong Dong, Xue Jiang, Zhi Jin, Ge Li. \textbf{Self-collaboration Code Generation via ChatGPT}. arXiv preprint arXiv:2304.07590 (2023).
\\ \hline
\hypertarget{S87}{[S87]}                    &Kechi Zhang, Zhuo Li, Jia Li, Ge Li, Zhi Jin. 2023. \textbf{Self-Edit: Fault-Aware Code Editor for Code Generation}. In Proceedings of the 61st Annual Meeting of the Association for Computational Linguistics (ACL), 769-787.
\\ \hline
\hypertarget{S88}{[S88]}                    &Gang Fan, Xiaoheng Xie, Xunjin Zheng, Yinan Liang, Peng Di. \textbf{Static Code Analysis in the AI Era: An In-depth Exploration of the Concept, Function, and Potential of Intelligent Code Analysis Agents}. arXiv preprint arXiv:2310.08837 (2023).
\\ \hline
\hypertarget{S89}{[S89]}                    &Carlos E. Jimenez, John Yang, Alexander Wettig, Shunyu Yao, Kexin Pei, Ofir Press, Karthik Narasimhan. 2024. \textbf{Swe-bench: Can language models resolve real-world GitHub issues?}. In Proceedings of the 12th International Conference on Representation Learning (ICLR), 1-51.
\\ \hline
\hypertarget{S90}{[S90]}                    &Xinyun Chen, Maxwell Lin, Nathanael Schärli, Denny Zhou. 2024. \textbf{Teaching Large Language Models to Self-Debug}. In Proceedings of the 12th International Conference on Representation Learning (ICLR), 1-80.
\\ \hline
\hypertarget{S91}{[S91]}                    &Chenyuan Yang, Yinlin Deng, Runyu Lu, Jiayi Yao, Jiawei Liu, Reyhaneh Jabbarvand, Lingming Zhang. \textbf{White-box Compiler Fuzzing Empowered by Large Language Models}. arXiv preprint arXiv:2310.15991 (2023).
\\ \hline
\hypertarget{S92}{[S92]}                    &Theo X. Olausson, Jeevana Priya Inala, Chenglong Wang, Jianfeng Gao, Armando Solar-Lezama. 2024. \textbf{Is Self-Repair a Silver Bullet for Code Generation?}. In Proceedings of the 12th International Conference on Representation Learning (ICLR), 1-49.
\\ \hline
\hypertarget{S93}{[S93]}                    &Chen Qian, Wei Liu, Hongzhang Liu, Nuo Chen, Yufan Dang, Jiahao Li, Cheng Yang, Weize Chen, Yusheng Su, Xin Cong, Juyuan Xu, Dahai Li, Zhiyuan Liu, Maosong Sun. 2024. \textbf{ChatDev: Communicative Agents for Software Development}. In Proceedings of the 62nd Annual Meeting of the Association for Computational Linguistics (ACL), 15174-15186.
\\ \hline
\hypertarget{S94}{[S94]}                    &Romal Thoppilan, Daniel De Freitas, Jamie Hall, Noam Shazeer, Apoorv Kulshreshtha, Heng-Tze Cheng, Alicia Jin, Taylor Bos, Leslie Baker, Yu Du, YaGuang Li, Hongrae Lee, Huaixiu Steven Zheng, Amin Ghafouri, Marcelo Menegali, Yanping Huang, Maxim Krikun, Dmitry Lepikhin, James Qin, Dehao Chen, Yuanzhong Xu, Zhifeng Chen, Adam Roberts, Maarten Bosma, Vincent Zhao, Yanqi Zhou, Chung-Ching Chang, Igor Krivokon, Will Rusch, Marc Pickett, Pranesh Srinivasan, Laichee Man, Kathleen Meier-Hellstern, Meredith Ringel Morris, Tulsee Doshi, Renelito Delos Santos, Toju Duke, Johnny Soraker, Ben Zevenbergen, Vinodkumar Prabhakaran, Mark Diaz, Ben Hutchinson, Kristen Olson, Alejandra Molina, Erin Hoffman-John, Josh Lee, Lora Aroyo, Ravi Rajakumar, Alena Butryna, Matthew Lamm, Viktoriya Kuzmina, Joe Fenton, Aaron Cohen, Rachel Bernstein, Ray Kurzweil, Blaise Aguera-Arcas, Claire Cui, Marian Croak, Ed Chi, Quoc Le. \textbf{LaMDA: Language Models for Dialog Applications}. arXiv preprint arXiv:2201.08239 (2022).
\\ \hline

\rev{\hypertarget{S95}{[S95]}}   &\rev{Malik Abdul Sami, Muhammad Waseem, Zheying Zhang, Zeeshan Rasheed, Kari Systä, Pekka Abrahamsson. \textbf{AI based Multiagent Approach for Requirements Elicitation and Analysis}. arXiv preprint arXiv:2409.00038 (2024).}
\\ \hline
\rev{\hypertarget{S96}{[S96]}}   &\rev{Jierui Li, Hung Le, Yingbo Zhou, Caiming Xiong, Silvio Savarese, Doyen Sahoo. \textbf{CodeTree: Agent-guided Tree Search for Code Generation with Large Language Models}. arXiv preprint arXiv:2411.04329 (2024).}
\\ \hline
\rev{\hypertarget{S97}{[S97]}}   &\rev{Huy Nhat Phan, Tien N. Nguyen, Phong X. Nguyen, Nghi D. Q. Bui. \textbf{HyperAgent: Generalist Software Engineering Agents to Solve Coding Tasks at Scale}. arXiv preprint arXiv:2409.16299 (2024).}
\\ \hline
\rev{\hypertarget{S98}{[S98]}}   &\rev{Hung Le, Yingbo Zhou, Caiming Xiong, Silvio Savarese, Doyen Sahoo. 2024. \textbf{INDICT: code generation with internal dialogues of critiques for both security and helpfulness}. In Proceedings of the 38th International Conference on Neural Information Processing Systems (NeurIPS), 85546-85582.}
\\ \hline
\rev{\hypertarget{S99}{[S99]}}   &\rev{Feng Lin, Dong Jae Kim, Tse-Hsun Chen. 2025. \textbf{SOEN-101: Code Generation by Emulating Software Process Models Using Large Language Model Agents}. In Proceedings of the 47th IEEE/ACM International Conference on Software Engineering (ICSE), 1527-1539.}
\\ \hline
\rev{\hypertarget{S100}{[S100]}}   &\rev{Noah Shinn, Federico Cassano, Ashwin Gopinath, Karthik Narasimhan, Shunyu Yao. 2023. \textbf{Reflexion: Language agents with verbal reinforcement learning}. In Proceedings of the 37th International Conference on Neural Information Processing Systems (NeurIPS), 8634-8652.}
\\ \hline
\rev{\hypertarget{S101}{[S101]}}   &\rev{Xiangyan Liu, Bo Lan, Zhiyuan Hu, Yang Liu, Zhicheng Zhang, Fei Wang, Michael Shieh, Wenmeng Zhou. 2025. \textbf{CodexGraph: Bridging Large Language Models and Code Repositories via Code Graph Databases}. In Proceedings of the 63rd Conference of the Nations of the Americas Chapter of the Association for Computational Linguistics (ACL), 142-160.}
\\ \hline
\rev{\hypertarget{S102}{[S102]}}   &\rev{Gelei Deng, Yi Liu, Víctor Mayoral-Vilches, Peng Liu, Yuekang Li, Yuan Xu, Tianwei Zhang, Yang Liu, Martin Pinzger, Stefan Rass. 2024. \textbf{PENTESTGPT: evaluating and harnessing large language models for automated penetration testing}. In Proceedings of the 33rd USENIX Conference on Security Symposium (USENIX Security), 847-864.}
\\ \hline
\rev{\hypertarget{S103}{[S103]}}   &\rev{Sihao Hu, Tiansheng Huang, Fatih İlhan, Selim Furkan Tekin, Ling Liu. 2023. \textbf{Large Language Model-Powered Smart Contract Vulnerability Detection: New Perspectives}. In Proceedings of the 5th IEEE International Conference on Trust, Privacy and Security in Intelligent Systems and Applications (TPS-ISA), 297–306.}
\\ \hline
\rev{\hypertarget{S104}{[S104]}}   &\rev{Bin Lei, Yuchen Li, Yiming Zeng, Tao Ren, Yi Luo, Tianyu Shi, Zitian Gao, Zeyu Hu, Weitai Kang, Qiuwu Chen. \textbf{Infant Agent: A Tool-Integrated, Logic-Driven Agent with Cost-Effective API Usage}. arXiv preprint arXiv:2411.01114 (2024).}
\\ \hline
\rev{\hypertarget{S105}{[S105]}}   &\rev{Samuel Holt, Max Ruiz Luyten, Mihaela van der Schaar. \textbf{L2MAC: Large Language Model Automatic Computer for Extensive Code Generation}. 2024. In Proceedings of the 12th International Conference on Learning Representations (ICLR), 1-61.}
\\ \hline
\rev{\hypertarget{S106}{[S106]}}   &\rev{Qian Wang, Tianyu Wang, Zhenheng Tang, Qinbin Li, Nuo Chen, Jingsheng Liang, Bingsheng He. 2025. \textbf{MegaAgent: A Large-Scale Autonomous LLM-based Multi-Agent System Without Predefined SOPs}. In Proceedings of the 63rd Findings of the Association for Computational Linguistics (ACL), 4998-5036.}
\\ \hline
  \end{longtable}
\end{center}

\end{document}